\documentclass[11pt]{article}
\pdfoutput=1
\usepackage{jcapmod}

\usepackage{amsfonts, amsmath}
\usepackage{booktabs}
\usepackage[english]{babel}
\usepackage{colortbl}
\usepackage{graphicx}
\usepackage{hyperref}
  \usepackage{framed}
\usepackage{simplewick}
\usepackage{siunitx}

\RequirePackage{color}

\usepackage{colortbl}
\definecolor{rp}{cmyk}{0.2, 1, 0.6, 0}
\definecolor{green2}{cmyk}{0, 1, 0.5, 0}
\definecolor{lightgreen}{cmyk}{0.2, 0, 0.2, 0.2}
\definecolor{lightgray}{cmyk}{0.1,0.2,0,0.1}
\definecolor{lightgray2}{cmyk}{0.4,0.4,0,0.8}
\definecolor{black}{cmyk}{1.0,1.0,1.0,1.0}

\allowdisplaybreaks[1]

\usepackage{colortbl}
\definecolor{lightgreen}{cmyk}{0.2, 0, 0.2, 0.2}
\definecolor{lightgray}{cmyk}{0.1,0.2,0,0.1}
\definecolor{lightgray2}{cmyk}{0.1,0.1,0,0.1}

\definecolor{NewRed}{RGB}{200,37,6}
\definecolor{NewOrange}{RGB}{222,106,16}
\definecolor{NewGreen}{RGB}{0,136,43}

\makeatletter
\newlength{\apb@width}
\newcommand{\autoparbox}[2][c]{\settowidth{\apb@width}{#2}\parbox[#1]{\apb@width}{#2}}
\newcommand{\includegraphicsbox}[2][]{\autoparbox{\includegraphics[#1]{#2}}}
\makeatother

\numberwithin{equation}{section}

\newcommand{\h}{\hskip 1pt}

\def\beq{\begin{equation}}
\def\eeq{\end{equation}}
\def\bea{\begin{eqnarray}}
\def\eea{\end{eqnarray}}

\def\ct{\tau}

\def\d{{\rm d}}

\def\d{{\rm d}}

\def\Mp{M_{\rm pl}}
\def\d{{\rm d}}

\def\H{{\cal H}}

\def\0{{\boldsymbol 0}}

\def\p{{\bf p}}
\def\v{{\boldsymbol{v}}}
\def\x{{{\bf x}}}

\def\e{{\hat{\bf p}}}
\def\n{{\bf n}}

\def\k{{\boldsymbol{k}}}
\def\p{{\bf{p}}}
\def\v{{\bf{v}}}

\def\x{{\bf{x}}}

\def\d{{\rm d}}
\def\Mp{M_{\rm pl}}

\def\k{{\bf k}}

\DeclareRobustCommand{\SkipTocEntry}[4]{}

\begin{document}

\begin{titlepage}

\baselineskip=15.5pt \thispagestyle{empty}

\bigskip\

\vspace{1cm}
\begin{center}

{\fontsize{21}{21} \selectfont TASI Lectures on}\\[16pt]
{\fontsize{32}{32} \bf Primordial Cosmology}

\end{center}

\vspace{0.2cm}
\begin{center}
{\fontsize{15}{30}\selectfont  Daniel Baumann} 
\end{center}

\begin{center}

\vskip 8pt
\textsl{Institute of Theoretical Physics, University of Amsterdam, \\
Science Park, 1098 XH Amsterdam, The Netherlands}
\end{center}

\vspace{1cm}
\beq
\includegraphicsbox[scale=.45]{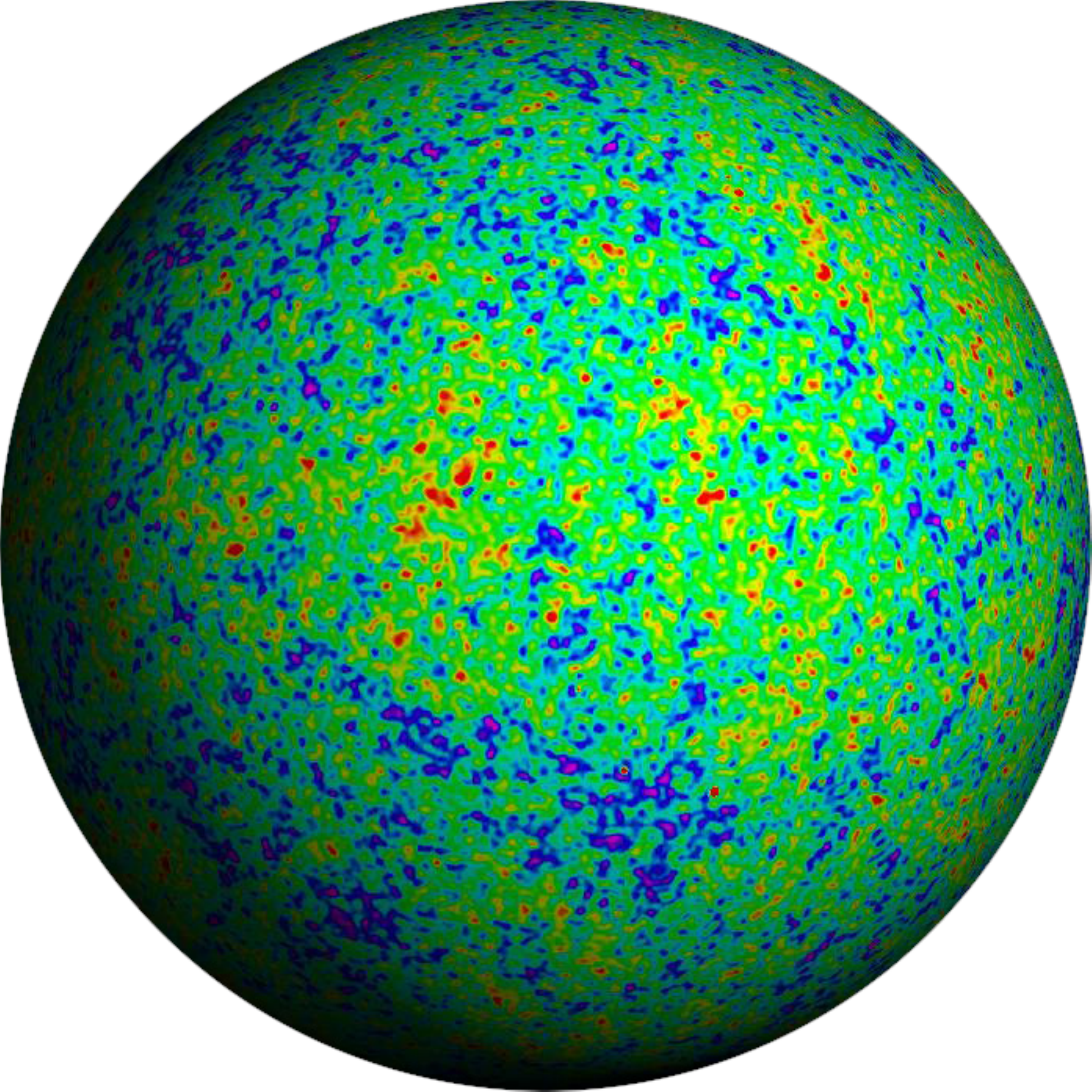} \nonumber
\eeq

\vspace{0.6cm}
 \end{titlepage}

\tableofcontents
\thispagestyle{empty}

\newpage
\setcounter{page}{1} 
\section{Introduction}
\label{sec:motivation}

There are many reasons to believe that our current understanding of fundamental physics is incomplete. For example, the nature of dark matter and dark energy are still unknown, the stability of the Higgs mass remains unsolved, the origin of neutrino masses is unexplained,  the strong CP problem is still there, the physics of inflation remains elusive, and the origin of the matter-antimatter asymmetry is still to be discovered. Attempts to address these shortcomings of the Standard Model (hereafter SM) often involve new degrees of freedom.
These new particles can escape detection in terrestrial experiments either because they are too {\it heavy} to be produced at the energies available or because they are too {\it weakly coupled} to be created in sufficiently large numbers.  However, in astrophysical systems and in the early universe the small cross sections can be compensated for by probing environments with large densities and by following the evolution over long time scales.
For example, the high densities at the cores of stars allow a significant production of new weakly coupled species. Integrated over the long lifetime of stars this can lead to large deviations from the standard stellar evolution.  Observations of the lifecycles of stars have therefore put interesting constraints on the couplings of new light species to the SM.  
Similarly, in the hot environment of the early universe, even extremely weakly coupled particles can be produced prolifically and their gravitational influence becomes detectable in the cosmic microwave background (CMB).  Moreover, during inflation even very massive particles can be created by the rapid expansion of the spacetime.  When these particles decay they produce distinctive signatures in cosmological observables.  Observing these effects could teach us a great deal about the physics driving the inflationary expansion.

\vskip 4pt
New physics can affect cosmological observables in two distinct ways: {\it i}\hskip 1pt) it may affect the {\it initial conditions} of the hot big bang, or {\it ii}\hskip 1pt) it may affect the {\it evolution} of these initial conditions through changes to the thermal history of the universe.
These opportunities for probing new physics with cosmological observations are illustrated in the following schematic:
\begin{align}
&\hspace{3.4cm}{\footnotesize \text{new physics?}} \nonumber \\
&\hspace{4cm} \downarrow \nonumber \\
&\hspace{3.6cm}{\footnotesize \text{evolution}} \nonumber \\[-0.35cm]
&\langle \zeta_{\k_1} \zeta_{\k_2} \ldots \rangle \ \xrightarrow{\ \hspace{4cm}\ }\ \langle O_1 O_2 \ldots \rangle \nonumber \\[-4pt]
&\hspace{-0.3cm}{\footnotesize \text{initial conditions}}  \hspace{3.9cm} {\footnotesize \text{cosmological correlators}} \nonumber \\
&\hspace{0.7cm}\uparrow \nonumber \\
&{\footnotesize \text{new physics?}} \nonumber
\end{align}
where $\zeta$ denotes the primordial curvature perturbations and $O$ stands for late-time observables, which may be the temperature variations $\delta T$ of the CMB or the density fluctuations $\delta \rho_g$ in the distribution of galaxies.
In these lectures, we will give examples from both of these avenues of tests for physics beyond the Standard Model (BSM).

\vskip 4pt
The initial fluctuations were drawn from a probability distribution $P[\zeta]$ and hence are characterized by the following correlation functions 
\beq
\langle \zeta_{\k_1} \zeta_{\k_2} \ldots \rangle = \int {\cal D} \zeta \, P[\zeta]\,\zeta_{\k_1} \zeta_{\k_2} \cdots \, . \nonumber
\eeq
By tracing the observed correlations  $\langle O_1 O_2 \ldots \rangle$ back in time, cosmologists try to measure the initial correlations $\langle \zeta_{\k_1} \zeta_{\k_2} \ldots \rangle$ and use them to extract information about the initial probability distribution $P[\zeta]$ and about the physics that gave rise to it. 
If the early universe went through a period of inflation then these initial correlations were produced dynamically {\it before} the hot big bang.  
New physics during inflation, such as the existence of new degrees of freedom $X$, can leave imprints in the spectrum of primordial perturbations, i.e.~$P[\zeta] \to P[\zeta,X]$.

\vskip 4pt
New physics may also affect how the initial correlations evolve into the late-time observables.  For example, the evolution equations for fluctuations in the primordial plasma can be affected by the presence of new degrees of freedom $X$:
\beq
G_{\mu \nu} = 8\pi G \left(T_{\mu \nu}^{\rm SM} + T_{\mu \nu}^X\right)\,, \qquad   \nabla^\mu\left(T_{\mu \nu}^{\rm SM} + T_{\mu \nu}^X\right) = 0\, . \nonumber
\eeq
As we will see, much of the evolution in the early universe is very well understood and probed by very precise observations. The possibilities for new physics are therefore highly constrained.

\vskip 4pt
The most conservative way to parameterize physics beyond the Standard Model is in terms of an 
{\it effective field theory} (EFT). The basic input of an EFT are the field content and the symmetries that are relevant at a given energy scale.
The effective Lagrangian is then the sum of all operators consistent with the symmetries,
\beq
{\cal L}_{\rm eff} \,\subset\, \sum g\, {\cal O}_X {\cal O}_{\rm SM}\, ,\nonumber
\eeq
where ${\cal O}_{\rm SM}$ denotes operators made from the SM degrees of freedom and ${\cal O}_X$ stands for operators constructed from any additional fields.
The couplings $g$ parameterize the strengths of the interactions between the fields $X$ and the SM. Deviations from the SM predictions then scale with the size of the couplings $g$. In these lectures, I will show how cosmological observations put constraints on these couplings.

\paragraph{Outline} The goal of these lectures is to show that cosmology is becoming an increasingly sensitive probe of BSM physics.  The presentation is divided into two parts: 
In Part I, we study the production of new light particles in the hot big bang and describe their effects on the anisotropies of the cosmic microwave background. In Part II, we investigate the possibility of very massive particles being created during inflation and determine their imprints in higher-order cosmological correlations.


\paragraph{Notation and conventions} Throughout these lectures, we will use natural units, $c=\hbar=1$, with reduced Planck mass $\Mp^2=1/8\pi G$. Our metric signature is ($-++\hskip 1pt +$).  Greek letters will stand for spacetime indices, $\mu, \nu, \ldots =0,1,2,3$, and Latin letters for spatial indices, $i,j,\ldots=1,2,3$. Three-dimensional vectors will be written in boldface, $\k$, and unit vectors will be hatted,~$\hat \k$.  Overdots and primes will denote derivatives with respect to conformal time $\ct$ and physical time~$t$, respectively.
The dimensionless power spectrum of a Fourier mode $\zeta_\k$ will be defined as 
\beq
{\cal P}_\zeta(k) \equiv \frac{k^3}{2\pi^2} \langle \zeta_\k \zeta_{-\k}\rangle' \, ,  \nonumber
\eeq
where the prime on the expectation value indicates that the overall momentum-conserving delta function has been dropped.

\paragraph{Acknowledgements} 
I am grateful to the organizers Igor Klebanov and Mirjam Cvetic for the invitation to lecture at TASI 2017.
Thanks to my students and collaborators for many helpful discussions on the material presented in these lectures.   I am especially grateful to Daniel Green, Joel Meyers, Benjamin Wallisch and Matias Zaldarriaga for collaboration on topics covered in Part I, and to Nima Arkani-Hamed, Valentin Assassi, Garrett Goon, Daniel Green, Hayden Lee, Liam McAllister and Guilherme Pimentel for collaboration on work described in Part II.   Finally, thanks to Benjamin Wallisch and Swagat Saurav Mishra for comments on a preliminary version of the manuscript.

\newpage
\part{Relics from the Hot Big Bang}

\vspace{1cm}
Consider a $2\to N$ scattering process producing a new particle species $X$.
Assume that $X$ is very weakly coupled to the Standard Model degrees of freedom. The production of particles $X$ then carries energy and momentum away from the interaction region.
Such missing energy, of course, famously led to the discovery of neutrinos.  However, particles that are more weakly coupled than neutrinos are produced very rarely in colliders and their missing energy signatures are hard to detect. Fortunately, the creation of new species may be enhanced in astrophysical systems and in the early universe. To see this, consider the change in the number densities of the particles involved in the above interaction. Schematically, this
 is given by
 \beq
\frac{\Delta n}{n} \sim n \sigma \times \Delta t\, , \nonumber
\eeq
i.e.~the fractional change in the number density is equal to the interaction rate, $\Gamma \sim n \sigma$, times the interaction time, $\Delta t$. 
We see that small cross sections $\sigma$ can be compensated for by high densities and long time scales.  This explains why observations of the lifecycles of stars have put interesting constraints on the existence of extra species.  Taking the typical number density inside of stars to be $n \sim (1\,{\rm keV})^3$ 
 and integrating over the typical lifetime of a star, $\Delta t \sim 10^{16}\,{\rm sec} \approx 1.5\times 10^{31}\,{\rm eV}^{-1}$, we find significant changes in the stellar evolution if $$\sigma >  (n\Delta t)^{-1} \sim (10^{10}\,{\rm GeV})^{-2} \, .$$
 Similar constraints can be derived from the observed energy loss in supernova explosions. 
 In that case, the relevant time scales are much shorter, $\Delta t \sim 10\,{\rm sec}$, but the densities are much higher, $n \sim (30\,{\rm MeV})^3$.
 
 \vskip 4pt
 These order of magnitude estimates also give us a sense for the power of cosmological constraints.
In the early universe, the interaction time scales are short, $\Delta t < 1\,{\rm sec}$, but the densities can be very high, $n \sim T^3 \gg (1\,{\rm MeV})^3$. For temperatures above $10^4\,{\rm GeV}$, we expect cosmological constraints to be stronger than those from astrophysics.
In this part of the lectures, we will show that primordial cosmology is indeed a highly sensitive probe of new light particles produced in the hot Big Bang.

\vskip 10pt
We will begin, in Section~\ref{sec:BigBang}, with a quick review of  FRW cosmology.  In Section~\ref{sec:CMB}, we will introduce the cosmic microwave background as a tool for precision cosmology. In Section~\ref{sec:Sound}, we will discuss the physics of the acoustic oscillations observed in the CMB anisotropy spectrum. Finally, in Section~\ref{sec:LightRelics}, we will show that extra relativistic particles leave a unique signature in the CMB spectrum.

\newpage
\section{Big Bang Cosmology}
\label{sec:BigBang}

We will begin with a lightning review of some elementary concepts in cosmology.
I will assume that you have seen most, if not all, of this material before, so I will cite many results without detailed derivations.  Further details can be found in my {\it Cosmology} course~\cite{PartIII-Cosmo} or in any of the standard textbooks (e.g.~\cite{dodelson2003modern, peter2013primordial}).

\subsection{Geometry and Dynamics}

The Friedmann-Robertson-Walker (FRW) metric of a homogenous and isotropic spacetime is
\beq
\d s^2 = - \d t^2 + a^2(t) \gamma_{ij} \h \d x^i \d x^j\, ,
\eeq
where $\gamma_{ij}$ denotes the metric of a maximally symmetric 3-space and $a(t)$ is the scale factor.  Throughout these lectures, we will restrict to the special case of flat spatial slices, i.e.~$\gamma_{ij}=\delta_{ij}$, and define $\d \x^2 \equiv \delta_{ij} \d x^i \d x^j$.
We will also introduce
conformal time, $\d\ct =\d t/a(t)$, so that the metric becomes 
\beq
\d s^2 = a^2(\ct)\left(-\d \ct^2 + \d \x^2\right) . \label{equ:FRW}
\eeq
We will first discuss the kinematics of particles in an FRW spacetime for an arbitrary scale factor~$a(\ct)$. After that, we will show how the Einstein equations determine $a(\ct)$ in terms of the matter content of the universe.

\subsubsection*{Kinematics}

Particles in the FRW spacetime evolve according to the geodesic equation
\beq
P^\nu \nabla_\nu P^{\mu} = P^\nu \left(\frac{\partial P^\mu}{\partial x^\nu} + \Gamma^\mu_{\nu \rho} \hskip1pt P^\rho \right)= 0\, , \label{equ:geo}
\eeq
where $P^\mu \equiv d x^\mu/d \lambda$ is the four-momentum of the particle. 
In an expanding spacetime, it is convenient to write the components of the four-momentum as
\beq
P^\mu  = a^{-1}  [E, \p]\, .\label{equ:pmu}
\eeq
For massless particles, such as photons, we have the constraint $g_{\mu \nu} P^\mu P^\nu = - E^2 + |\p|^2  = 0$, so we can write $\p=E\hskip 1pt\e$, where $\e$ is a unit vector in the direction of propagation.
\begin{framed}
\noindent
{\small {\it Exercise}.---Show that the non-zero connection coefficients associated with the metric (\ref{equ:FRW}) are
\beq
\Gamma^0_{00} = \H\, , \quad \Gamma^0_{ij}=\H\delta_{ij}\, , \quad \Gamma^i_{j0}=\H\delta^i_j\, , \label{equ:Gamma}
\eeq
where $\H \equiv  \dot a/a$ is the conformal Hubble parameter.}
\end{framed}
\noindent
The $\mu=0$ component of the geodesic equation (\ref{equ:geo}) becomes
\beq
P^0 \frac{d P^0}{d\ct} = - \Gamma^{0}_{\alpha \beta} P^\alpha P^\alpha\, .
\eeq
Using (\ref{equ:pmu}) and (\ref{equ:Gamma}), we get
\beq
(a^{-1}E) \frac{d}{d\ct}(a^{-1}E) = - \H a^{-2}E^2 -  \H a^{-2}|\p|^2\, ,
\eeq
which simplifies to
\beq
\frac{1}{E}\frac{dE}{d\ct} = - \frac{1}{a}\frac{da}{d\ct}\, .  \label{equ:redshift} 
\eeq
This describes the {\it redshifting} of the photon energy in an expanding spacetime, $E \propto a^{-1}$.

\subsubsection*{Dynamics}

The evolution of the scale factor is determined by the Friedmann equations
\begin{align}
3\H^2 &= 8\pi G a^2 \bar \rho\, , \label{equ:F1}\\[2pt]
2\dot \H + \H^2 &= - 8\pi G a^2\, \bar P \, , \label{equ:F2}
\end{align}
where 
$\bar \rho$ and $\bar P$ are the background density and pressure, respectively. 
\begin{framed}
\noindent
{\small {\it Exercise}.---By substituting (\ref{equ:Gamma}) into \beq
R_{\mu \nu} \equiv \partial_\lambda \Gamma^\lambda_{\mu \nu} - \partial_\nu \Gamma^\lambda_{\mu \lambda}  + \Gamma^\lambda_{\lambda \rho} \Gamma^\rho_{\mu \nu}  - \Gamma^\rho_{\mu \lambda} \Gamma^\lambda_{\nu \rho}\ , \label{equ:Rmunu}
\eeq 
show that 
\beq
R_{00}=-3\dot \H\,, \quad R_{ij} = (\dot \H+2\H^2) \delta_{ij}\quad \Rightarrow \quad a^2 R \equiv a^2 g^{\mu \nu} R_{\mu \nu}= - 6 (\dot \H+ \H^2)\, .
\eeq
Hence, show that the non-zero components of the Einstein tensor, $G_{\mu \nu} \equiv R_{\mu \nu} - \frac{1}{2}R g_{\mu \nu}$, are
\beq
G_{00} = 3\H^2\, , \ \quad G_{ij}=-(2\dot \H+\H^2) \delta_{ij}\, . \phantom{\quad \Rightarrow \hspace{3.9cm}\quad \qquad x}
\eeq
Use this to confirm that the $00$-Einstein equation, $G_{00}=8\pi G \hskip 1pt T_{00}$, implies (\ref{equ:F1}) and the $ij$-Einstein equation, $G_{ij}=8\pi G  \hskip 1pt T_{ij}$, leads to (\ref{equ:F2}).
}
\end{framed}
\noindent
Combining (\ref{equ:F1}) and (\ref{equ:F2}), we can write an evolution equation for the density
\beq
\dot{\bar \rho} = -3\H(\bar \rho+\bar P)\, . \label{equ:CONT}
\eeq
For pressureless matter ($\bar P_m \approx 0$) this implies $\bar \rho_m \propto a^{-3}$, while for radiation ($\bar P_r = \frac{1}{3} \bar \rho_r$) we have $\bar \rho_r \propto a^{-4}$.

\begin{framed}
\noindent
{\small {\it Exercise}.---Derive the continuity equation~(\ref{equ:CONT}) from the conservation of the stress tensor, $\nabla^\mu T_{\mu \nu} = 0$. By integrating the Friedmann equation (\ref{equ:F1}) for matter and radiation show that
\beq
a(\ct)=   \left\{ \begin{array}{ll} \ct^2 & \ {\rm matter} \\[4pt] \ct & \ { \rm radiation}\end{array} \right. \, .
\eeq
\vspace{-0.5cm}}
\end{framed}

\subsubsection*{Cosmic inventory}

The universe is filled with several different species of particles:
$$
\underbrace{\ \small {\rm photons}\ (\gamma) \quad {\rm neutrinos}\ (\nu)\ }_{\displaystyle {\small \rm radiation}\ (r)}\quad \underbrace{\overbrace{\ {\rm electrons}\ (e)  \quad {\rm protons}\ (p)\ }^{\displaystyle {\rm baryons}\ (b)} \quad \text{cold dark matter}\ (c)\ }_{\displaystyle {\rm matter}\ (m)}\ .
$$
 The number density, energy density and pressure of each species $a$ can be written as
\begin{align}
n_a &=  g_a \int \frac{\d^3 p}{(2\pi)^3} \,f_a(\x,\p) \, ,\\
\rho_a &= g_a \int \frac{\d^3 p}{(2\pi)^3} \,f_a(\x,\p) \hskip 1pt E(p)\, , \label{equ:rhoa}\\
P_a &= g_a \int \frac{\d^3 p}{(2\pi)^3} \,f_a(\x,\p)\hskip 1pt \frac{p^2}{3E(p)}\, , \label{equ:Pa}
\end{align}
 where $f_a(\x,\p)$ is the (phase space) distribution function of the species $a$, and $g_a$ is the number of internal degrees of freedom.  In the unperturbed universe, the distribution functions should not depend on the position and the direction of the momentum, i.e.~$f_a(\x,\p)  \to  \bar f_a(E(p))$.
 
 \vskip 4pt
At early times, particle interactions were efficient enough to keep the different species in local equilibrium.  They then shared a common temperature $T$ and the distribution functions take the following maximum entropy form
 \beq
\bar f_a(E)= \frac{1}{e^{(E_a-\mu_a)/T} \pm 1}\, , \label{equ:f}
\eeq
with $+$ for fermions and $-$ for bosons.  The chemical potential $\mu_a$ vanishes for photons and is (likely) small for all other species. We will henceforth set it to zero.
 When the temperature drops below the mass of a particle species,
$T \ll m_a$,  they become non-relativistic and their distribution function receives an exponential (Boltzmann) suppression, $f_a \to e^{-m_a/T}$.
This means that relativistic particles (`radiation') dominate the density and pressure of the primordial plasma.
By performing the integrals (\ref{equ:rhoa}) and (\ref{equ:Pa}) in the limit $E \to p$, one finds 
\beq
\bar \rho_a = \frac{\pi^2 }{30} \, g_a \hskip 1pt T^4\,   \left\{ \begin{array}{ll} 1 & \ {\rm bosons} \\[4pt] \frac{7}{8} & \ { \rm fermions}\end{array} \right. \quad {\rm and} \quad \bar P_a = \frac{1}{3} \bar \rho_a\, .
\eeq
The total radiation density is
\beq
\bar \rho_r = \frac{\pi^2 }{30} \, g_* \hskip 1pt T^4\, , \quad \text{where}\quad g_* \equiv  \sum_{a=b} g_{a} + \frac{7}{8}\, \sum_{a=f} g_{a}\, .
\eeq
If equilibrium had persisted until today, all species with masses greater than $10^{-3}$\,eV would be exponentially suppressed. This would not be a very interesting world.
Fortunately, many massive particle species (e.g.~dark matter) are weakly interacting and decoupled from the primordial plasma at early times. 

\subsection{Thermal History}
\label{ssec:ThermalHistory}

\begin{figure}[h!]
   \centering
      \includegraphics[scale =0.5]{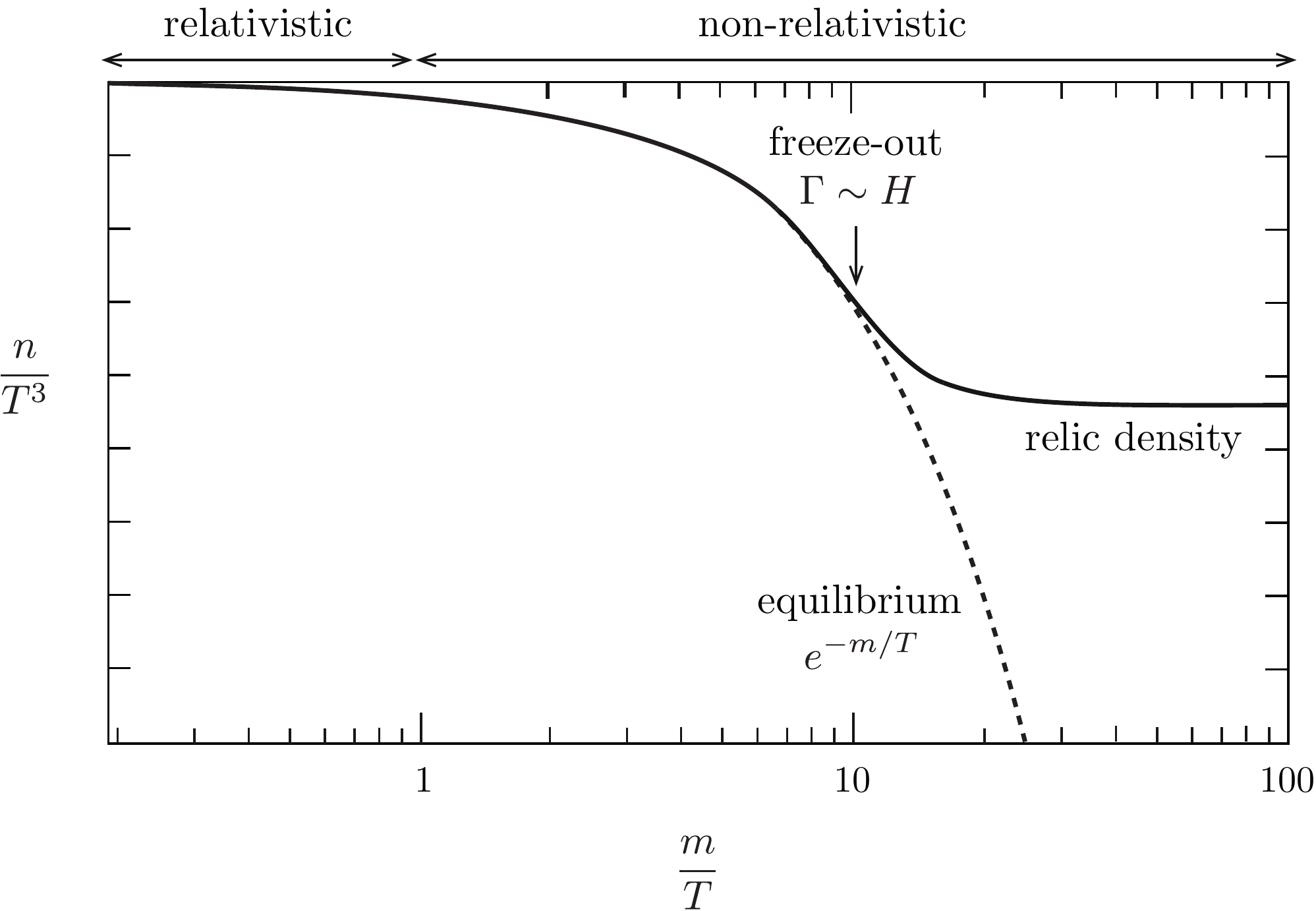}
   \caption{A schematic illustration of particle freeze-out. At high temperatures, 
   the particle abundance tracks its equilibrium value. At low temperatures,  
   the particles freeze out and maintain a relic density that is much larger than the Boltzmann-suppressed equilibrium abundance. }
  \label{fig:freeze}
\end{figure}

The key to understanding the thermal history of the universe is understanding the competition between the interaction rate of particles, $\Gamma$, and the expansion rate of the universe, $H$. Particles maintain equilibrium as long as $\Gamma \gg H$ and freeze out when $\Gamma \lesssim H$ (see Fig.~\ref{fig:freeze}). 

\vskip 4pt
Neutrinos are the most weakly interacting particles of the Standard Model and therefore decoupled first (around $0.8$\,MeV or $1$\,sec after the Big Bang).  
Shortly after neutrino decoupling, electrons and positrons annihilated.  The energies of the electrons and positrons got transferred to the photons, but not the neutrinos. The temperature of the photons today is therefore slightly larger than that of the neutrinos (see insert below).
At around the same time, neutron-proton interactions became inefficient, leading to a relic abundance of neutrons. These neutrons were essential for the formation of the light elements during Big Bang nucleosynthesis (BBN), which occurred around 3 minutes after the Big Bang.

\begin{figure}[h!]
   \centering
      \hspace{-1cm} \includegraphics[scale =0.6]{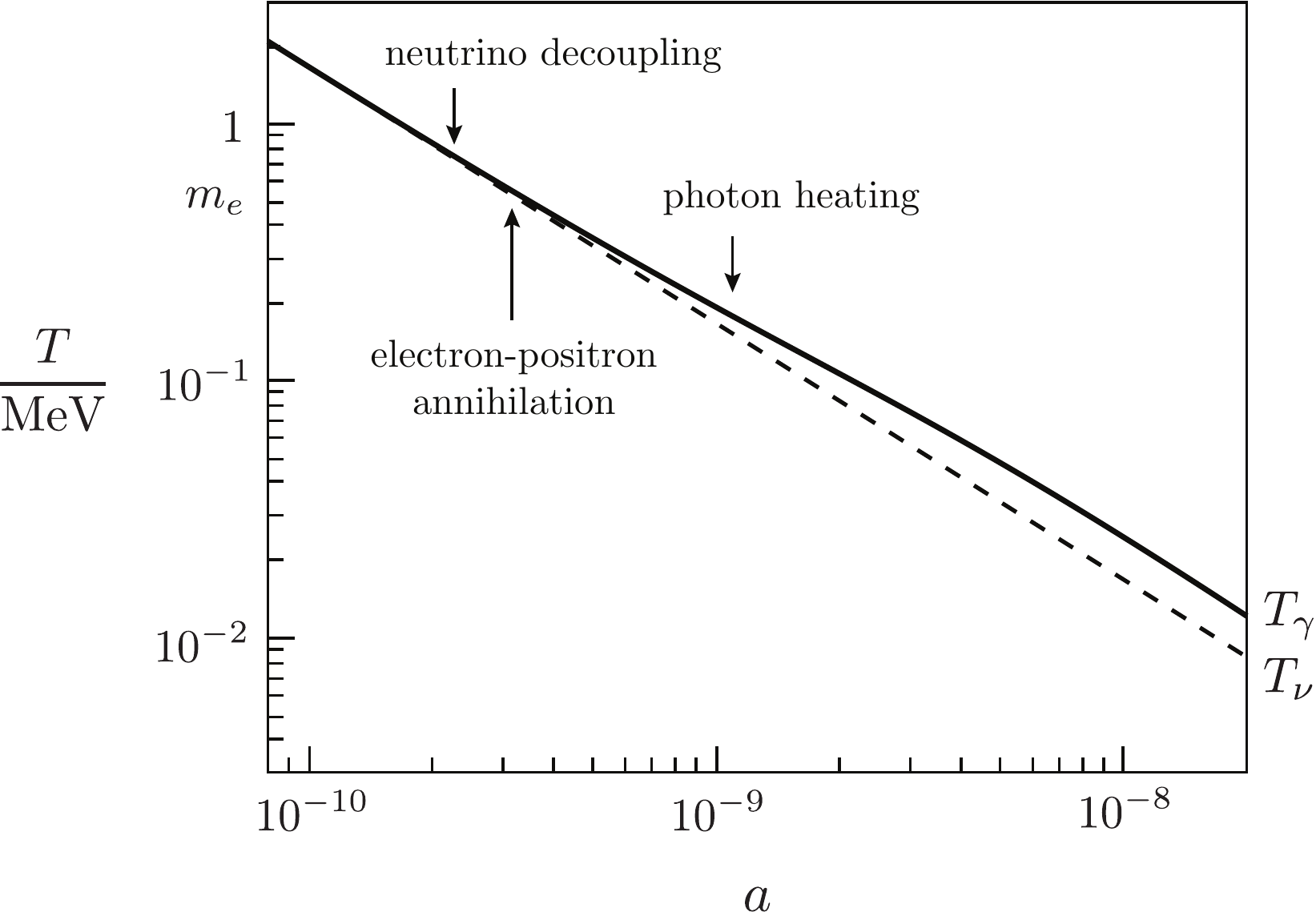}
   \caption{Thermal history through electron-positron annihilation. Neutrinos are decoupled and their temperature redshifts simply as $T_\nu \propto a^{-1}$. The energy density of the electron-positron pairs is transferred to the photon gas whose temperature therefore redshifts more slowly, $T_\gamma \propto g_{*}^{-1/3} a^{-1}$.}
  \label{fig:heating}
\end{figure}


\begin{framed}
\noindent
{\small {\it Cosmic neutrino background.} After the neutrinos decouple, their momenta redshift, $p_\nu \propto a^{-1}$, and their distribution functions $f_\nu$ maintain their shape. The combination of these two facts requires that the neutrino temperature evolves as $T_\nu \propto a^{-1}$.  
We would like to compare this to the evolution of the photon temperature $T_\gamma$, since this is what has been measured by observations of the CMB.
We will use the fact that the {\it comoving entropy} is conserved in equilibrium~\cite{PartIII-Cosmo}:
\beq
s a^3 = \frac{\rho+P}{T} a^3 = \frac{2\pi^2}{45} g_* (aT)^3 = const\, .
\eeq
Since entropy is separately conserved for the thermal bath and the decoupled species, we only need to consider the change in the effective number of relativistic degrees of freedom in equilibrium with the photons.
Before  $e^+ e^-$ annihilation, i.e.~at $T_+ > m_e$, we have
\beq
g_*(T_+) = 2 + \frac{7}{8} \times (2\times 2) = \frac{11}{2} \, ,
\eeq
where we have counted photons, electrons and positrons.
After  $e^+ e^-$ annihilation, i.e.~at $T_- < m_e$, only the two polarization degrees of freedom of the photons contribute,
\beq
\hspace{-3.1cm} g_*(T_-) = 2 \, .
\eeq
Comparing $g_* (aT_\gamma)^3 = {const.}$ to $aT_\nu= {const.}$, we get
\beq
T_\nu = \left(\frac{g_*(T_-)}{g_*(T_{+})}\right)^{1/3} T_\gamma =  \left( \frac{4}{11}\right)^{1/3} T_\gamma \, . \label{equ:NuTemp}
\eeq
Given the measured temperature of the CMB today, $T_{\gamma,0}=2.7\,{\rm K}$, this tells us that the present temperature of the cosmic neutrino background (C$\nu$B) is $T_{\nu,0} = 1.9\,{\rm K}$.

\vskip 4pt
While the neutrinos are still relativistic (i.e.~for most of the history of the universe), each species carries the following energy density
\beq
\frac{\rho_\nu}{\rho_\gamma} = \frac{\frac{7}{8} \times 2\times T_\nu^4}{\phantom{\frac{7}{8} \times\ } 2\times T_\gamma^4} = \frac{7}{8}   \left( \frac{4}{11}\right)^{4/3} \equiv a_\nu \approx 0.227\, .
\eeq
The $N_\nu=3$ neutrino species of the Standard Model therefore contribute a significant amount to the total radiation density in the early universe:
\begin{align}
\frac{\sum\rho_\nu }{\rho_r} = \frac{N_\nu \rho_\nu}{N_\nu \rho_\nu + \rho_\gamma} = \frac{N_\nu a_\nu}{N_\nu a_\nu+1}  \,\approx\, 41\%\, .
\end{align}
Although the neutrinos are decoupled, their gravitational effects are significant and have recently been observed in the CMB~\cite{Follin:2015hya, Baumann:2015rya}.  }
\end{framed}

Below about 1\hskip 2pt eV, or 380,000 years after the Big Bang, the temperature  became low enough for neutral 
hydrogen atoms to form through the reaction $e^- + p^+ \to {\rm H} + \gamma$.  
This is the moment of {\it recombination}. At this point the density of free electrons dropped dramatically. 
Before recombination the strongest coupling between the photons and the rest of the plasma was through Thomson scattering, $e^- + \gamma \to e^- + \gamma$.  The sharp drop in the free electron density after recombination means that this process became inefficient and the photons decoupled. After {\it decoupling} the photons streamed freely through the universe and are observed today as the CMB. 

 \begin{figure}[t!]
    \centering
      \includegraphics[width=.8\textwidth]{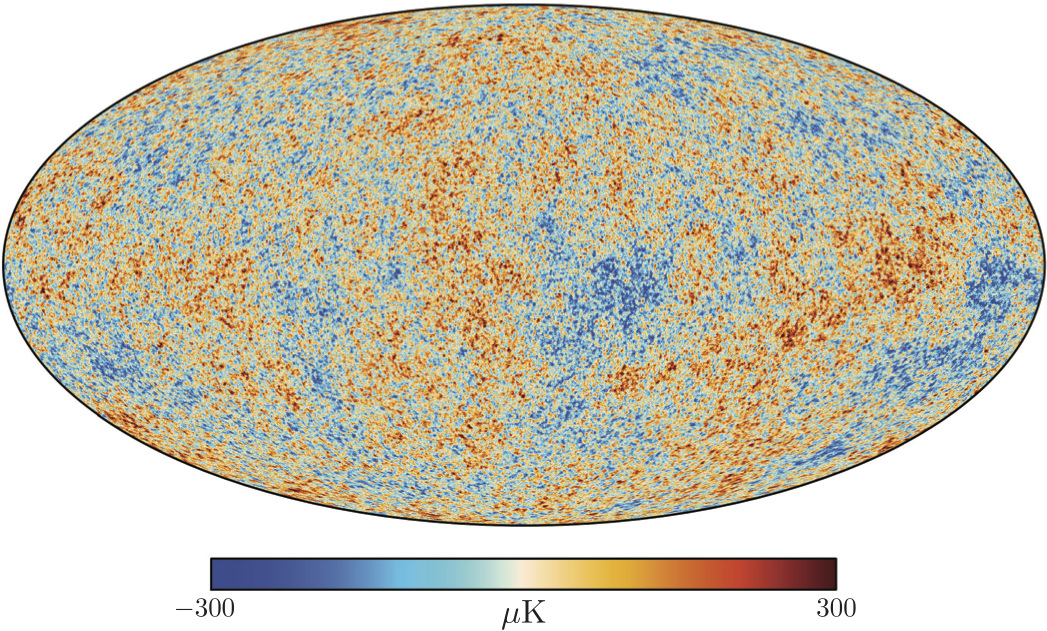}
        \caption{Planck measurement of the temperature variations in the CMB sky.}
    \label{fig:PlanckMap}
\end{figure}

\subsection{Structure Formation}

The CMB is an almost perfect blackbody with an average temperature of 2.7 K.  
Small variations in the CMB temperature
across the sky (see Fig.~\ref{fig:PlanckMap}), reflect spatial variations in the density of the primordial plasma, $\delta \rho_a \equiv \rho_a -\bar \rho_a$ (see Section~\ref{sec:CMB}), and related perturbations of the spacetime geometry, $\delta g_{\mu \nu} \equiv g_{\mu \nu} -\bar g_{\mu \nu}$.  During the radiation-dominated phase of the early universe,
the growth of matter perturbations was inhibited by the large pressure provided by the radiation.  Perturbations in the coupled photon and baryon fluids were oscillating with constant amplitude (see Section~\ref{sec:Sound}).
Shortly before recombination, however, the universe became matter dominated and the radiation pressure disappeared, so that density fluctuations could start to grow under the influence of gravity.  This growth of the matter perturbations eventually led to the large-scale structure (LSS) of the universe.

\subsection{Initial Conditions}

At sufficiently early times, all scales of interest to current observations were outside the Hubble radius, $k < {\cal H}$.  On super-Hubble scales, the evolution of perturbations becomes very simple, especially for adiabatic initial conditions.

 \subsubsection*{Adiabatic perturbations}
 
 Adiabatic perturbations have the property that the local state of matter (determined, for example, by the energy density $\rho$ and the pressure $P$) at some spacetime point ($\ct,\x$) of the perturbed universe is the same as in the {\it background} universe at some slightly different time $\ct+\delta \ct(\x)$. (Notice that the time shift varies with position $\x$.)  If the universe is filled with multiple fluids, adiabatic perturbations correspond to perturbations induced by a {\it common, local shift in time} of all background quantities; e.g.~adiabatic density perturbations are defined as
\beq
\delta \rho_a(\ct,\x) \equiv \bar \rho_a(\ct+\delta \ct(\x)) - \bar \rho_a(\ct) \approx \dot{\bar \rho}_a \hskip 1pt \delta \ct (\x) \, ,
\eeq
where $\delta \ct$ is the same for all species $a$.  This implies that all matter perturbations can be characterized by a single degree of freedom. It also means that we can perform a local time reparameterization to set all matter perturbations to zero, e.g.~$\delta \rho_a \equiv 0$.  In that gauge, the information about fluctuations is carried by the following perturbation of the metric
\beq
g_{ij}(\ct,\x) = a^2(\ct) \, e^{2\zeta(\ct,\x)} \,\delta_{ij}\, ,
\eeq
where $\zeta$ is called the {\it curvature perturbation}.  An attractive property of the curvature perturbation is that it is conserved on super-Hubble scales. 

 \subsubsection*{Statistics}

The initial conditions for the hot Big Bang are believed to have been created by quantum fluctuations during a period of inflationary expansion~\cite{Baumann:2009ds}. 
As we will see in Section~\ref{sec:Quantum}, this mechanism 
predicts the statistics of the initial conditions, i.e.~it predicts the correlations between the CMB fluctuations in different directions in the sky, rather than the specific value of the temperature fluctuation in a specific direction.  For Gaussian initial conditions, these correlations are completely specified by the two-point correlation function
\beq
\langle \zeta(\x) \zeta(\x') \rangle \equiv \xi_\zeta(\x,\x') = \xi_\zeta(|\x'-\x|)\, ,
\eeq
where the last equality holds as a consequence of statistical homogeneity and isotropy.  The Fourier transform of $\zeta$ then satisfies
\beq
\langle \zeta(\k) \zeta^*(\k') \rangle = \frac{2\pi^2}{k^3} {\cal P}_\zeta(k)\, \delta_D(\k-\k')\, , \label{equ:PR}
\eeq
where ${\cal P}_\zeta(k)$ is the (dimensionless) {\it power spectrum}. 
\begin{framed}
\noindent
{\small {\it Exercise}.---Show that
\beq
\xi_\zeta(\x,\x') \,=\,  \int \frac{\d k}{k}\, {\cal P}_\zeta(k)\, {\rm sinc}(k|\x-\x'|)\, .
\eeq
\vspace{-0.5cm}}
\end{framed}

\noindent
In Section~\ref{sec:Quantum}, we will explicitly compute the form of ${\cal P}_\zeta(k)$ predicted by inflation. However, before we do that, we will show, in Sections~\ref{sec:CMB} and~\ref{sec:Sound}, how generic scale-invariant initial conditions, ${\cal P}_\zeta(k) \approx const.$,  evolve into the anisotropies of the cosmic microwave background.
We will first do this in the context of the Standard Model of particle physics, before asking, in  Section~\ref{sec:LightRelics}, what kind of deviations can arise in theories beyond the SM.

\newpage
\section{Afterglow of the Big Bang}
\label{sec:CMB}

Observations of the temperature fluctuations in the cosmic microwave background have played a pivotal role in establishing the standard cosmological model. We now have a detailed understanding of the geometry and composition of the universe, and there is growing evidence that the primordial fluctuations originated from quantum fluctuations during a period of inflation.
In this section and the next, we will give a simplified analytical treatment of the physics of the CMB anisotropies.  Our goal is to present just enough details to be able to explain how the CMB can be used as a probe of BSM physics.

\subsection{CMB Anisotropies}

\begin{figure}[h!]
   \centering
       \includegraphics[scale =0.75]{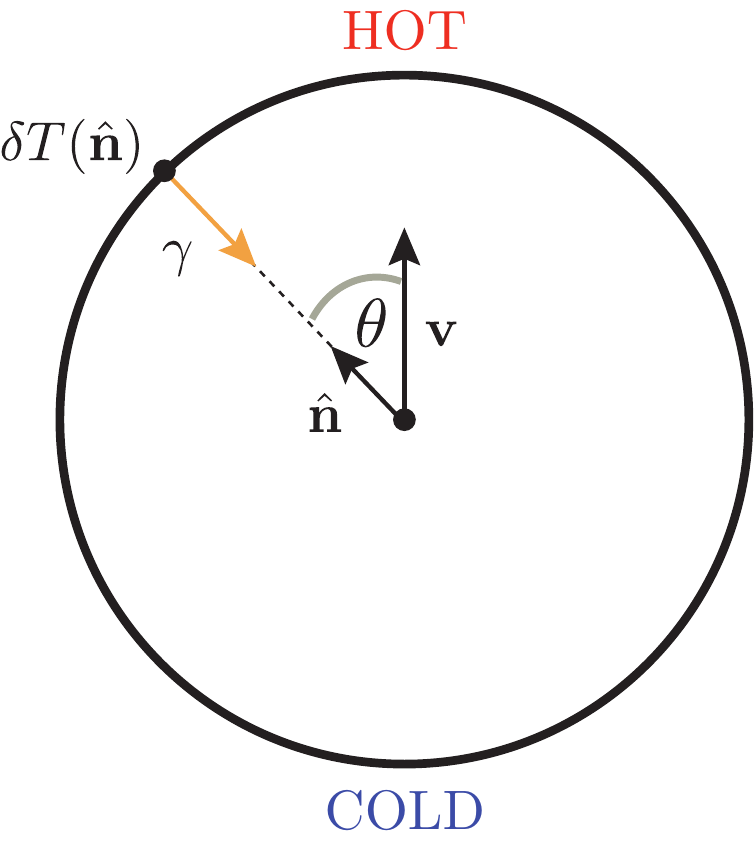}
   \caption{The motion of the Solar System relative to the CMB rest frame produces a dipolar pattern in the observed CMB temperature.}
  \label{fig:CMB-Dipole}
\end{figure}

The first thing one sees when looking at the microwave scy is the motion of the Solar System with respect to the rest frame of the CMB (cf.~Fig.~\ref{fig:CMB-Dipole}).

\noindent
Consider a photon entering our detectors from a direction $\hat \n$.
In the rest frame of the CMB, it has momentum $\p = - p\hskip 1pt \hat \n$.
Due to the Doppler effect, the observed momentum is
\beq
p_{0}(\hat \n) = \frac{p}{\gamma(1-\hat \n \cdot \v)} \approx p \left( 1 + \hat \n \cdot \v \right)  ,
\eeq
where $\v$ is our velocity relative to the CMB rest frame, $p$ is the momentum of the photon in the CMB rest frame and $\gamma = (1-v^2)^{-1/2}$ is the Lorentz factor. 
We have also shown an approximation at leading order in $|\v| \ll 1$.
As expected, the momentum is higher if we move towards the photon ($\hat \n \cdot \v = v $) and smaller if we move away from it ($\hat \n \cdot \v = - v $). Since the CMB has a blackbody spectrum, we can relate the change in the observed momentum of photons to a change in the observed temperature:
\beq
\frac{\delta T(\hat \n)}{T}  \equiv \frac{T_{0}(\hat \n)  - T}{T} = \frac{p_{0}(\hat \n) - p}{p} = \hat \n \cdot \v = v \cos \theta \, .
\eeq
Fitting this dipolar anisotropy to the data, we find that the speed of the Solar System relative to the CMB is
\beq
v = 368\, {\rm km}/{\rm s}\ .
\eeq
After subtracting the dipole, we are left with the {\it primordial anisotropy}.  

\subsubsection*{Perturbed photon geodesics}

Let us trace the life of a photon after decoupling.
Its (physical) momentum will redshift due to the expansion of the universe.
In addition, the momentum will change in response to the inhomogeneities of the universe. 
We will study these effects by solving the geodesic equation~(\ref{equ:geo}) in the perturbed spacetime.

\vskip 4pt
We will treat perturbations in the metric in (conformal) Newtonian gauge
\beq
\d s^2 = a^2(\ct)\big[-(1+2\Phi) \d \ct^2 + (1-2\Psi) \delta_{ij} \d x^i \d x^j\big] \, ,\label{equ:gN}
\eeq
where the perturbation $\Phi \approx \Psi$ plays the role of the gravitational potential.

\begin{framed}
\noindent
{\small {\it Exercise.}---Show that the connection coefficients associated with the metric (\ref{equ:gN}) are
\beq
\begin{aligned}
\Gamma^0_{00} &= \H + \dot \Phi\, ,  \\
\Gamma^0_{i0} &=\partial_i \Phi\, , \\
\Gamma^i_{00}&=\delta^{ij} \partial_j\Phi\, ,\\
\Gamma^0_{ij} &= \H \delta_{ij} - \big[\dot \Psi+2\H(\Phi+\Psi) \big] \delta_{ij} \, , \\
\Gamma^i_{j0} &= \big[\H - \dot \Psi\big] \delta^i_j \, , \\
\Gamma^i_{jk} &= - 2\delta^i_{(j} \partial_{k)} \Psi + \delta_{jk} \delta^{il} \partial_l \Psi\, .
\end{aligned}
\eeq
\vspace{-0.3cm}}
\label{equ:ChrisP}
\end{framed}

As we will show in the following insert, the geodesic equation then leads to the following evolution equation for the photon momentum:
\beq
\fbox{$\displaystyle  \frac{d}{d \ct}  \ln (ap) =  - \frac{d \Phi}{d \ct} + \frac{\partial (\Phi+ \Psi)}{\partial  \ct}  $} \ . \label{equ:PP}
\eeq
In the absence of the source terms on the right-hand side, this implies $p \propto a^{-1}$, which is the expected redshifting due to the expansion of the universe, cf.~eq.~(\ref{equ:redshift}). The inhomogeneous source terms describe how photons lose (gain) energy as they move out of (into) potential wells. 


\begin{framed}
\noindent
{\small {\it Derivation.}---We will derive eq.~(\ref{equ:PP}) from the geodesic equation for photons,
\beq
\frac{d P^0}{d\lambda} = - \Gamma^0_{\alpha \beta} P^\alpha P^\beta\, , \label{equ:GX}
\eeq
where $P^\mu = d x^\mu/d \lambda$ is the four-momentum of the photons.
We need expressions for the components of the four-momentum in the presence of metric perturbations.
\begin{itemize}
\item We first consider the $P^0$ component.
Since photons are massless we have
\begin{align}
P^2 = g_{\mu \nu} P^\mu P^\nu &= 0 \nonumber \\
&= - a^2(1+2\Phi) (P^0)^2 + p^2 \, , \label{equ:P0}
\end{align}
where we have substituted the metric (\ref{equ:gN}) and defined $p^2 \equiv g_{ij} P^i P^j$. 
Solving~(\ref{equ:P0}) for $P^0$, we find 
\beq
P^0 = \frac{p}{a} (1 -\Phi)\, .
\eeq
\item We then write the spatial component of the four-momentum as
\beq
P^i \equiv \alpha\, \hat p^i\, .
\eeq
To determine the constant of proportionality $\alpha$, we use
\begin{align}
p^2 =  g_{ij} P^i P^j 
&= a^2(1-2\Psi) \delta_{ij}  \hat p^i \hat p^j \alpha^2 \nonumber \\
&= a^2(1-2\Psi) \alpha^2\, , \label{equ:pSqr}
 \end{align}
 where the last equality holds because the direction vector is a unit vector.
 Solving~(\ref{equ:pSqr}) for $\alpha$, we get $\alpha = p(1+\Psi)/a$, or
 \beq
P^i = \frac{p \hat p^i}{a} (1+\Psi) \, . 
\eeq
\end{itemize}

\vskip 4pt
\noindent
Substituting these results into the geodesic equation (\ref{equ:GX}), we get
\beq
\frac{p}{a}(1-\Phi) \frac{d}{d\ct} \left[\frac{p}{a}(1-\Phi)\right] = - \Gamma^0_{\alpha \beta} P^\alpha P^\beta \, ,
\eeq
where we have used the standard trick of rewriting the derivative with respect to $\lambda$ as a derivative with respect to time multiplied by $d\ct/d\lambda = P^0$. We expand out the time derivative to get
\beq
\frac{dp}{d\ct} (1-\Phi) = \H p(1-\Phi) + p \frac{d \Phi}{d\ct} - \Gamma^0_{\alpha \beta} P^\alpha P^\beta\, \frac{a^2}{p}(1+\Phi)\, .
\eeq
Multiplying both sides by $(1+\Phi)/p$ and dropping all quadratic terms in $\Phi$, we find
\beq
\frac{1}{p}\frac{dp}{d\ct} = \H+ \frac{d \Phi}{d\ct} - \Gamma^0_{\alpha \beta} P^\alpha P^\beta\, \frac{a^2}{p^2} (1+2\Phi)\, . \label{equ:GeoY}
\eeq
To evaluate the last term on the right-hand side, we use the perturbed Christoffel symbols (\ref{equ:ChrisP}).  After a bit of algebra, we get
\beq
- \Gamma^0_{\alpha \beta} \frac{P^\alpha P^\beta}{p^2} (1+2\Phi) = - 2\H + \frac{\partial \Psi}{\partial \ct} - \frac{\partial \Phi}{\partial \ct} - 2 \hskip 1pt \hat p^i \frac{\partial \Phi}{\partial x^i }  \, . \label{equ:Psi2}
\eeq
Equation (\ref{equ:GeoY}) then becomes
\begin{align}
\frac{1}{p}\frac{dp}{dt}  
&= - {\cal H} + \frac{d\Phi}{d\ct}  - 2\left( \frac{\partial\Phi}{\partial\ct} + \hat p^i \frac{\partial \Phi}{\partial x^i}\right) + \frac{\partial (\Phi+\Psi)}{\partial \ct}\, . \label{equ:GeoZ}
 \end{align}
At leading order, the term in brackets is equal to the total time derivative of $\Phi$:
\beq
\frac{d \Phi}{d\ct} = \frac{\partial \Phi}{\partial \ct} + \frac{d x^i}{d\ct} \frac{\partial \Phi}{\partial x^i} = \frac{\partial \Phi}{\partial \ct} + \hat p^i \frac{\partial \Phi}{\partial x^i} \, , \label{equ:Psi1}
\eeq
where we used
\beq
\frac{d x^i }{d\ct} = \frac{dx^i}{d\lambda} \frac{d\lambda}{d\ct} =  \frac{P^i}{P^0} = \hat p^i (1+\Psi + \Phi) =\hat p^i + {\cal O}(1)\, .
\eeq

\vskip 4pt
\noindent
Substituting (\ref{equ:Psi1}) into (\ref{equ:GeoZ}), we get
\beq
\frac{1}{p} \frac{d p}{d\ct} = - \frac{1}{a}\frac{da}{d\ct} -  \frac{d\Phi}{d\ct}+ \frac{\partial (\Phi +\Psi)}{\partial \ct}\, ,
\eeq
which confirms the result~(\ref{equ:PP}).}
\end{framed}

\subsubsection*{Line-of-sight solution}

By integrating the geodesic equation (\ref{equ:PP}) along a line-of-sight, we can relate the observed CMB temperature anisotropies to the fluctuations at recombination.  
To simplify matters, we will therefore work with the idealised approximation of {\it instantaneous recombination}.
The CMB photons were then emitted at a fixed time $\ct_{*} $. This moment is often called 
{\it last scattering}. 
Integrating ~(\ref{equ:PP})  from the time of emission $\ct_{*}$ to the time of observation $\ct_{0}$, we then get
\beq
\ln(ap)_{0} = \ln(ap)_{*}  + (\Phi_{*} - \Phi_{0} ) + \int_{\ct_{*}}^{\ct_{0}} \d \ct \, \frac{\partial}{\partial \ct} (\Phi + \Psi) \, . \label{equ:ap}
\eeq
To relate this to the temperature anisotropy, we note that
\beq
ap \propto a \bar T \left( 1 + \frac{\delta T}{\bar T} \right) ,
\eeq
where $\bar T(\ct)$ is the mean temperature.
Taylor-expanding the logarithms in~(\ref{equ:ap}) to first order in $\delta T/\bar T$, and keeping in mind that $a_{0} \bar T_{0} = a_{*} \bar T_{*}$, we find
\beq
\left. \frac{\delta T}{\bar T} \right|_{0} = \left. \frac{\delta T}{\bar T} \right|_{*} + (\Phi_{*} - \Phi_{0}) + \int_{\ct_{*}}^{\ct_{0}} \d \ct \, \frac{\partial}{\partial \ct} (\Phi + \Psi) \, . \label{equ:dT/T}
\eeq
The term $ \Phi_{0}$ only affects the monopole perturbation, so it is unobservable and therefore usually dropped from the equation.
The fractional temperature perturbation at last-scattering can be expressed in terms of the density contrast of photons, $\delta_\gamma \equiv \delta \rho_\gamma/\rho_\gamma$, as
\beq
 \left. \frac{\delta T}{\bar T} \right|_{*} = \frac{1}{4} (\delta_\gamma)_{*}\, ,
\eeq
where the factor of $\frac{1}{4}$ arises because $\rho_\gamma \propto T^4 $.
Equation~(\ref{equ:dT/T}) then reads
\beq
\left. \frac{\delta T}{\bar T} \right|_{0} = \left(\frac{1}{4} \delta_\gamma + \Phi \right)_{*}  + \int_{\ct_{*}}^{\ct_{0}} \d \ct \, \frac{\partial}{\partial \ct} (\Phi + \Psi) \, .
\eeq

\vskip 4pt
\noindent
Each term on the right-hand side has a simple physical interpretation:
\begin{itemize}
\item The term $\frac{1}{4}\delta_\gamma$ can be
thought of as the intrinsic temperature variation over the background
last-scattering surface. 
\item The term $\Phi$ arises from the gravitational redshift that the photons experience when climbing out of a potential well at last-scattering. 
The combination
$\frac{1}{4}\delta_\gamma + \Phi$ is often called the \emph{Sachs-Wolfe} (SW) term. 
\item Finally, the \emph{integrated Sachs-Wolfe} (ISW) term
describes the effect of gravitational redshifting from evolution of the
potentials along the line-of-sight. During matter domination, $\dot \Phi \approx \dot \Psi = 0$ and this term vanishes.
\end{itemize}

\begin{figure}[t!]
   \centering
       \includegraphics[scale =0.75]{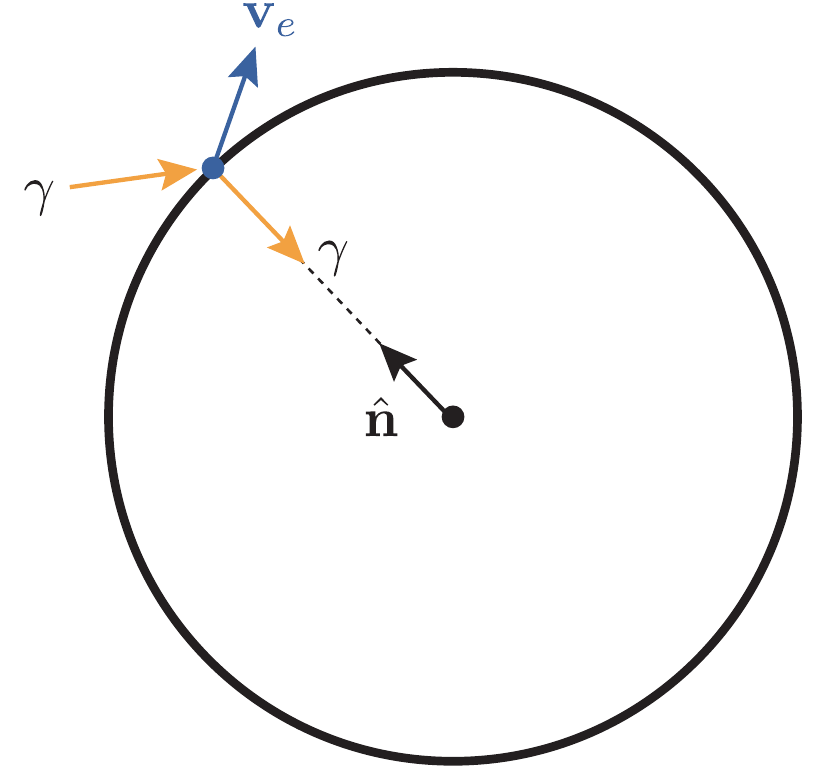}
       \caption{The motion of electrons at the surface of last-scattering produces an additional temperature anisotropy. }\label{fig:Doppler}
\end{figure}

\begin{figure}[t!]
\centerline{\includegraphics[width=10cm,angle=0]{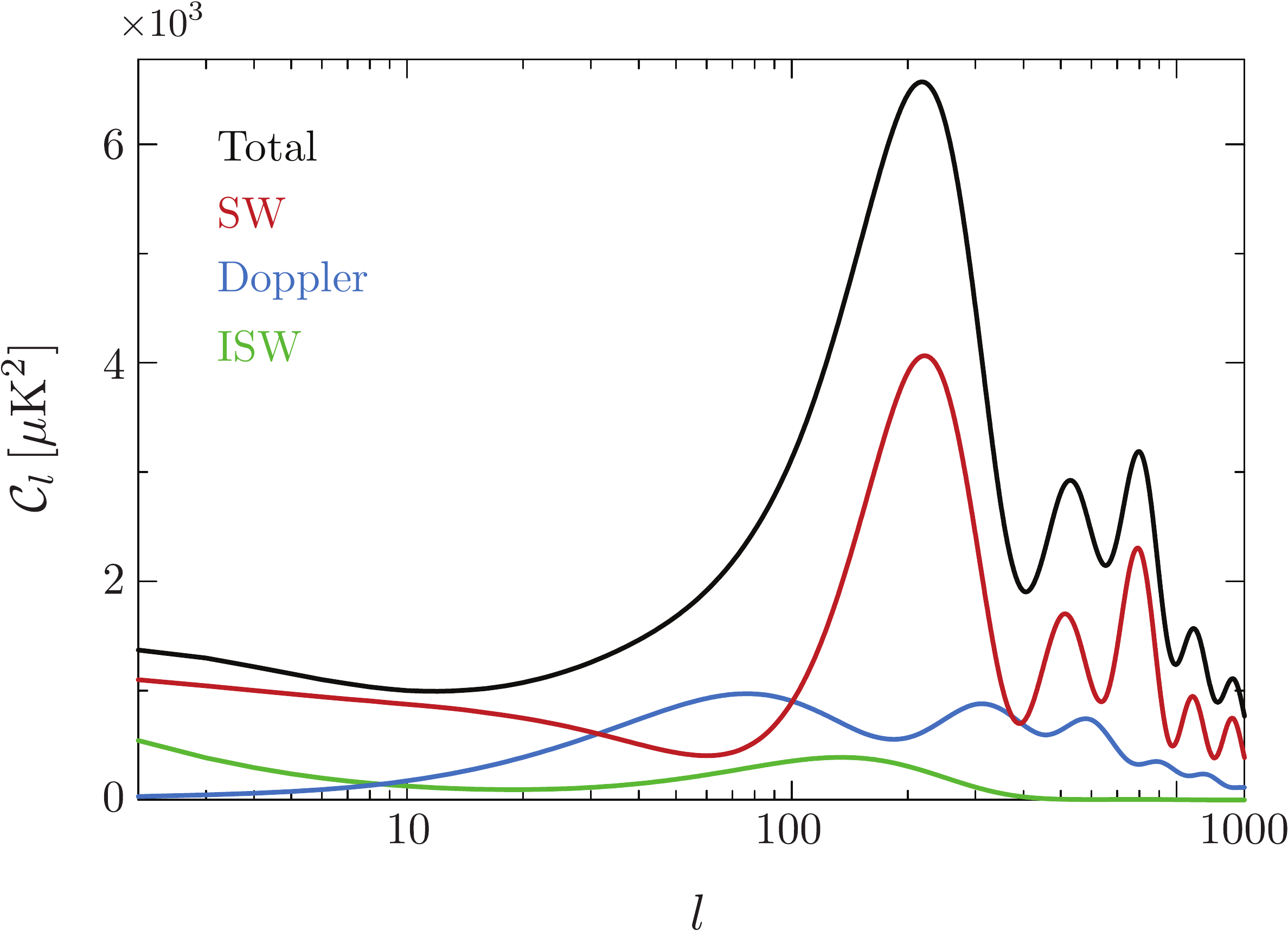}}
\caption{Contributions of the various terms in~(\ref{equ:CMBmaster}) to the (rescaled) power spectrum of CMB anisotropies, ${\cal C}_l \equiv l(l+1) C_l$. }
\label{fig:adiabatic_spectrum}
\end{figure}

\vskip 4pt
So far, we have ignored the motion of the electrons at the surface of last-scattering.
Including this effect leads to an extra Doppler shift in the received energy of photons when referenced to an observer comoving with the electrons at last-scattering (see Fig.~\ref{fig:Doppler}),
\beq
\left. \frac{\delta T}{\bar T} \right|_{0} \ \subset \  (\hat \n \cdot \v_e)_{*} \, .
\eeq
Putting everything together, we obtain the following important result
\beq
\fbox{$\displaystyle \frac{\delta T}{\bar T}(\hat \n) =  \left(\frac{1}{4} \delta_\gamma + \Phi + \hat \n \cdot \v_e\right)_{*} + \int_{\ct_*}^{\ct_0} \d \ct \,(\dot \Phi + \dot \Psi) $} \ , \label{equ:CMBmaster}
\eeq
where we have dropped the subscript `$0$' on the observed $\delta T/\bar T$ to avoid clutter.   Figure~\ref{fig:adiabatic_spectrum} illustrates the contributions that each of the terms in~(\ref{equ:CMBmaster}) makes to the power spectrum of the CMB temperature anisotropies (see \S\ref{sec:PS}).  We see that the ISW contribution is subdominant and that the shape of the power spectrum is mostly determined by the Sachs-Wolfe and Doppler contributions.

\begin{framed}
{\small \noindent {\it Large scales.}---For adiabatic initial conditions, the superhorizon initial condition is $\delta_\gamma \approx \frac{4}{3} \delta_m \approx - \frac{8}{3} \Psi$. The Sachs-Wolfe term then becomes
\beq
\frac{1}{4} \delta_\gamma + \Phi = - \frac{2\Psi}{3} + \Phi \approx  \frac{1}{3} \Phi\, .
\eeq
This shows that, on large scales, an overdense region ($\Psi \approx \Phi < 0$) appears as a cold spot in
the sky. While the temperature at the bottom of the potential well is hotter than the average ($- \frac{2}{3}\Psi$), photons lose more energy ($\Phi$) as they climb out of the potential well, resulting in a cold spot ($\frac{1}{3}\Phi <0$).}
\end{framed}

\subsection{CMB Power Spectrum}
\label{sec:PS}

\begin{figure}[t!]
\begin{center}
\includegraphics[width=2.3in]{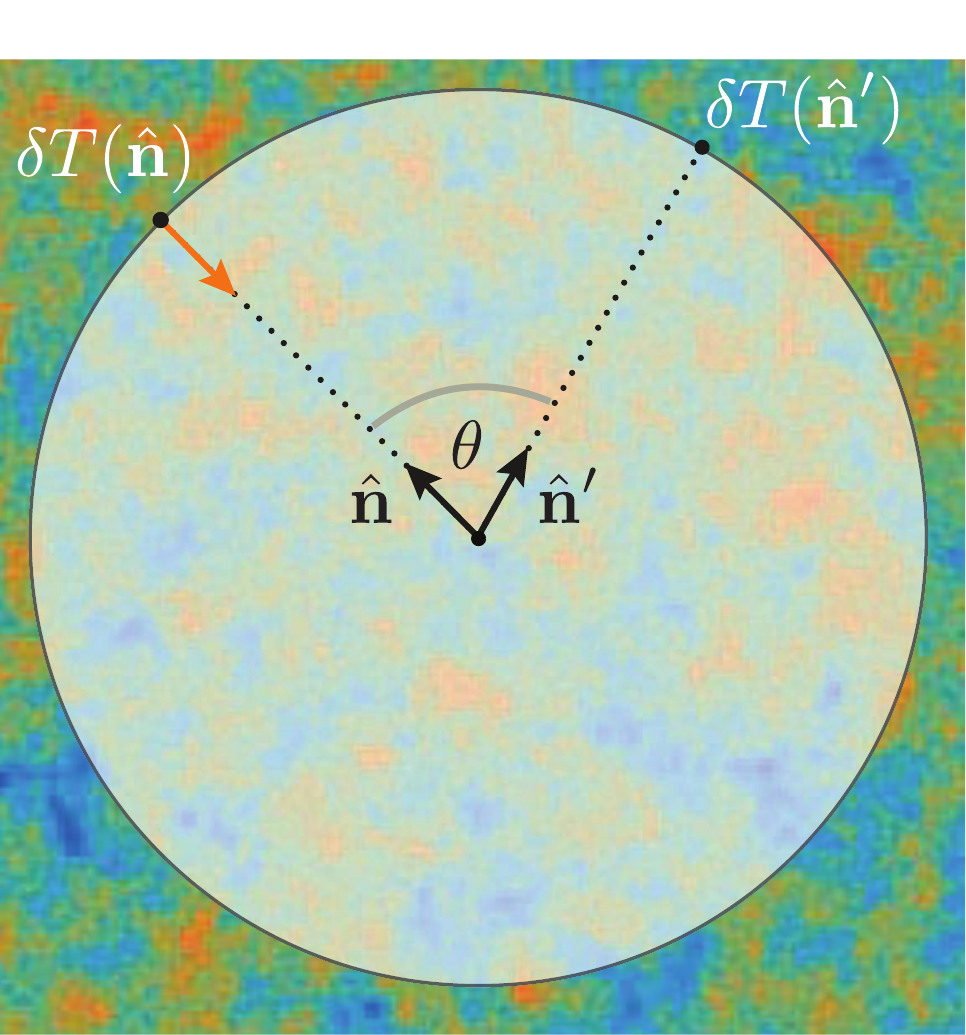}
   \includegraphics[scale =0.75]{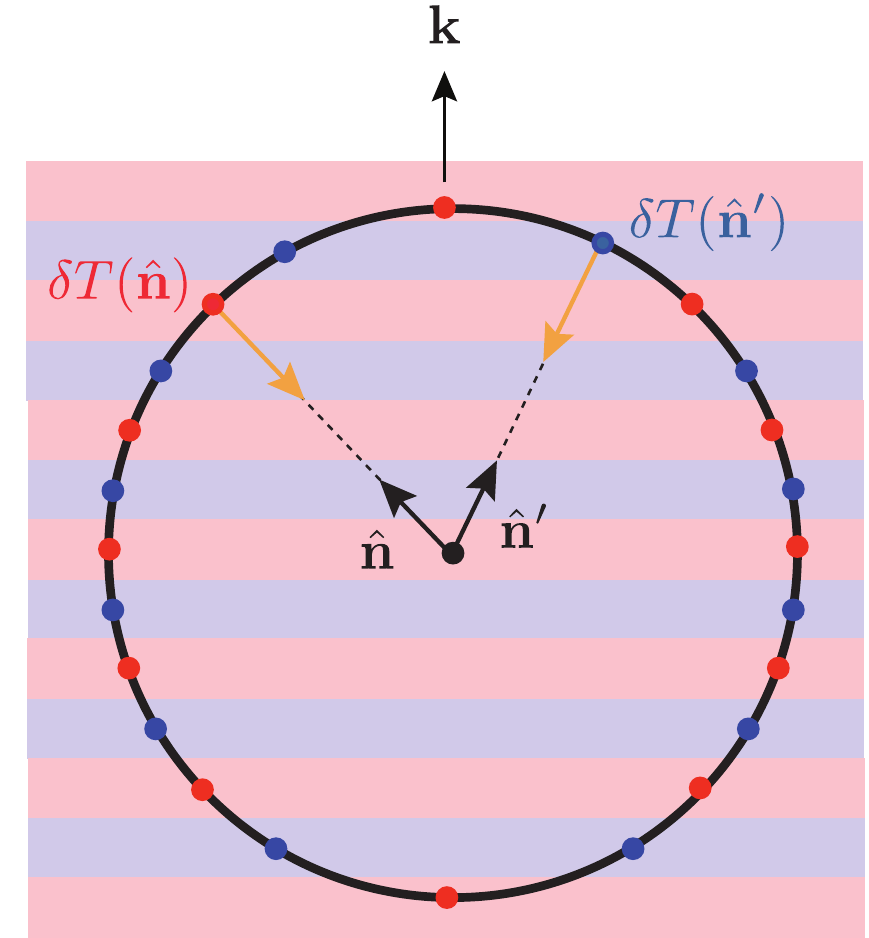}
\caption{{\it Left}: Illustration of the two-point correlation function of the temperature anisotropy $\delta T(\hat \n)$. {\it Right}: Illustration of the temperature  anisotropy created by a single plane wave inhomogeneity are recombination.} \label{fig:projection}
\end{center}
\end{figure}

A map of the cosmic microwave background radiation describes the variation of the CMB temperature as a function of direction, $\delta T(\hat\n)$.
We will be interested in the statistical correlations between temperature fluctuations in two different directions $\hat\n$ and $\hat\n'$ (see Fig.~\ref{fig:projection}), averaged over the entire sky.  

If the initial conditions are statistically isotropic, then we expect these correlations only to depend on the relative orientation of $\hat\n$ and $\hat\n'$.  In that case, we can write the two-point correlation function as
\beq
\big\langle \delta T(\hat\n) \hskip 1pt \delta T(\hat\n') \big\rangle = \sum_l \frac{2l+1}{4\pi} \hskip 1pt C_l \hskip 1pt P_l(\cos \theta)\, , \label{equ:2pt}
\eeq
where $\hat\n \cdot \hat\n' \equiv \cos \theta$ and $P_l$ are Legendre polynomials.  The expansion coefficients $C_l$ are the {\it angular power spectrum}. 
If the fluctuations are Gaussian, then the power spectrum contains the entire information of the CMB map.

\vskip 4pt
The right panel of Fig.~\ref{fig:projection} illustrates the temperature variations created by a single plane wave inhomogeneity.  The CMB anisotropies observed on the sky are a {\it superposition} of many such plane waves with amplitudes that are weighted by the spectrum of primordial curvature perturbations~${\cal P}_\zeta(k)$. In Section~\ref{sec:Quantum}, we will show that the initial conditions of the primordial perturbations are expected to be featureless, ${\cal P}_\zeta(k) \approx const$.  The observed features in the CMB anisotropy spectrum arise from the subhorizon evolution of perturbations in the photon density and the metric.  This evolution takes the form of cosmic sound waves  (see Section~\ref{sec:Sound}). These  waves are
{\it captured} at recombination and {\it projected} onto the sky.  The observed oscillations in the CMB power spectrum are therefore a snapshot of primordial sound waves caught at different phases in their evolution at the time when photons last scattered off electrons. The beautiful physics of the CMB fluctuations is described in  detail in my {\it Advanced Cosmology} course~\cite{Adv-Cosmo}.

\vskip 4pt
Substituting the line-of sight solution (\ref{equ:CMBmaster}) into the definition of the angular power spectrum (\ref{equ:2pt}), we find (see insert below)
\beq
\boxed{C_l = \frac{4\pi}{(2l+1)^2} \int \d\ln k\ T_l^2(k)\, {\cal P}_\zeta(k)}\ . \label{equ:Cl}
\eeq
The {\it transfer function} $T_l(k)$ captures both the {\it evolution} of the initial fluctuations until recombination and the {\it projection} onto the surface of last-scattering.  Ignoring the subdominant ISW contribution, we can write the transfer function as follows:
\beq
\boxed{T_l(k) = T_{\rm SW}(k)\hskip 1pt j_l(kr_*) + T_{\rm D}(k) \hskip 1pt j'_l(kr_*)}\,  , \quad {\rm where} \quad \begin{array}{rl} T_{\rm SW}(k) &\displaystyle \equiv \frac{(\frac{1}{4}\delta_\gamma +\Phi)_*}{\zeta(\k)}\, , \\[10pt] 
 T_{\rm D}(k) &\displaystyle \equiv  -\frac{(v_e)_*}{\zeta(\k)}\, . \label{equ:Tfcts}
\end{array}
\eeq
The subscript $*$ denotes quantities evaluated at recombination, with $r_*$ being the distance to last-scattering.  The Bessel function $j_l(kr_*)$ and its derivative $j_l'(kr_*)$ act almost like delta functions and map the Fourier modes $k$ to the harmonic moments $l \sim kr_*$.  Given that ${\cal P}_\zeta(k)$ is expected to be nearly constant, the angular power spectrum $C_l$ therefore measures the square of the transfer function~$T_l(k)$ evaluated at $k=l/r_*$:
\beq
C_l \,\sim\, \frac{4\pi}{(2l+1)^2} \Big[T_{\rm SW}^2(k) + T_{\rm D}^2(k)\Big] \bigg|_{k \sim l/r_*}\, , \label{equ:Cell}
\eeq
where we have dropped the cross term  $T_{\rm SW}(k) \,T_{\rm D}(k)$ because it is negligible.
  In the next section, we will discuss the evolution effects that determine the transfer function and hence the CMB power spectrum.

\begin{framed}
{\small \noindent {\it Projection.}---To understand the origin of the Bessel functions in (\ref{equ:Tfcts}), let us consider the projection of the Sachs-Wolfe term onto the surface of last-scattering. Assuming instantaneous recombination, we can write
\begin{align}
\delta T(\hat \n) &= \int \d r\, \delta T(\x,\n)\, \delta_D(r-r_*)\\
&= \int \frac{\d^3 k}{(2\pi)^3} \, e^{i(kr_*) \hat \k \cdot \hat \n} \, \delta T(\k)\, , \label{equ:dT}
\end{align}
where we substituted the Fourier expansion of the temperature field in the second line.
The exponential in (\ref{equ:dT}) can be written in a Rayleigh plane wave expansion,
\beq
e^{i(kr_*) \hat \k \cdot \hat \n} = \sum_l (-i)^l (2l+1) j_l(kr_*) P_l(\hat \k \cdot \hat \n)\, .
\eeq
The two-point function of temperature anisotropies then becomes
\begin{align}
\big\langle \delta T(\hat\n) \hskip 1pt \delta T(\hat\n') \big\rangle = \int \frac{\d^3 k}{(2\pi)^3} \int \frac{\d^3 k'}{(2\pi)^3}  \sum_l \sum_{l'} &\,(-i)^{l+l'} (2l+1)(2l'+1)\, j_l(kr_*) j_{l'}(k'r_*) \nonumber \\[-4pt] 
&\times P_l(\hat \k \cdot \hat \n) P_{l'}(\hat \k' \cdot \hat \n')\, \langle \delta T(\k)\hskip 1pt \delta T(\k') \rangle\, .  \label{equ:2ptX}
\end{align}
The power spectrum of the temperature field can be written in terms of the power spectrum of the primordial curvature perturbations, cf.~(\ref{equ:PR}),
\begin{align}
\langle \delta T(\k)\hskip 1pt \delta T(\k') \rangle &= T_{\rm SW}(k) T_{\rm SW}(k') \langle \zeta(\k)\hskip 1pt \zeta(\k') \rangle \nonumber \\
&= T_{\rm SW}^2(k) \, \frac{2\pi^2}{k^3} \, {\cal P}_\zeta(k) \, \delta_D(\k-\k')\, .
\end{align}
The delta function allows us to trivially perform one of the momentum integrals in (\ref{equ:2ptX}). To evaluate the angular part of the second momentum integral, we use the following identity
\beq
\int \d^2 \hat \k \, P_l(\hat \k\cdot \hat \n) P_{l'}(\hat \k \cdot \hat \n') = \frac{4\pi}{2l+1} P_l(\hat \n \cdot \hat \n') \,\delta_{ll'}\, .
\eeq
The two-point function then takes the form (\ref{equ:2pt}) with the angular power spectrum given by
\begin{align}
C_l^{\rm SW} &= \frac{4\pi}{(2l+1)^2} \int \d\ln k\ j_l^2(kr_*)\,T_{\rm SW}^2(k)\, {\cal P}_\zeta(k)\, . \\[-0.2cm]
\nonumber 
&\hspace{3.5cm} \uparrow \hspace{1.cm} \uparrow \hspace{1cm} \uparrow\\[-0.1cm]
&\hspace{2.5cm} \text{\footnotesize projection} \hspace{0.3cm} \text{\footnotesize evolution} \hspace{0.3cm} \text{\footnotesize initial conditions}\nonumber
\end{align}
A similar derivation gives the form of the Doppler contribution.
}
\end{framed}



\newpage
\section{Cosmic Sound Waves}
\label{sec:Sound}

\begin{figure}[h!]
    \centering
      \includegraphics[width=.4\textwidth]{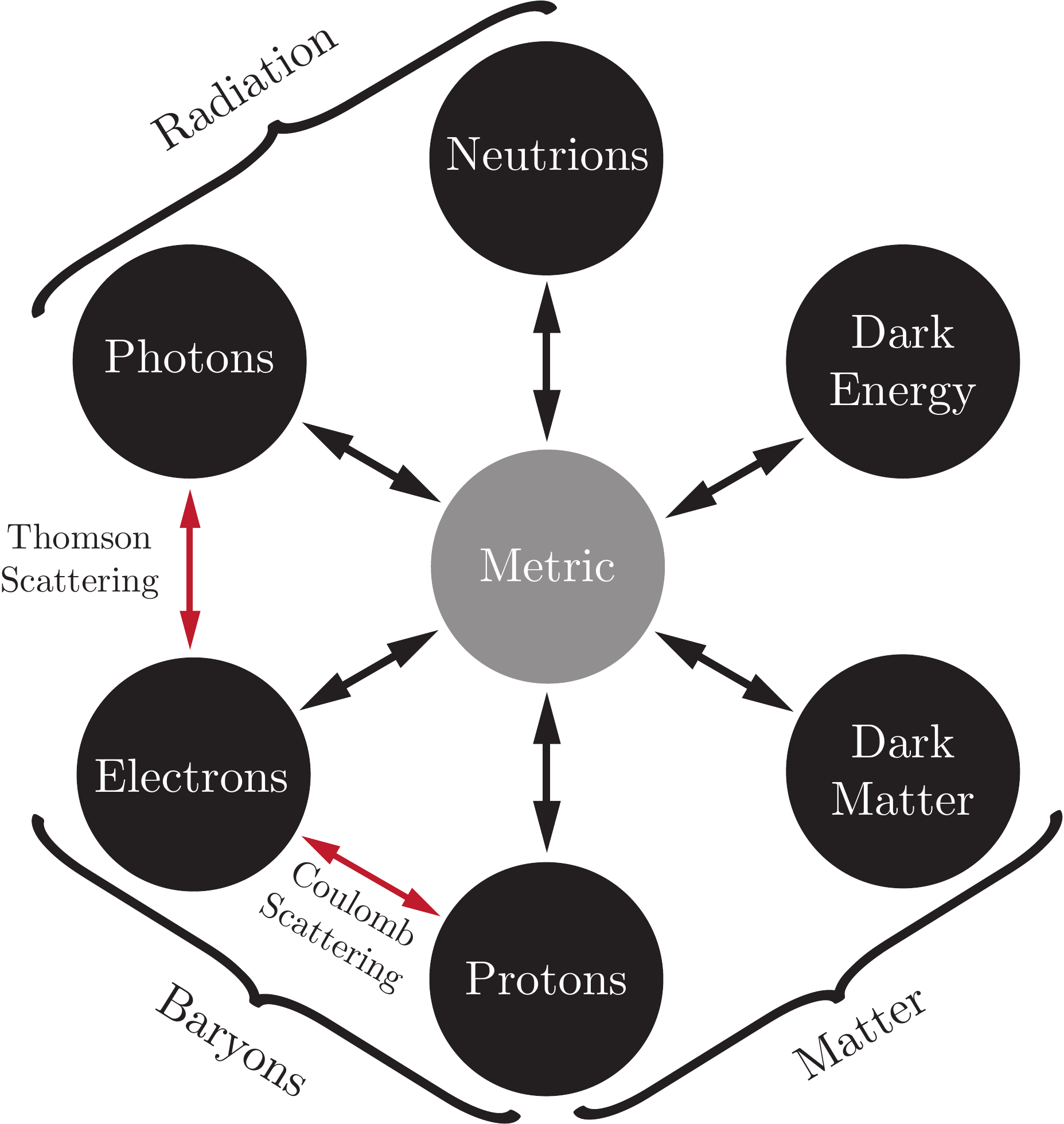}
        \caption{Interactions between the different forms of matter in the universe.}
    \label{fig:interactions}
\end{figure}

In the early universe, photons and electrons were strongly interacting, while the electrons were strongly coupled to protons.
The combined system is often called the {\it photon-baryon fluid}.  In this section, we will study the evolution of sound waves in this medium. 
These waves will evolve in an inhomogeneous spacetime whose perturbations are sourced by all forms of matter in the universe~(see Fig.~\ref{fig:interactions}).

Our treatment in this section will be rather telegraphic and is just meant to give a flavor for the beautiful physics underlying the CMB. Further details can be found in the notes of my {\it Advanced Cosmology} course~\cite{Adv-Cosmo}, or in the following textbooks~\cite{dodelson2003modern, Durrer:2008eom} and review articles~\cite{Hu:1995em, Hu:2001bc, Hu:2008hd, Hu:Concept}.

\subsection{Photon-Baryon Fluid}

Combining the continuity and Euler equations for the photon-baryon fluid leads to an evolution equation for the photon density perturbations~\cite{Adv-Cosmo}:
\begin{align}
&\boxed{\,\ddot \delta_\gamma + \frac{\H R}{1+R} \, \dot \delta_\gamma - c_s^2 \nabla^2 \delta_\gamma \,=\,   \frac{4}{3} \nabla^2 \Phi + 4 \ddot \Psi + \frac{4\H R}{1+R} \, \dot \Psi \,}\ , \label{equ:master} \\[-2pt]
&\hspace{1.7cm} \uparrow \hspace{1.5cm} \uparrow \hspace{1.6cm} \uparrow  \hspace{2cm} \uparrow \nonumber  \\[-4pt]
&\hspace{1.3cm} \text{\small friction} \hspace{0.5cm} \text{\small pressure} \hspace{0.7cm} \text{\small gravity} \hspace{0.7cm} \text{\small time dilation}\nonumber 
\end{align}
where $R \equiv \frac{3}{4} \bar \rho_b/\bar \rho_\gamma$ is the ratio of the momentum densities of baryons and photons, and the sound speed of the photon-baryon fluid is defined as
\beq
c_s^2 \equiv \frac{1}{3(1+R)}\, . \label{equ:cs}
\eeq
Equation~(\ref{equ:master}) is the master equation describing the entire CMB phenomenology.  The most important terms in the equation are the photon pressure term on the left-hand side and the gravitational forcing term on the right-hand side.  In addition, we have a friction term proportional to the baryon density $R$ on the left-hand side and two terms related to time dilation effects on the right-hand side.  The metric potentials $\Phi$ and $\Psi$ are determined by the Einstein equations (which include important contributions from dark matter). 

\begin{figure}[t!]
\begin{center}
\includegraphics[width=11cm,angle=0]{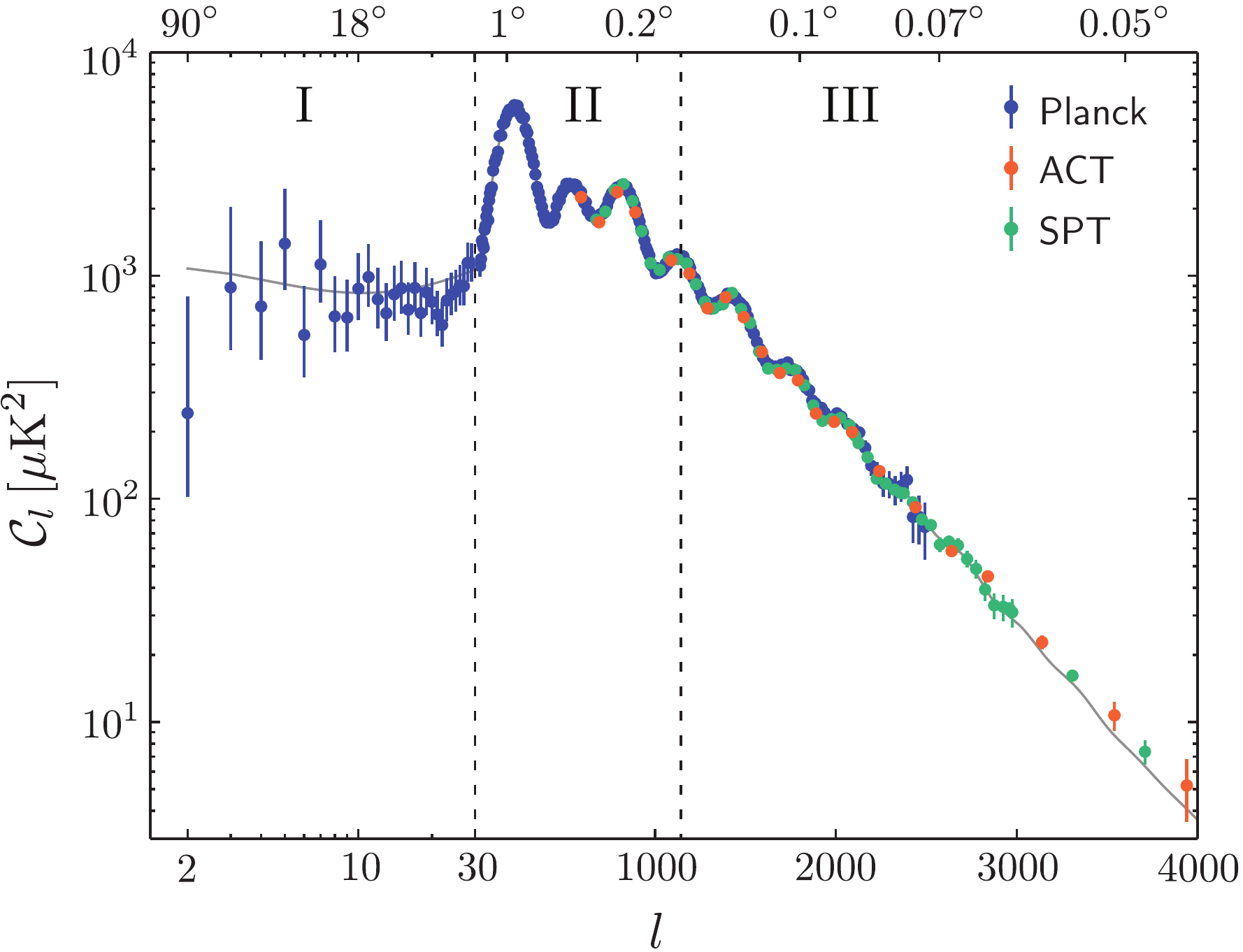}
\end{center}
\vspace{-0.5cm}
\caption{The angular variations of the CMB power spectrum are consequence of the dynamics of sound waves in the photon-baryon fluid. On large scales (region I), the fluctuations are frozen and we directly see the spectrum of the initial conditions. At intermediate scales (region II), we observe the oscillations of the fluid as captured at the moment of last-scattering. Finally, on small scales (region III), fluctuations are damped because their wavelengths are smaller than the mean free path of the photons.}
\label{fig:TASI0}
\end{figure}

\vskip 4pt
In practice, the equations describing the many coupled fluctuations in the primordial plasma have to be solved numerically.
To gain some intuition, however, it is nevertheless useful to obtain approximate analytic results.
In the following, we will solve equation~(\ref{equ:master}) by making several (more or less justified) approximations.
Our goal is to understand the main features of the CMB power spectrum shown in Fig.~\ref{fig:TASI0}.

\subsection{Acoustic Oscillations}
\label{sec:acoustic}

During radiation domination, the baryon density is subdominant, $R\ll 1$, so for the moment we will set $R=0$. 
For simplicity, we will also ignore the time dilation terms in (\ref{equ:master}); we will include their effects in \S\ref{sec:Imprints}. 
Equation~(\ref{equ:master}) can then be written as
\begin{align}
\ddot \Theta - c_s^2 \nabla^2 \Theta  \,=\, 0\, , \label{equ:Master2b} 
\end{align}
where $c_s^2 \approx \frac{1}{3}$ and $\Theta \equiv \frac{1}{4}\delta_\gamma + \Phi$ is precisely the combination of the Sachs-Wolfe term appearing in~(\ref{equ:CMBmaster}).
Solving (\ref{equ:Master2b}) for a single Fourier mode, we get 
\beq
\Theta(\k,\ct) \,=\, A_\k \cos(c_s k \ct) + B_\k \sin(c_s k \ct)\, ,
\eeq
where $A_\k$ and $B_\k$ are parameters that are fixed by the initial conditions.
For adiabatic initial conditions, all perturbations in the limit $\ct \to 0$ are analytic functions of  $k^2$, which is only the case for the cosine part of the solution above. We therefore set $B_\k=0$.
Moreover, the matching to the superhorizon initial conditions implies $A_\k= 3\hskip 1pt \zeta_\k$~\cite{Adv-Cosmo} and we get 
 \beq
\Theta(\k,\ct) \,=\, 3\hskip 1pt \zeta_\k \cos(c_s k \ct) \, .
\eeq
Evaluating this solution at recombination, $\ct=\ct_*$, gives the Sachs-Wolfe transfer function,
$T_{\rm SW}(k) \equiv \Theta(\k,\ct_*)/\zeta_\k = 3 \cos(c_s k \ct_*)$.  As we have seen in \S\ref{sec:PS}, the CMB power spectrum is roughly given by the square of the transfer function.  The oscillatory $k$-dependence of the transfer function then maps to the observed oscillations in the CMB power spectrum in harmonic space, cf.~eq.~(\ref{equ:Cell}).

\paragraph{Sound horizon}  Modes caught at extrema of their oscillations will have enhanced fluctuations
\beq
k_n = n \pi/s_{*}\, ,
\eeq
where $s_{*} \equiv c_s \ct_{*} \approx \frac{1}{\sqrt{3}} \ct_*$ is the sound horizon at recombination.
We see that the peaks occur at multiples of the fundamental scale $k_* \equiv \pi/s_{*} \approx \sqrt{3}\pi/\ct_*$.  This scale becomes a characteristic angular scale by simple projection 
\begin{align}
\theta_* &= \frac{\lambda_*}{D_A} \, ,  \\ 
l_* &= k_* D_A \approx \frac{\ct_*}{\ct_0}\, ,
\end{align}
where $D_A$ is the angular diameter distance (which in a flat universe is $D_A = \ct_0-\ct_* \approx \ct_0$). Assuming a purely matter-dominated universe after recombination, we have $\ct \propto a^{1/2}$ and therefore find 
\begin{align}
\theta_* &\approx \left(\frac{1}{1100}\right)^{1/2} \approx 2^\circ\, , \\
l_* &\approx 200\, .
\end{align}
The presence of dark energy and spatial curvature would slightly change this result.  Measurements on the CMB spectrum have now determined $\theta_*$ to better than $0.05\%$, which puts strong constraints on the geometry and composition of the universe. Keeping the physical densities of dark matter,
baryons, photons and neutrinos fixed, i.e.~$\Omega_i h^2 = const.$, the scale $\theta_*$ is a measure of the curvature parameter $\Omega_k$ through its effect on the angular diameter distance to last-scattering. The observed value of $\theta_*$ is found to be consistent with a flat universe. Allowing the matter density and the Hubble parameter to vary, one finds $\theta_* \approx f(\Omega_m h^3)$, i.e.~there is a specific degeneracy between variations in $\Omega_m$ and $h$.
This degeneracy is broken by measurements of the peak morphology of the CMB spectrum and by external data sets (BAO, supernovae, etc.).

\paragraph{Baryon loading}  
Let me briefly comment on the effects of baryons on the CMB spectrum.
The baryon-to-photon ratio increases with time, $R \propto a$, reaching an order one value at recombination. This decreases the sound speed, cf.~(\ref{equ:cs}).
Instead of (\ref{equ:Master2b}), we now have
\beq
\frac{d^2}{d\ct^2} (\Theta + R\hskip 1pt  \Phi) - \frac{1}{3} \nabla^2(\Theta + R\hskip 1pt \Phi) = 0\, ,
\eeq
where we have ignored the time variation of $R$ relative to the much faster evolution of the acoustic oscillations.
We see that the finite baryon density, $R\ne 0$,
 changes the equilibrium point of the oscillations from $\Theta=0$ to $\Theta = -R\hskip 1pt \Phi$.
Since the CMB spectrum depends on the square of the solution, the shift of the equilibrium of the oscillating solution leads to odd and even peaks in the CMB having unequal heights. The relative heights of the CMB spectrum therefore provide a measure of the baryon density~$\Omega_b$.

\begin{figure}[h!]
\begin{center}
\includegraphics[width=0.75\textwidth]{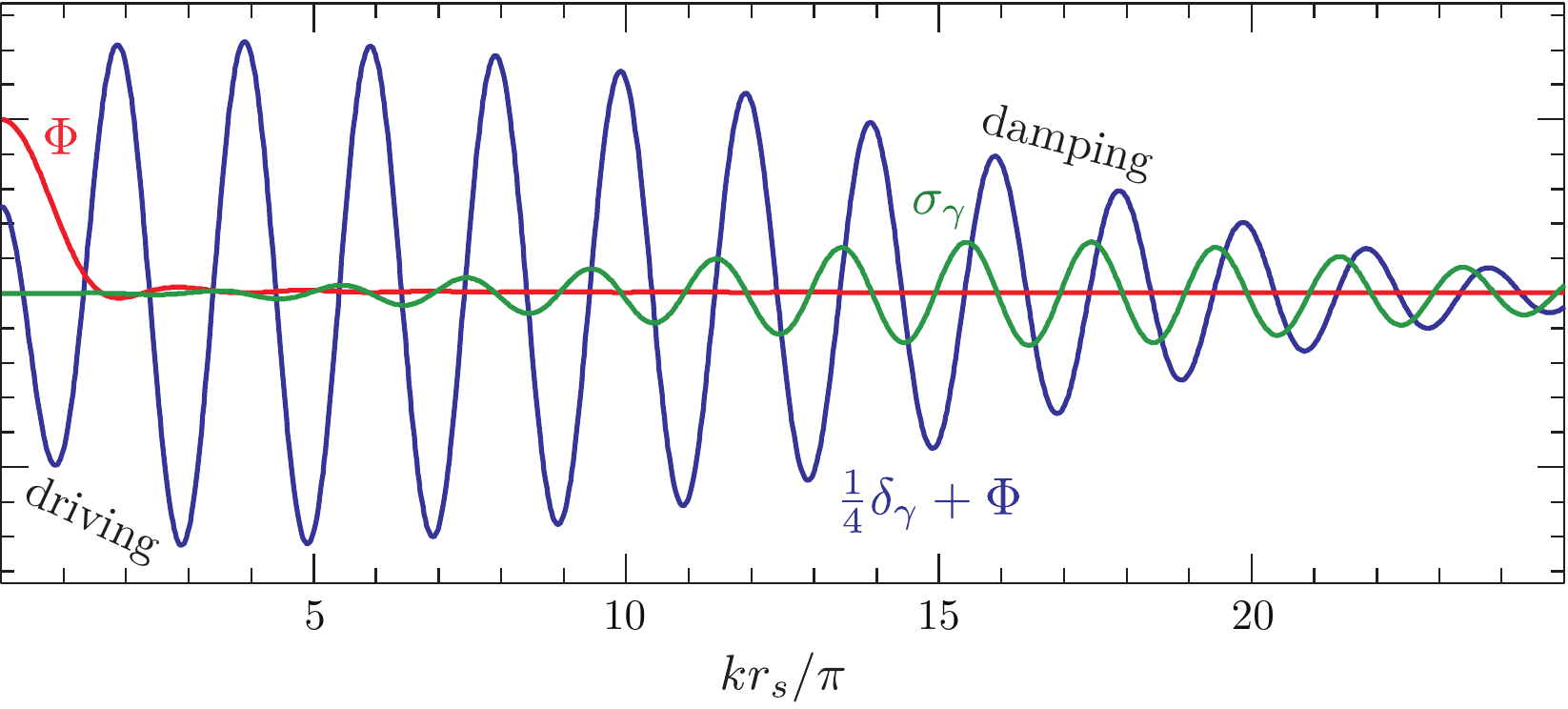}
\caption{Acoustic oscillations with gravitational forcing and diffusion damping.  For
a mode that enters the sound horizon during radiation domination, the gravitational potential
decays after horizon crossing and drives the acoustic amplitude higher.   As the photon 
diffusion length increases and becomes comparable to the wavelength, viscosity
$\sigma_\gamma$ is generated and small-scale fluctuations are washed out.} \label{fig:osc3}
\end{center}
\end{figure}

\paragraph{Radiation driving} One important effect is not included in our highly simplified treatment. During the radiation era 
the gravitational potential $\Phi$ decays inside the horizon. Counterintuitively, the decaying potential actually enhances temperature fluctuations through a subtle resonance effect (see Fig.~\ref{fig:osc3}).
Since the potential $\Phi$ decays after sound horizon crossing, it drives the first compression of the photon-baryon fluid without a counterbalancing effect on the subsequent rarefaction stage. 
The higher peaks in the CMB spectrum correspond to fluctuations that began their oscillations in the radiation-dominated era and therefore have enhanced amplitudes. This effect is sensitive to the ratio of radiation to matter, $\Omega_r/\Omega_m$.
Since the radiations density $\Omega_r$ is fixed by the observed CMB temperature, measuring the peak heights relative to the Sachs-Wolf plateau determines the matter density of the universe $\Omega_m$.

\paragraph{Diffusion damping}  So far, we have treated photons and baryons as a single perfect fluid, i.e.~we took the  mean free path of the photons to be zero.  In reality, the coupling between electrons and photons is imperfect and the photons have a finite mean free path:
\beq
\lambda_C = \frac{1}{n_e \sigma_T a}\, , \label{equ:lambdaC}
\eeq
where $n_e$ is the electron density and $\sigma_T$ is the Thomson cross section.
Accounting for this leads to the damping of small-scale fluctuations:
see Fig.~\ref{fig:TASI0}. 
As the photons random walk through the baryons, hot and cold regions are mixed.
By this process, fluctuations will be erased below the diffusion length:
\beq
\lambda_D = \sqrt{N}\, \lambda_C = \sqrt{\ct/\lambda_C}\, \lambda_C = \sqrt{\ct \lambda_C}\, , \label{equ:diff}
\eeq
which is the geometric mean between the horizon scale and the mean free path. 
As we will show in the following insert, the transfer function for the photon density fluctuations will receive an exponential suppression for modes with $k> k_D\equiv 2\pi/\lambda_D$. As shown in the insert below, the photon transfer function receives the following correction:
\beq
T(k) \to {\cal D}(k)\hskip 1pt T(k)\, ,
\eeq
where ${\cal D}(k) = e^{-k^2/k_D^2}$.

\begin{framed}
{\small \noindent {\it Imperfect fluid.}---Diffusion causes heat conduction and generates viscosity in the fluid.   Incorporating these effects into the dynamics leads to a modified oscillator equation~\cite{Adv-Cosmo},
\beq
\ddot \Theta + \mu\,c_s^2 k^2\,  \dot  \Theta +  c_s^2k^2\,  \Theta = 0\, ,
\eeq
where we have ignored the gravitational source terms and defined
\begin{align}
\mu &\equiv  \left[ \frac{16}{15}  + \frac{R^2}{1+R}\right] \lambda_C\, .
\end{align}
Using the WKB ansatz
\beq
 \Theta \propto \exp\left(i\int \omega \, \d \ct\right) , \label{equ:WKB}
\eeq
we get 
\beq
-\omega^2 + \mu\, c_s^2 k^2 \, i \omega +  c_s^2 k^2 = 0\, ,
\eeq 
which we can write as
\begin{align}
\omega &= \pm c_s k \Big[1+i \omega \mu \Big]^{1/2} \approx \pm c_sk \Big[1\pm \frac{i}{2}\mu\, c_s k\Big] \, .
\end{align}
Substituting this back into (\ref{equ:WKB}), we get
\beq
 \Theta \propto = e^{\pm i k r_s} \exp\left[- \frac{1}{2}(k/k_D)^2 \right] ,
\eeq
where we have defined the diffusion wavenumber as
\beq
k_D^{-2} = \int \d \eta \, \mu\, c_s^2 = \int \d \ct \, \frac{1}{3(1+R)} \left[ \frac{16}{15}  + \frac{R^2}{1+R}\right] \lambda_C\, .
\eeq
In the limit $R\to 0$, this becomes
\beq
k_D^{-2} \approx \frac{16}{45} \int \d \ct\, \lambda_C \sim \ct\lambda_C\, ,
\eeq
which agrees with our previous estimate (\ref{equ:diff}).}
\end{framed}

\noindent
Within the Standard Model, the physics of the CMB anisotropies is understood extremely well.
In the next section, we will explore what can be learned about physics beyond the Standard Model by looking for subtle deviations in the CMB spectrum.

\newpage
\section{Light Relics}
\label{sec:LightRelics}

Future cosmological observations will measure the radiation density of the early universe at the percent level.  
In this section, I will show how these observations will probe the physics of neutrinos, as well as the possible existence of extra light particles that are more weakly coupled to the SM than neutrinos. Examples of light relics that can be constrained in this way are the QCD axion~\cite{1977PhRvL..38.1440P, Weinberg:1977ma, 1978PhRvL..40..279W}, axion-like particles (ALPs)~\cite{Arvanitaki:2009fg}, dark photons~\cite{Holdom:1985ag} and light sterile neutrinos~\cite{Abazajian:2012ys}.  These particles may be so weakly coupled that they are hard to detect in terrestrial experiments, but the large number densities in the early universe make their gravitational imprints observable.

\subsection{Dark Radiation}

Let us assume that some physics beyond the Standard Model adds an extra radiation density $\rho_X$ to the early universe. It is conventional to measure this density relative to the density of a SM neutrino species:
\beq
\Delta N_{\rm eff} \equiv \frac{\rho_X}{\rho_\nu} = \frac{1}{a_\nu} \frac{\rho_X}{\rho_\gamma}\, ,
\eeq
and define $N_{\rm eff} = N_\nu + \Delta N_{\rm eff}$ as the {\it effective number of neutrinos}, although $\rho_X$ may have nothing to do with neutrinos.
 Current measurements of the CMB anisotropies and the light element abundances find
\begin{align}
N_{\rm eff}^{\rm CMB} = 3.04 \pm 0.18\, , \label{equ:NeffCMB} \\[4pt]
N_{\rm eff}^{\rm BBN} = 2.85 \pm 0.28\, ,
\end{align}
which is consistent with the SM prediction,\footnote{The predicted value of $N_{\rm eff} = 3.046$ accounts for plasma corrections of
quantum electrodynamics, flavour oscillations and, in particular, the fact that neutrinos have not
fully decoupled when electrons and positrons annihilated.} $N_{\rm eff} = 3.046$.
Future CMB observations have the potential to improve these constraints by an order of magnitude~\cite{Abazajian:2016yjj}.

\begin{figure}[t]
\begin{center}
\includegraphics[width=0.75\textwidth]{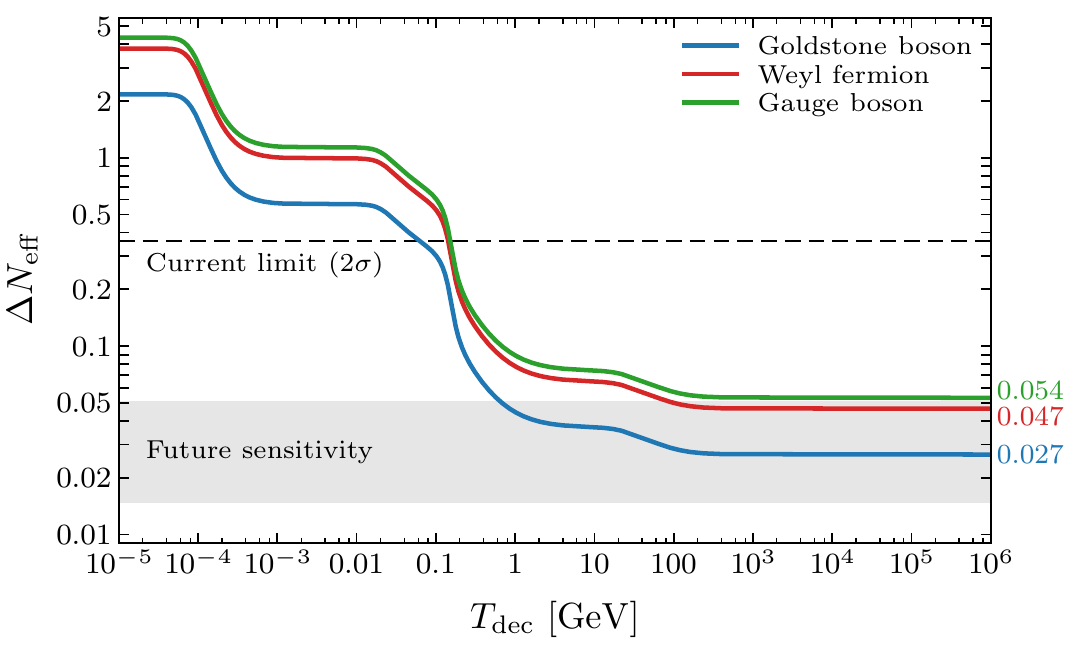}
\vspace{-0.1cm}
\caption{Contributions of a single thermally-decoupled Goldstone boson, Weyl fermion or massless gauge boson to the effective number of neutrinos, $\Delta N_{\rm eff}$, as a function of its decoupling temperature $T_{\rm dec}$. }
\label{fig:freezeout}
\end{center}
\end{figure}

\vskip 4pt
A natural source for $\Delta N_{\rm eff} \ne 0$ are extra relativistic particles.
Let us therefore consider a light species $X$ as the only additional particle in some BSM theory. Assuming that this species was
in thermal equilibrium with the SM at some point in the history of the universe, we
can compute its contribution to $N_{\rm eff}$ in the same way as we derived the relic density of neutrinos in~\S\ref{ssec:ThermalHistory}.  For concreteness, let us assume that the particles of the species $X$ decouple before neutrino decoupling, $T_{{\rm dec},X} > 10$ MeV. 
Particle-antiparticle annihilations until neutrino decoupling will lead to a difference between the temperature associated with the species~$X$ and that of neutrinos:
\begin{align}
T_X = \left(\frac{g_*(T_{{\rm dec},\nu})}{g_*(T_{{\rm dec},X})}\right)^{1/3} T_\nu &=   \left(\frac{10.75}{106.75}\right)^{1/3}  \left(\frac{106.75}{g_*(T_{{\rm dec},X})}\right)^{1/3} T_\nu \nonumber \\[4pt]
&= 0.465 \left(\frac{106.75}{g_*(T_{{\rm dec},X})}\right)^{1/3} T_\nu\, .
\end{align} 
After neutrino decoupling, $T_X$ and $T_\nu$ evolve in the same way, with both receiving the same suppression relative to $T_\gamma$ from $e^+ e^-$ annihilation.
As long as both  $X$ and $\nu$ are relativistic, they therefore maintain a constant energy ratio
\beq
\Delta N_{\rm eff} \equiv \frac{\rho_X}{\rho_\nu} = \frac{g_{*,X}}{g_{*,\nu}} \left(\frac{T_X}{T_\nu}\right)^4 = 0.027 \,g_{*,X} \left(\frac{106.75}{g_*(T_{{\rm dec},X})}\right)^{4/3} \, , \label{equ:55}
\eeq
where $g_{*,\nu}=\frac{7}{4}$ and $g_{*,X} = \{1, \frac{7}{4}, 2,\ldots\}$ are the internal degrees of freedom for spin $\{0,\frac{1}{2},1, \ldots\}$ particles.
Figure~\ref{fig:freezeout} shows the contribution of a single decoupled species as a function of its decoupling temperature.
We see that the contributions asymptote to fixed values for decoupling temperatures above the mass of the top quark (the heaviest particle of the SM).
 Using $g_*(T_{{\rm dec},X})\le 106.75$ in expression (\ref{equ:55}), we find that the extra species $X$ contributes the following minimal amount\footnote{In deriving this bound, we assumed an extension of the SM in which there is no significant entropy production after decoupling and that the species $X$ is the only addition to the SM particle content. Additional particles may both increase $\Delta N_{\rm eff}$ is there are light enough, or decrease it if they are relativistic at the decoupling of $X$, but become non-relativistic before neutrino decoupling. While entropy production typically dilutes the effects of extra relativistic species, it can also lead to additional effects that can be looked for in cosmological observations. For a more detailed discussion of these issues, see~\cite{BenThesis}. \label{foot}} 
\beq
\boxed{\Delta N_{\rm eff} > 0.027 \,g_{*,X}} \ =\ \left\{ \begin{array}{ll} 0.054 & \hspace{0.5cm}\text{gauge boson} \\[6pt] 0.047 &  \hspace{0.5cm}\text{Weyl fermion} \\[6pt] 0.027 &  \hspace{0.5cm}\text{Goldstone boson} \end{array}\right. 
\eeq
As we will see, this is an interesting target for future CMB experiments.

\newpage
\subsection{Imprints in the CMB}
\label{sec:Imprints}

Adding extra relativistic species will change the shape of the CMB spectrum.
Precise measurements of the spectrum therefore make the CMB an accurate tool for probing this type of BSM physics.

\begin{figure}[b!]
\begin{center}
\includegraphics[width=0.7\textwidth]{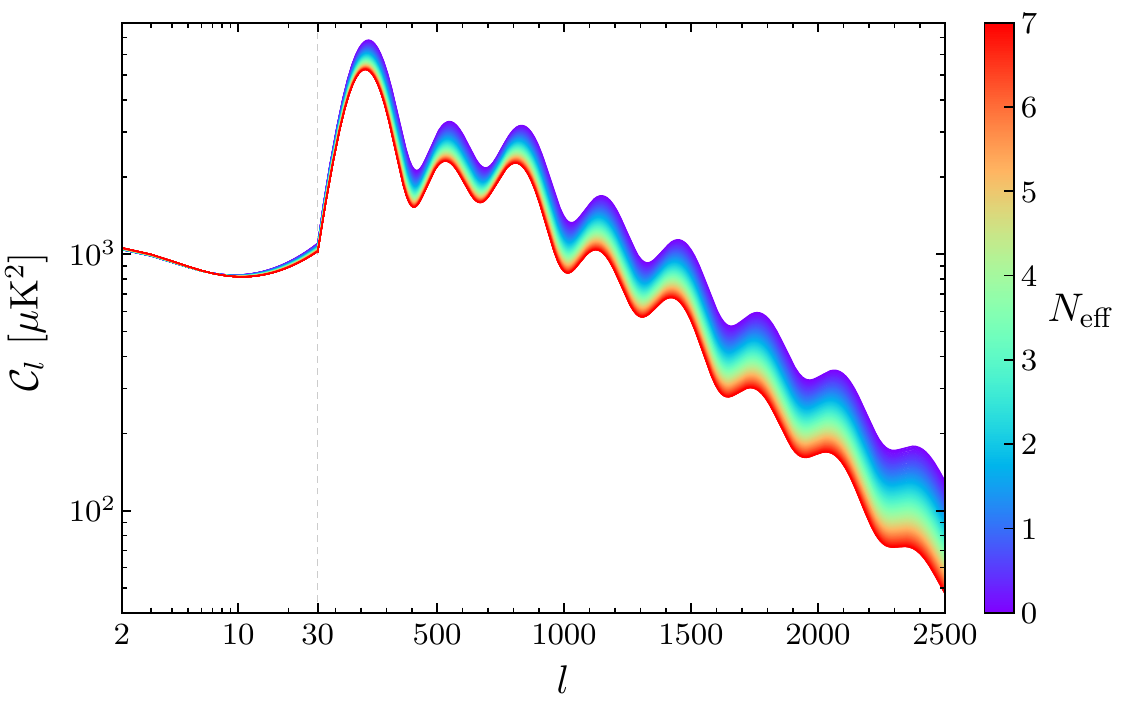}
\caption{Variation of the CMB spectrum ${\cal C}_l \equiv l(l+1) C_l$ as a function of $N_{\rm eff}$ for fixed $\theta_*$.}
\label{fig:damping2}
\end{center}
\end{figure}

\paragraph{Diffusion damping} The main effect of adding radiation density to the early universe is to increase the damping of the CMB spectrum~\cite{Hou:2011ec} (see Fig.~\ref{fig:damping2}). Increasing $N_{\rm eff}$ increases $H_*$, the expansion rate at recombination.  This would change both the damping scale $\theta_D$ and the peak location $\theta_*$. 
Using the estimates presented in \S\ref{sec:acoustic}, the ratio of $\theta_D$ and $\theta_*$ scales as
\beq
\frac{\theta_D}{\theta_*} = \frac{1}{r_{s,*} \,k_D} \propto \frac{1}{H_*^{-1} H_*^{1/2}} = H_*^{1/2}\, .
\eeq
Since $\theta_*$ is measured very accurately, we need to keep it fixed. This can be done, for example, by simultaneously increasing the Hubble constant $H_0$.
Increasing $N_{\rm eff}$ (and hence $H_*$) at fixed $\theta_*$ then implies larger $\theta_D$, i.e.~the damping kicks in at larger scales reducing the power in the damping tail (see Fig.~\ref{fig:damping2}).  By accurately measuring the small-scale CMB anisotropies, observations therefore put a constraint on the number of relativistic species at recombination, cf.~eq.~(\ref{equ:NeffCMB}). 

The main limiting factor in these measurements is a degeneracy with the primordial Helium fraction $Y_P \equiv n_{\rm He}/n_b$. At fixed $\omega_b \equiv \Omega_b h^2$, increasing $Y_P$ decreases the number density of free electrons. This increases the diffusion length, cf.~eq.~(\ref{equ:lambdaC}), and hence reduces the power in the damping tail. The parameters $Y_P$ and $N_{\rm eff}$ are therefore anti-correlated.  As we will discuss next, this degeneracy is broken by a more subtle effect of free-streaming relativistic species on the CMB spectrum.

\begin{figure}[t!]
\begin{center}
\includegraphics[width=0.7\textwidth]{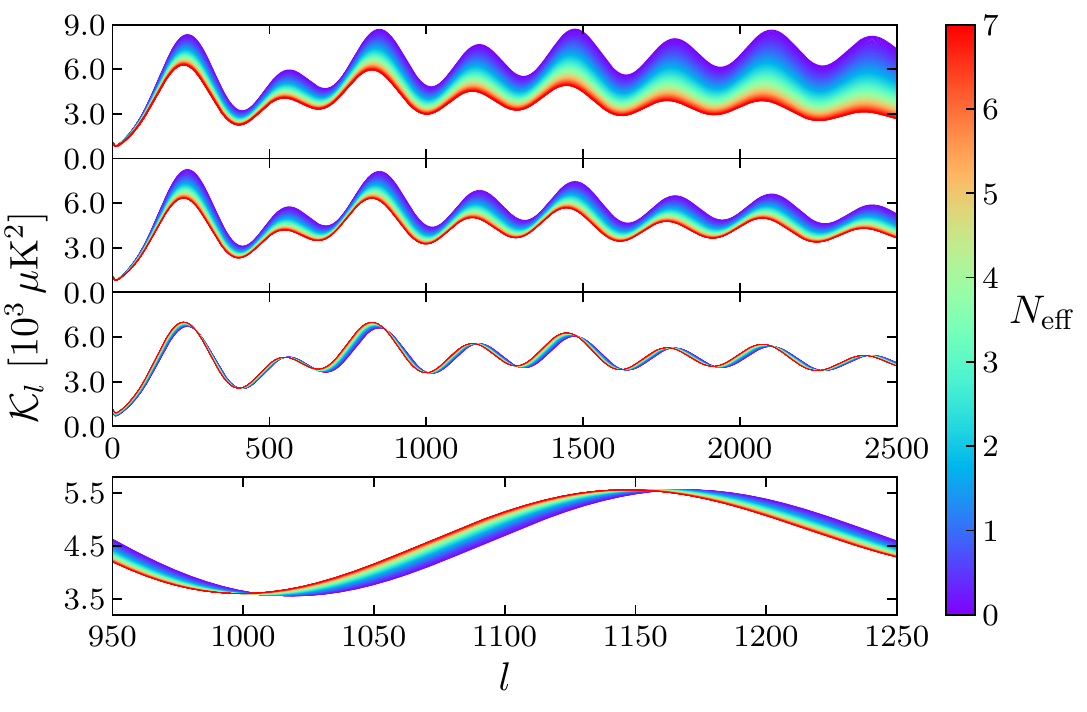}
\caption{Variation of the undamped power spectra, ${\cal K}_l \equiv {\cal D}_l^{-1} {\cal C}_l$, as a function of $N_{\rm eff}$. The physical baryon density $\omega_b$, the matter-to-radiation ratio $\rho_m/\rho_r$ and the angular size of the sound horizon $\theta_*$ are held fixed in all panels. The dominant effect in the first panel is the variation of the damping scale $\theta_D$. In the second panel, we fixed $\theta_D$ by adjusting the Helium fraction $Y_P$. The dominant variation is now the amplitude perturbation $\delta A$. In the third panel, the spectra are normalized at the fourth peak. The remaining variation is the phase shift $\varphi$ (see the zoom-in in the fourth panel).}
\label{fig:PhaseShift}
\end{center}
\end{figure}

\paragraph{Phase shift} Recently, CMB experiments have started to become sensitive to perturbations in the gas of relativistic particles. As we will see, perturbations in the density of free-streaming relativistic particles (e.g.~neutrinos) leave a small imprint in the temporal phase of the acoustic oscillations and hence a coherent shift in the peak locations of the CMB spectrum (see Fig.~\ref{fig:PhaseShift}). 
 To get some intuition for the physical origin of this effect, let us return to the master equation~(\ref{equ:master}), but now solve it slightly more accurately. 
Since the effect occurs during the radiation-dominated era, we can still assume perfect radiation domination with $R=0$. 
Equation (\ref{equ:master}) then becomes~\cite{Bashinsky:2003tk}
\beq
\ddot d_\gamma - c_s^2 \nabla^2 d_\gamma  \,=\ \nabla^2 \Phi_+\ , \label{equ:Master2} 
\eeq
where we have defined $d_\gamma \equiv \frac{3}{4}\delta_\gamma - 3 \Phi$ and $\Phi_\pm \equiv \Phi \pm \Psi$. 
The solution for a single Fourier mode can be written as
\beq
d_\gamma(y) = A \cos(y) - c_s^{-2} \int_0^y \d y'\, \Phi_{+}(y')\, \sin(y-y')\, ,
\eeq
where $y\equiv c_s k \ct$ and we have dropped the argument $\k$ on $d_\gamma$, $A$, and $\Phi_+$ to avoid clutter. Using $\sin(y-y') = \sin(y) \cos(y')-\cos(y)\sin(y')$, we can write this as
\beq
d_\gamma(y) = \left[A + c_s^{-2} \alpha(y) \right] \cos(y) - c_s^{-2} \beta(y)\sin(y)\, , \label{equ:INHOM}
\eeq
where 
\begin{align}
\alpha(y) &\equiv \int_0^y \d y' \, \Phi_+(y')\, \sin(y') \, ,\\
\beta(y) & \equiv \int_0^y \d y' \, \Phi_+(y')\, \cos(y') \, .
\end{align}
To obtain the CMB spectrum, we need to evaluate the solution at recombination.
For the high-$l$ modes of the CMB, it is a good approximation to use $y\to \infty$ for the limit of integration.
The solution (\ref{equ:INHOM}) can then be written as
\beq
d_\gamma(y) = \tilde A \cos(y+\varphi)\, ,
\eeq
where $\tilde A \equiv A + c_s^{-2} \alpha$ and
\beq
\sin \varphi \equiv \frac{\beta}{\sqrt{\beta^2+(\alpha + c_s^2 A)^2}}\, .
\eeq
We see that $\beta \ne 0$ corresponds to a constant phase shift of the acoustic oscillations.

\vskip 4pt
To diagnose when such a phase shift can occur, it is useful to write the parameter $\beta$ as follows
\begin{align}
\beta &\,=\, \int_0^\infty \d y'\, \Phi_+(y') \cos(y') \nonumber \\
 &\,=\, \frac{1}{2} \int_{-\infty}^{+\infty} \d y'\, \Big[\Phi_+^{({\rm s})}(y')\Big]\, e^{iy'}\, ,
\end{align}
where, in the second line, we have analytically continued the integrand and defined the symmetric part of the potential, $\Phi_+^{({\rm s})}(y) \equiv  \Phi_+(y) + \Phi_+(-y)$.
For adiabatic modes, we expect $\Phi_+^{({\rm s})}(y)$ to be an analytic function.
Closing the contour in the upper-half plane, we find $\beta=0$ if the contour at infinity vanishes.  This is the case if $\Phi_+^{({\rm s})}$ is sourced by fluctuations that travel at $c< c_s$.  
Neutrinos or other free-streaming particles, on the other hand, travel at the speed of light. 
This induces a mode in $\Phi_+^{({\rm s})}$ of the form $e^{-ic k \eta} = e^{-i(c/c_s)y}$, with $c=1 > c_s$, and therefore leads to a finite phase shift.

\begin{framed}
{\small \noindent {\it Neutrino free-streaming.}---We will briefly sketch how free-streaming relativistic particles, such as neutrinos, produce a phase shift in the CMB anisotropy spectrum. Details can be found in~\cite{Bashinsky:2003tk,Baumann:2015rya}.

\vskip 4pt
The evolution of $\Phi_+$ is related to that of $\Phi_-$ via the following Einstein equation:
\begin{align}
\Phi_+'' + \frac{4}{y} \Phi_+' + \Phi_+ \ =\  \Phi_-'' + \frac{2}{y}\Phi_-' + 3 \Phi_- \ \equiv\ {\cal S}[\Phi_-]\, ,
\end{align}
where $\Phi_-$ is sourced by the anisotropic stress $\pi_\nu$ created by the neutrinos:
\beq
\Phi_-(y) = - \frac{2k^2}{y^2} f_\nu \pi_\nu(y)\, . \label{equ:PhiM}
\eeq
Here, we have introduced the fractional neutrino density $f_\nu \equiv \sum \rho_\nu/\rho_r \approx 0.41$.
The evolution of~$\pi_\nu$ follows from the Boltzmann equation for the neutrino distribution function. The solution can be written as~\cite{Bashinsky:2003tk} 
\beq
\frac{k^2}{2} \pi_\nu(y) \approx - \zeta\, j_2(c_s^{-1}y) + c_s^{-1} \int_0^y \d y'\, \Phi_+(y')\left[\frac{2}{5}j_1(c_s^{-1}(y-y')) - \frac{3}{5}j_3(c_s^{-1}(y-y'))\right] .
\eeq
We see that the solution involves an integral over $\Phi_+$. Moreover, the solution depends on $c_s^{-1}y = k \ct$, i.e.~it contains modes travelling at the speed of light.
  Following~\cite{Bashinsky:2003tk}, the system of equations can be solved order by order in the fractional neutrino density: 
  \begin{itemize}
  \item At zeroth order in $f_\nu$, the potential $\Phi_-$ vanishes and the homogeneous solution is a function only of $y$, i.e.~it doesn't contain modes travelling faster than the sound speed of the photon-baryon fluid. No phase shift is generated.
  \item At first order in $f_\nu$, the potential $\Phi_-$ is non-zero. Note that the right-hand side of (\ref{equ:PhiM}) is proportional to $f_\nu$, so only the zeroth-order solution for $\pi_\nu$ (and $\Phi_+$) is needed to determine the first-order solution for $\Phi_-$.
Computing the induced first-order correction to $\Phi_+$, one finds~\cite{Bashinsky:2003tk,Baumann:2015rya}
\beq
\beta \approx 0.60\, \zeta\, f_\nu \quad {\rm and} \quad \varphi \approx 0.19\hskip 1pt \pi\, f_\nu\, .
\eeq
As expected, a finite phase shift is generated.  This phase shift has recently been detected in the CMB data~\cite{Follin:2015hya, Baumann:2015rya}. It has also been measured in the clustering of galaxies via its imprint in the spectrum of baryon acoustic oscillations~\cite{Baumann:2018qnt}.
\end{itemize}
\vspace{-0.2cm}
}
\end{framed}

\begin{figure}[h!]
\begin{center}
\includegraphics[scale=0.57]{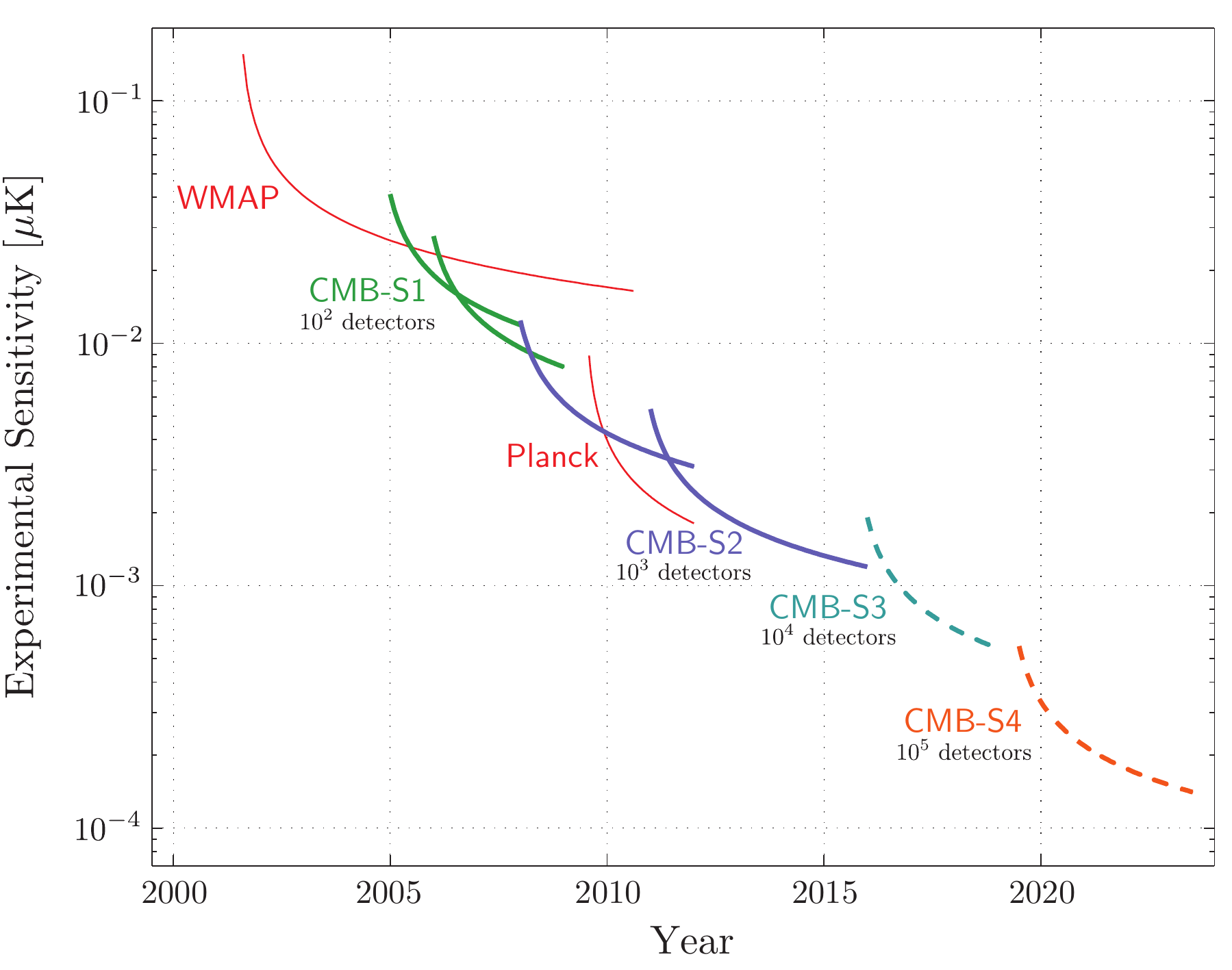}
\caption{Evolution of the sensitivity of past and future CMB experiments (figure adapted from~\cite{Abazajian:2016yjj}).}
\label{fig:Moore}
\end{center}
\end{figure}

\paragraph{CMB Stage 4} The sensitivity of ground-based CMB experiments can be 
characterized by the number of detectors that are mounted onto the telescope.  The current generation of experiments has about $10^3$ detectors, but there are plans to increase the number of detectors by up to two orders of magnitude~\cite{Abazajian:2016yjj}.  
This would lead to a significant improvement in the sensitivity of CMB experiments (see Fig.~\ref{fig:Moore}).  What is particularly exciting about this is that it will lead to an order of magnitude improvement in constraints on $N_{\rm eff}$, allowing us to probe particles that decoupled before the QCD phase transition (cf.~Fig.~\ref{fig:freezeout}).  If these so-called CMB Stage 4 experiments can reach the threshold $\Delta N_{\rm eff}=0.027$, they would be sensitive to any light relics that have ever been in thermal equilibrium with the Standard Model (modulo the constraints described in footnote~\ref{foot}).
As we will show in the next section, even the absence of a detection would be informative since it would put strong constraints on the couplings of extra light species to the SM.

\newpage
\subsection{EFT of Light Species}

Extra light species arise in many models of physics beyond the Standard Model. Rather than studying these models one by one, it is more efficient to pass directly to an effective field theory for the couplings of the new light fields $X$ to the SM,
\beq
{\cal L} \,\subset\, \sum g \,{\cal O}_X {\cal O}_{\rm SM}\, . \label{equ:Left}
\eeq
The strength of the couplings $g$ determines the decoupling temperature, $T_{\rm dec}(g)$, and hence the thermal abundance of the species $X$, cf.~Fig.~\ref{fig:freezeout}.

\vskip 4pt
Following~\cite{Brust:2013xpv}, we consider EFTs that are minimal
and technically natural. Minimality here means that the additional particle content is as small as
possible, usually consisting of only one additional elementary particle. 
Naturalness requires the small masses of the extra particles to be protected by symmetries, which also restricts the allowed interactions in (\ref{equ:Left}). Since the available symmetries depend on the spin of the new particles, it is convenient to organize the EFT according to spin.

\paragraph{Spin-0} A particularly well-motivated example of light particles are {\it Goldstone bosons} created by the spontaneous breaking of additional global symmetries. Goldstone bosons are either massless (if the broken symmetry was exact) or naturally light (if it was approximate).
 Examples of light pseudo-Nambu-Goldstone bosons (pNGBs) are \textit{axions}~\cite{1977PhRvL..38.1440P,Weinberg:1977ma,1978PhRvL..40..279W}, \textit{familons}~\cite{PhysRevLett.49.1549,reiss1982can,kim1987light}, and \textit{majorons}~\cite{Chikashige:1980ui,Chikashige:1980qk}, associated with spontaneously broken Peccei-Quinn, family and lepton-number symmetry, respectively. Axion, familon and majoron models are characterized by different couplings in~(\ref{equ:Left}).

Axions arise naturally in many areas of high-energy physics, the QCD axion being a famous example. They are a compelling example of a new particle that is experimentally elusive because of its weak coupling rather than due to kinematic constraints. What typically distinguishes axions from other pNGBs are their unique couplings to the SM gauge fields. Below the scale of electroweak symmetry breaking, the shift-symmetric couplings of the axion to the SM gauge fields are
\begin{equation}
\mathcal{L} = -\frac{1}{4} \left( \frac{a}{\Lambda_\gamma} \hskip1pt F_{\mu\nu} \tilde{F}^{\mu\nu} + \frac{a}{\Lambda_g} \hskip1pt \hskip1pt G_{\mu\nu}^a \tilde{G}^{\mu\nu,a} \right) , \label{equ:LEW0}
\end{equation}
where $X_{\mu \nu}\equiv\{F_{\mu\nu},G_{\mu\nu}^a\}$ are the field strengths of photons and gluons, and $\tilde{X}^{\mu\nu}\equiv\frac{1}{2}\epsilon^{\mu\nu\rho\sigma}X_{\rho\sigma}$ are their duals. Axion models will typically include couplings to all SM gauge fields, but only the coupling to gluons is strictly necessary to solve the strong CP problem.

\paragraph{Spin-$\boldsymbol{\frac{1}{2}}$} Light {\it fermions} are a natural possibility because Weyl and Dirac mass terms are protected by chiral and axial symmetries, respectively. A hidden Weyl fermion $\chi$ then couples to the SM through an anapole moment, $\chi^\dagger \bar \sigma^\mu \chi \partial^\nu B_{\mu \nu}$, or a four-fermion interaction, $\chi^\dagger \bar \sigma^\mu \chi \psi \gamma_\mu \psi$, while a
Dirac fermion $\Psi$ can couple through a dipole interaction, $\bar \Psi \sigma^{\mu \nu} \Psi B_{\mu \nu}$.

\paragraph{\bf Spin-1}Massless spin-1 particles are technically natural because they have fewer degrees of freedom than their massive counterparts. The dominant coupling of {\it dark photons} $A^\prime_\mu$ to the SM is through the dipole interaction $H \bar \psi \sigma^{\mu \nu} \psi F_{\mu \nu}^\prime$.

\paragraph{Spin-$\boldsymbol{\frac{3}{2}}$} The {\it gravitino} is a universal prediction of supergravity. Its mass is set by the SUSY breaking scale, $m_{3/2}= F/M_{\rm pl}$, and can be very small in low-scale SUSY breaking scenarios.  The coupling of the longitudinal component of the gravitino to the SM is equivalent to the Goldstino coupling $\chi^\dagger \sigma_\mu \partial_\nu \chi \,T^{\mu \nu}$. The strength of the coupling is set by the SUSY-breaking scale $F$ rather
than $M_{\rm pl}$.

\paragraph{Spin-2} The {\it graviton} interacts only through Planck-suppressed gravitational interactions and hence has never been in thermal equilibrium with the SM.  Its thermal abundance is therefore negligible.

\vskip 4pt
\paragraph{Cosmic axions} Let us illustrate the power of future CMB observations through the example of axions~\cite{Baumann:2016wac}.   To simplify the narrative, I will assume that a future CMB-S4 mission will be sensitive enough to exclude the minimal abundance of thermal axions, $\Delta N_{\rm eff} > 0.027$. In practice, this will probably require additional data from large-scale structure surveys~\cite{Baumann:2017gkg}.

At high energies, the rate of axion production is through the gauge field interactions (\ref{equ:LEW0}) and can be expressed as~\cite{Salvio:2013iaa} 
\begin{equation}
\Gamma(\Lambda_n,T) = \sum_n\gamma_n(T)\, \frac{T^3}{\Lambda_n^2}\, .\label{eqn:Gamma}
\end{equation}
The prefactors $\gamma_n(T)$ have their origin in the running of the couplings and are only weakly dependent on temperature. We will ignore this temperature dependence in the following.
We see that the production rate, $\Gamma\propto{T^3}$, decreases faster than the expansion rate during the radiation era, $H\propto{T^2}$. 
To avoid producing a thermal axion abundance requires that the interaction rate was never larger than the expansion rate.  Denoting the reheating temperature of the universe by~$T_R$, this implies
\begin{equation}
\Gamma(\Lambda_n,T_R) < H(T_R) = \frac{\pi}{\sqrt{90}} \sqrt{g_{*,R}}\,\frac{T_R^2}{\Mp}\, ,
\end{equation}
where $g_{*,R}\equiv{g_*(T_R)}$.
For a given reheating temperature, this is a constraint on the couplings~$\Lambda_n$ in~(\ref{eqn:Gamma}). Treating the different axion couplings separately, we can write
\begin{equation}
\Lambda_n \, >\, \left(\frac{\pi^2}{90} g_{*,R}\right)^{\!-1/4} \sqrt{\gamma_{n,R} \hskip1pt T_R\hskip1pt \Mp}\, ,
\label{eq:freezeoutconstraint}
\end{equation}
where $\gamma_{n,R}\equiv\gamma_n(T_R)$. 

\begin{figure}[t!]
\begin{center}
\includegraphics[width=0.65\columnwidth]{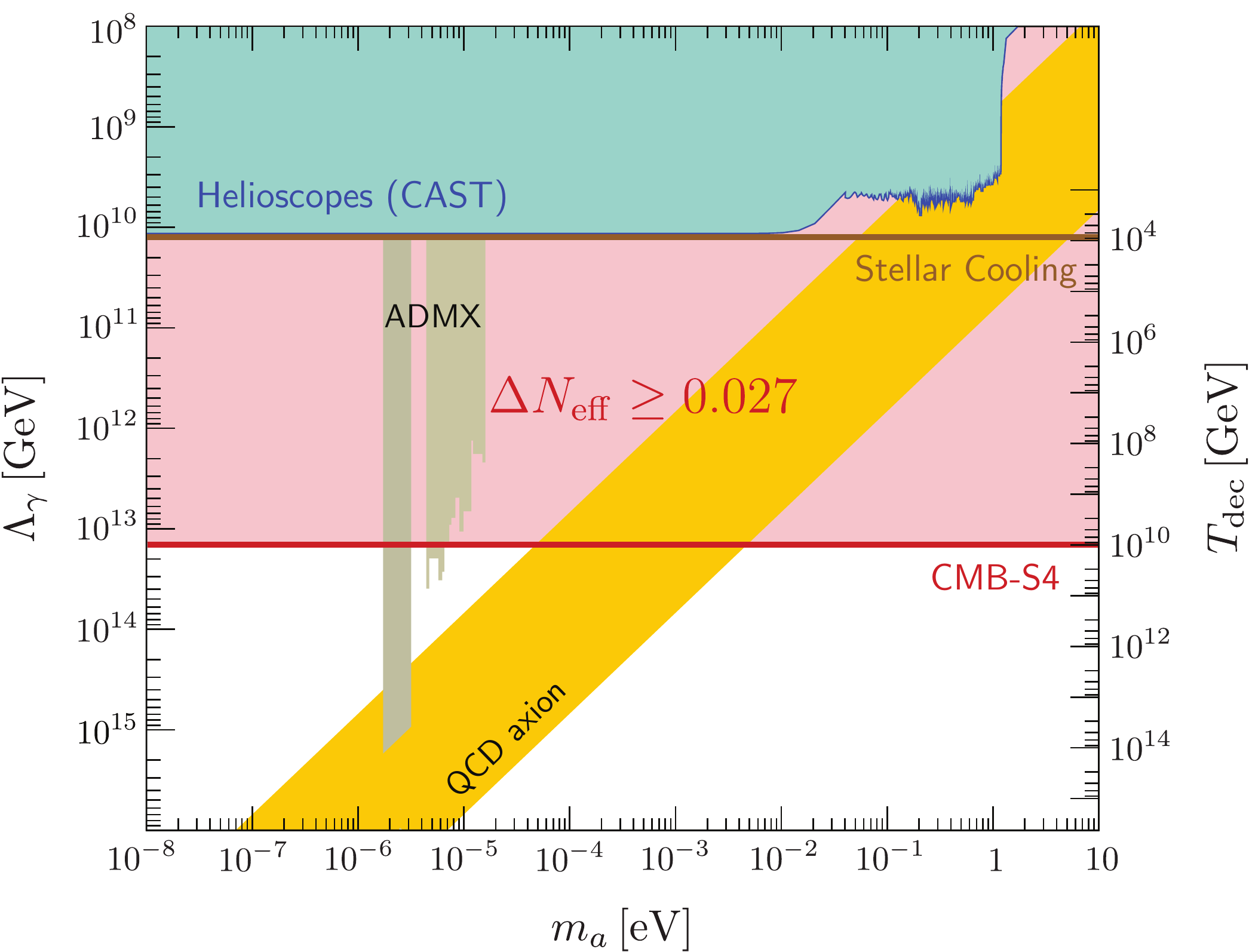}
\caption{Comparison between current constraints on the axion-photon coupling and the sensitivity of a future CMB-S4 mission (figure adapted from~\cite{Carosi:2013rla}). The yellow band indicates a range of representative models for the QCD axion. The future CMB bound is a function of the reheating temperature~$T_R$. We note that ADMX assumes that the axion is all of the dark matter, while all other constraints do not have this restriction.}
\label{fig:S4axion}
\end{center}
\end{figure}

\vskip 4pt
The operator that has been most actively investigated experimentally is the coupling to photons. Photons are easily produced in large numbers in both the laboratory and in many astrophysical settings which makes this coupling a particularly fruitful target for axion searches. The couplings in the high-energy theory prior to electroweak symmetry breaking are related to the photon coupling $\Lambda_\gamma$ through the Weinberg mixing angle. To be conservative, I will present the weakest constraint which arises when the axion only couples to the $U(1)_Y$ gauge field. A specific axion model is likely to also couple to the $SU(2)_L$ sector and the constraint on $\Lambda_\gamma$ would then be stronger. Using $\gamma_{\gamma,R}\approx\gamma_\gamma(\SI{e10}{GeV})=0.029$ and $g_{*,R}=106.75+1$, we find
\begin{equation}
\Lambda_\gamma \,>\, \SI{1.4e13}{GeV}\, \sqrt{T_{R,10}}\, ,
\label{equ:PhotonBound}
\end{equation}
where $T_{R,10}\equiv T_R/\SI{e10}{GeV}$. For a reheating temperature of about \SI{e10}{GeV}, the bound in~(\ref{equ:PhotonBound}) is three orders of magnitude stronger than the best current constraints (cf.~Fig.~\ref{fig:S4axion}). Even for a reheating temperature as low as \SI{e4}{GeV} the bound from the CMB would still marginally improve over existing constraints. 

\begin{figure}[t!]
\begin{center}
\includegraphics[width=0.75\columnwidth]{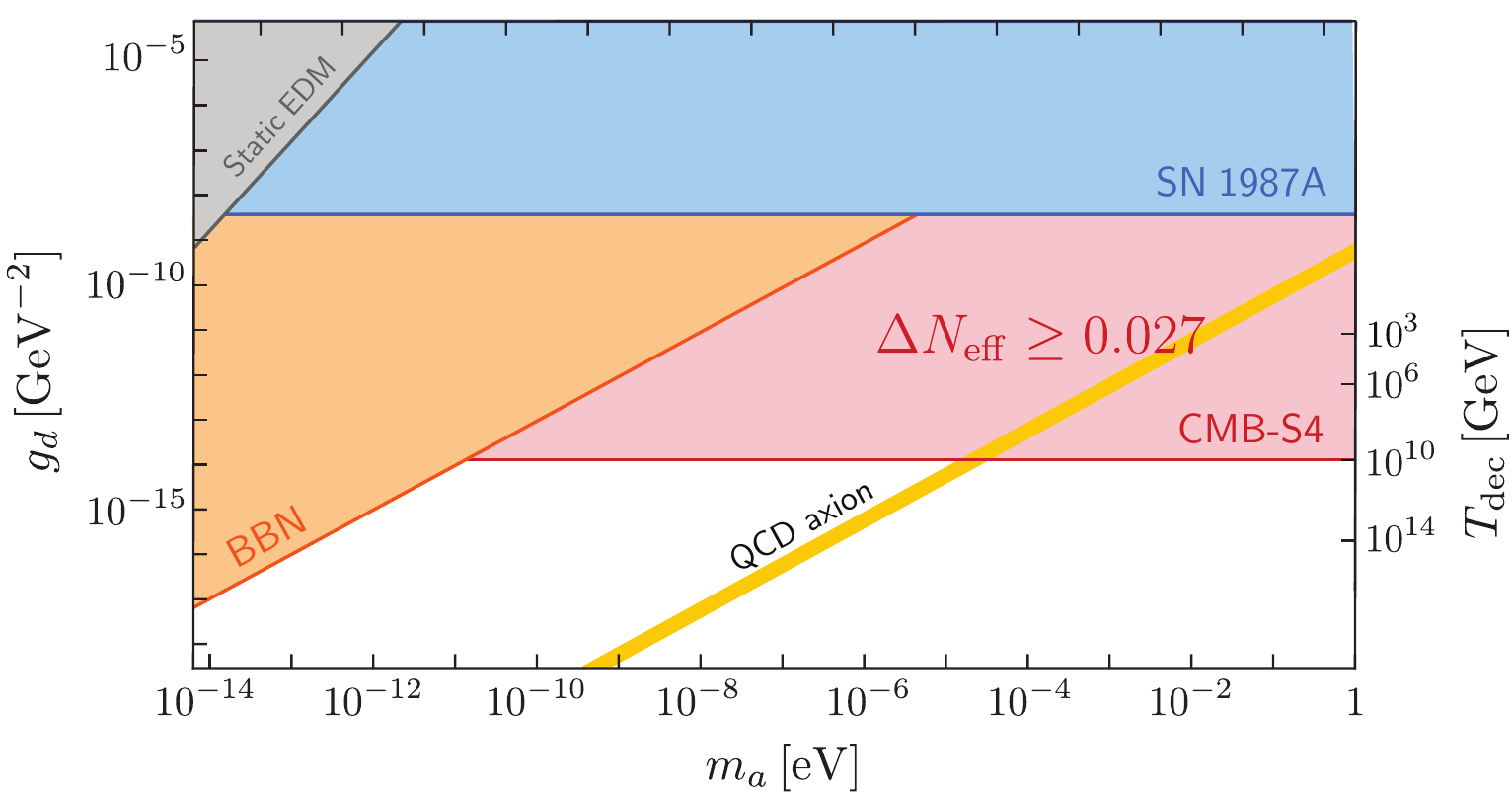}
\caption{Comparison between current constraints on the axion-gluon coupling and the sensitivity of a future CMB-S4 mission (figure adapted from~\cite{Graham:2013gfa,Blum:2014vsa}).  We note that the static EDM~\cite{Graham:2013gfa} and BBN constraints~\cite{Blum:2014vsa} assume that the axion is all of the dark matter, while SN~1987A~\cite{Raffelt:1996wa} and the future CMB constraint do not have this restriction. }
\label{fig:S4dipole}
\end{center}
\end{figure}

\vskip 4pt
The coupling to gluons is especially interesting for the QCD axion since it has to be present in order to solve the strong CP problem. The axion production rate associated with the gluon interaction in (\ref{equ:LEW0}) is $\Gamma_g\simeq0.41\hskip1pt{T^3}/\Lambda_g^2$~\cite{Salvio:2013iaa}. The bound~(\ref{eq:freezeoutconstraint}) then implies
\begin{equation}
\Lambda_g > \SI{5.4e13}{GeV} \,\sqrt{T_{R,10}}\, .
\end{equation}
Laboratory constraints on the axion-gluon coupling are usually phrased in terms of the induced electric dipole moment (EDM) of nucleons: $d_n=g_d a_0$, where $a_0$ is the value of the local axion field. For the QCD axion, the coupling $g_d$ is given  by~\cite{Pospelov:1999ha,Graham:2013gfa}
\begin{equation}
g_d \,\approx\, \frac{2\pi}{\alpha_s} \times \frac{\SI{3.8e-3}{\per\GeV}}{\Lambda_g} \, .
\end{equation}
Constraints on $g_d$ (and hence $\Lambda_g$) are shown in Fig.~\ref{fig:S4dipole}. We see that future CMB-S4 observations can improve over existing constraints on $\Lambda_g$ by up to six orders of magnitude if $T_R =\mathcal{O}(\SI{e10}{GeV})$. Even if the reheating temperature is as low as \SI{e4}{GeV}, the future CMB constraints will be tighter by three orders of magnitude. 

\vskip 4pt
Deriving similar constraints for the other axion couplings and for the couplings of fields with spin is left as an exercise for the reader.

\newpage
\part{Relics from Inflation}

\vspace{1cm}
Inflation predicts that the quantum fluctuations of any massless fields get amplified by the rapid expansion of the spacetime. Two massless fields that are guaranteed to exist in all inflationary models are the curvature perturbation $\zeta$ and the tensor fluctuations $\gamma_{ij}$.  
In Part I of these lectures, we assumed a nearly scale-invariant spectrum of curvature perturbations as a source for the density fluctuations in the late universe.  In the following, I will show that these initial conditions indeed naturally arise from inflation. I will also demonstrate that the same effect produces tensor fluctuations. The search for this stochastic background of primordial gravitational waves is a very active area of observational cosmology.

\vskip 4pt
Inflation also excites massive particles as long as their masses aren't far above the inflationary Hubble scale. Since the Hubble scale during inflation may be as high as
$10^{14}$ GeV, this gives us the opportunity to probe the particle spectrum at energies far beyond the reach of conventional particle colliders.
Once produced, these massive particles quickly decay into the massless modes $\zeta$ and $\gamma_{ij}$, creating higher-order correlations in the inflatonary fluctuations:
\beq
\includegraphicsbox[scale=0.7]{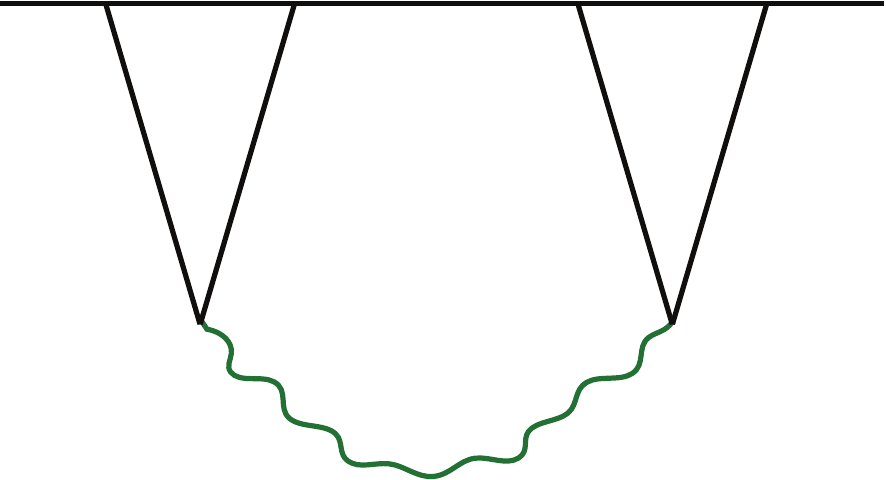}  \nonumber
\eeq
As we will see, this effect leads to a characteristic non-locality in cosmological correlators.

\vskip 10pt
We will start with a brief review of slow-roll inflation.
In Section~\ref{sec:Inflation}, we will discuss the dynamics of the inflationary background, while in Section~\ref{sec:Quantum} we will explicitly compute the spectrum of quantum fluctuations.  In Section~\ref{sec:NG}, we will extend this treatment to include interactions and show that they lead to a characteristic non-Gaussianity in cosmological correlators.  In Section~\ref{sec:HeavyRelics}, we will discuss the imprints of extra fields, showing how the masses and spins are encoded in the momentum dependence of higher-order correlation functions.

\newpage
\section{Inflationary Cosmology}
\label{sec:Inflation}

A key fact about the universe is that on large scales it is described by the FRW metric~(\ref{equ:FRW}). But why? 
A naive extrapolation of the radiation-dominated Big Bang cosmology suggests
 that the early universe consisted of many causally disconnected regions of space. The fact that these apparently disjoint patches of space are observed to have nearly the same densities and temperatures is called the {\it horizon problem}. In this section, I will explain how inflation---an early period of accelerated expansion---drives the primordial universe towards homogeneity and isotropy, even if it started in a more generic initial state. 

\vspace{0.5cm}
\begin{figure}[h!]
    \centering
     \includegraphics[width=0.9\textwidth]{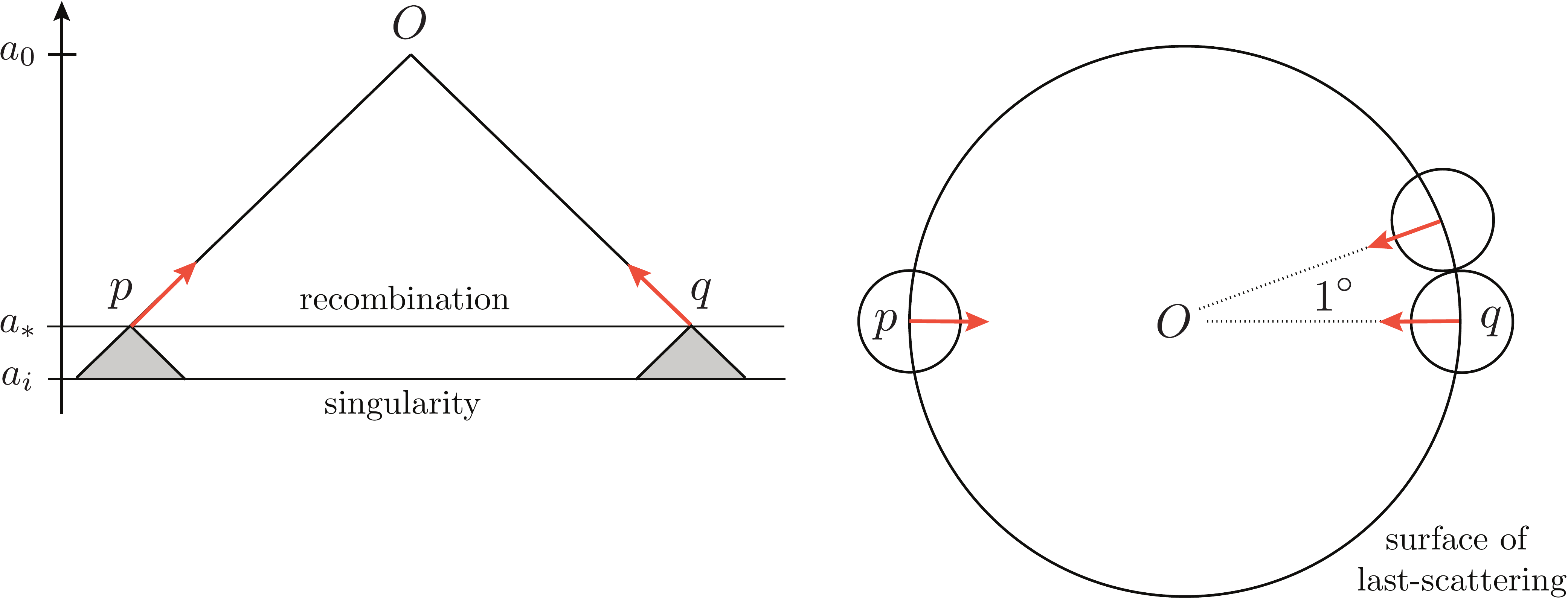}
       \caption{Illustration of the horizon problem in the conventional Big Bang model. All events that we currently observe are on our past light cone. The intersection of our past light cone with the spacelike slice labelled ``recombination" corresponds to the ``surface of last-scattering". 
       Any two points on the surface of last-scattering that are separated by more than 1 degree, appear never to have been in causal contact.  This means that their past light cones do not overlap before the singularity. This is illustrated for opposite points on the sky labelled $p$ and $q$.}
       \label{equ:horizon1}
\end{figure}

\subsection{Horizon Problem}

The {\it particle horizon} is the maximal distance that a signal can travel between 
the time corresponding to the initial singularity, $t_i\equiv 0$, and a later time $t$. In physical coordinates, this distance is given by
\beq
D(t) \,=\, a(t) \int_0^t \frac{\d t}{a(t)} = a(t) \int_0^{a} \frac{\d \ln a}{a'} \, . \label{equ:horizon}
 \eeq
 If the early universe was filled by ordinary matter, then $a'' < 0$. In that case,
the integral in (\ref{equ:horizon}) is dominated by late times and converges to a finite value:
\beq
{\rm e.g.}~\quad a(t) \propto \left\{ \begin{array}{l} t^{2/3} \\[8pt] t^{1/2} \end{array}\right.  \quad \to \quad D(t) \,=\,  \left\{ \begin{array}{ll} 3t & \quad{\rm matter} \\[8pt] 2t & \quad{\rm radiation}\end{array}\right.  
\eeq
This leads to a puzzle: because the age of the universe ($t_0$) is much larger than the time of recombination ($t_*$), the CMB naively consists of many causally disconnected patches (see~Fig.~\ref{equ:horizon1}).  The following questions immediately arise: Why is the CMB so homogeneous? And, more importantly, why are the observed CMB fluctuations correlated on large scales and not just random noise?

\begin{figure}[h!]
    \centering
 \hspace{1cm} \includegraphics[width=0.65\textwidth]{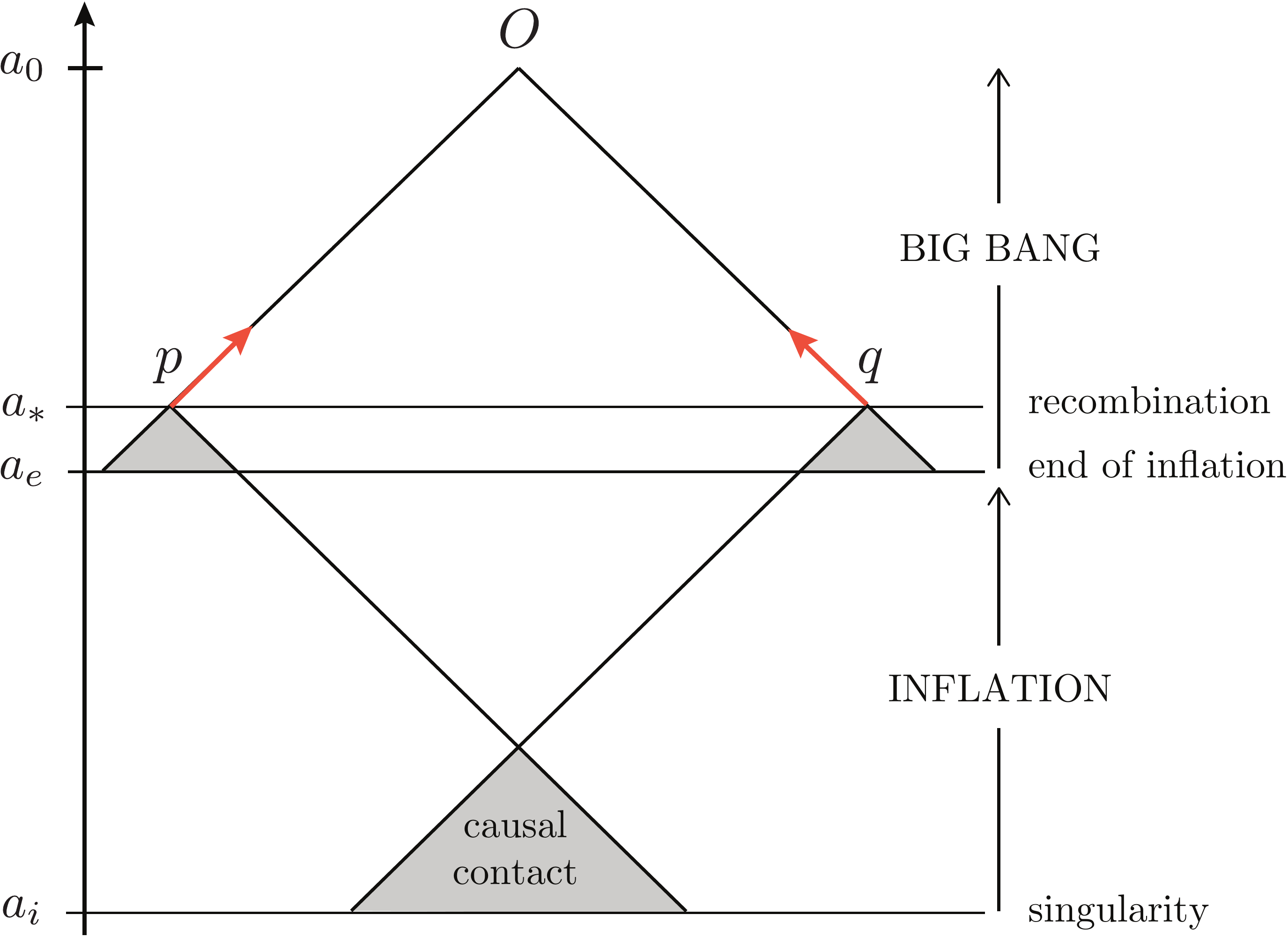}
     \caption{Illustration of the inflationary solution to the horizon problem in comoving coordinates (using conformal time on the vertical axis). The spacelike singularity of the standard Big Bang is replaced by the reheating surface, i.e.~rather than marking the beginning of time it now simply corresponds to the transition from the end of inflation to the standard Big Bang evolution. All points in the CMB have overlapping past light cones and therefore originated from a causally connected region of space.}
     \label{fig:horizon2}
\end{figure}

\vskip 4pt
The horizon problem is solved if the early universe experienced a sustained period of accelerated expansion (=\,inflation), $a'' >0$. In that case, the integral in (\ref{equ:horizon}) is dominated by early times and
the particle horizon diverges in the past.  Signals were therefore able to travel a much larger distance than suggested by the naive extrapolation of the standard FRW expansion (see Fig.~\ref{fig:horizon2}).

\begin{framed}
\noindent
{\small {\it Exercise}.---A special case of accelerated expansion is the quasi-de Sitter limit, which is characterized by a nearly constant expansion rate, $H =a'/a \approx const.$, so that $a(t)=e^{H(t-t_0)}$, where $t_0$ is some fiducial time at which $a(t_0) \equiv 1$. Show that
\beq
a(\ct)= - \frac{1}{H\ct}\, ,
\eeq
for $\ct < 0$. Notice that the initial singularity has been pushed to $\ct=-\infty$ (cf.~Fig.~\ref{fig:horizon2}).
}
\end{framed}

\begin{framed}
\noindent
{\small {\it Exercise}.---Show that $a'' > 0$ is equivalent to a slow variation of the Hubble parameter
\beq
\varepsilon \equiv - \frac{H'}{H^2} < 1\, . \label{equ:eps}
\eeq
Notice that $\varepsilon \approx 0$ corresponds to a quasi-de Sitter spacetime with a nearly constant expansion rate, $H \approx const$. 
Using the Friedmann equations, show that 
\beq
\varepsilon = \frac{3}{2} \left( 1 + \frac{P}{\rho}\right)  \, <\, 1 \quad \ \Leftrightarrow \quad \
w \equiv \frac{P}{\rho} < - \frac{1}{3} \, .
\eeq
The last condition corresponds to a violation of the strong energy condition.}
\end{framed}

\subsection{Slow-Roll Inflation}

\begin{figure}[h!]
    \centering
        \includegraphics[width=0.5 \textwidth]{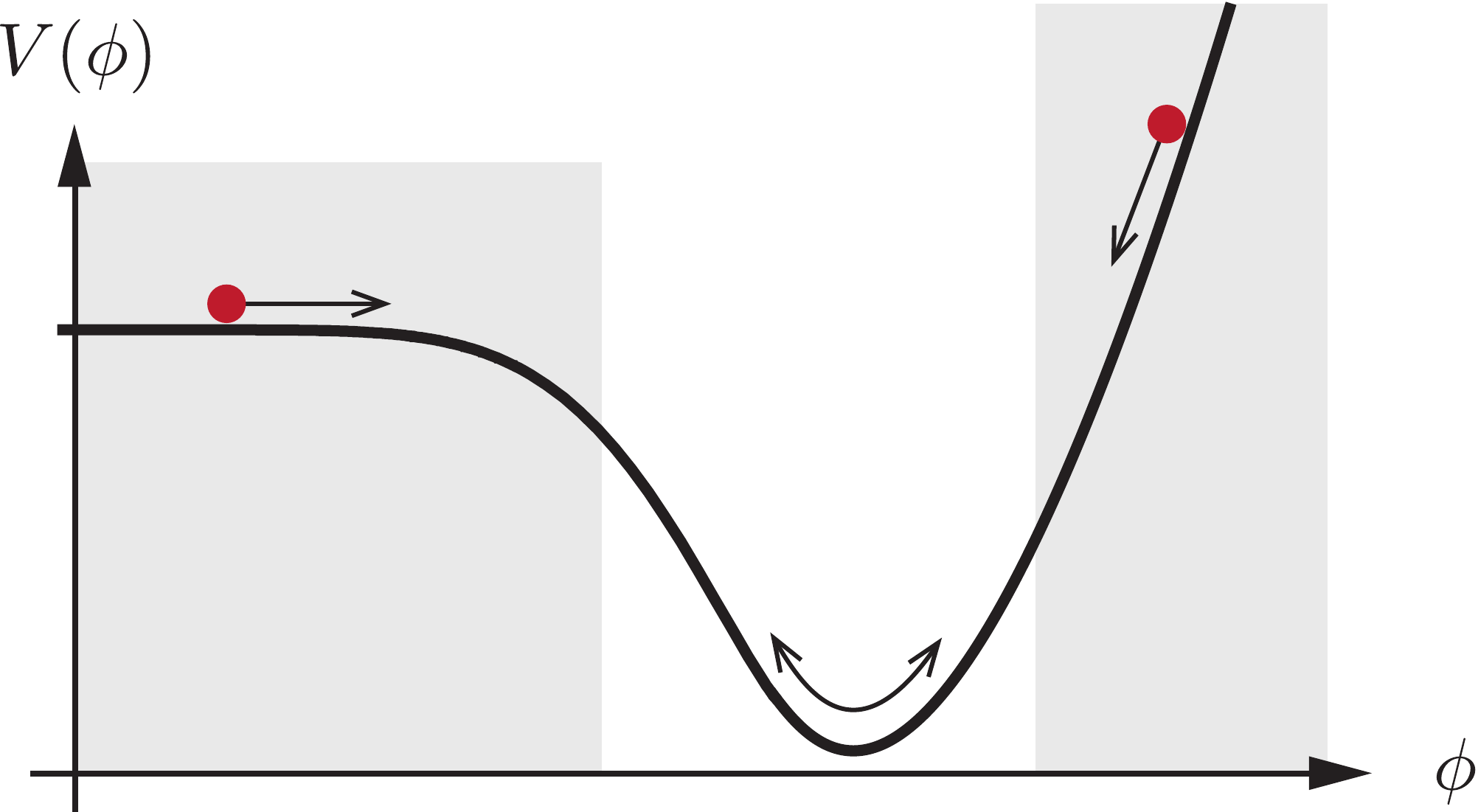}
\caption{Example of a slow-roll potential.  Inflation occurs in the shaded parts of the potential.}   \label{fig:small}
\end{figure}

As a simple toy model for inflation, let us consider the dynamics of a scalar field, the {\it inflaton}\, $\phi(t,\x)$.
As indicated by the notation, the value of the field can depend on time $t$ and the position in space $\x$.  Associated with each field value is a potential energy density $V(\phi)$ (see Fig.~\ref{fig:small}).  If the field is dynamical (i.e.~changes with time) then it also carries a kinetic energy density.
If the energy density associated with the scalar field dominates the universe, it sources the evolution of the FRW background.
We want to determine under which conditions this can lead to accelerated expansion.

\vskip 4pt
The stress-energy tensor of the scalar field is
\begin{equation}
T_{\mu \nu}  \, =\, \partial_\mu \phi \hskip 1pt \partial_\nu \phi - g_{\mu \nu} \left( \frac{1}{2} g^{\alpha \beta} \partial_\alpha \phi \hskip 1pt \partial_\beta \phi - V(\phi) \right) . \label{equ:Tphi}
\end{equation}
Consistency with the symmetries of the FRW spacetime requires that the background value of the inflaton only depends on time, $\phi =  \phi(t)$.
From the time-time component $T^0{}_0 = -\rho_\phi$, we infer that the energy density of the field is
\beq
\rho_\phi \, =\, \frac{1}{2} (\phi')^2 + V( \phi) \, . \label{equ:rhoP}\\
\eeq  
We see that this is simply the sum of the kinetic energy density, $\frac{1}{2} (\phi')^2$, and the potential energy density, $V(\phi)$.
From the space-space component $T^i{}_j = P_\phi \, \delta^i_j$, we find that the pressure is the {\it difference} of kinetic and potential energy densities,
\beq 
P_\phi \, =\, \frac{1}{2} (\phi')^2 - V( \phi) \, . \label{equ:PPP}
\eeq
A field configuration therefore leads to inflation, $P_\phi < -\frac{1}{3} \rho_\phi$, 
if the potential energy dominates over the kinetic energy. i.e.~if the field {\it rolls slowly}.

\begin{framed}
\noindent
{\small 
{\it Exercise.}---Using the Einstein equations (\ref{equ:F1}) and (\ref{equ:F2}), show that
\beq \left. \begin{array}{c} \ 3 \Mp^2 H^2 = \frac{1}{2} (\phi')^2 + V\\[10pt]
 \Mp^2 H' = - \frac{1}{2} (\phi')^2 \end{array} \right\} \quad \Rightarrow  \quad  \varepsilon = \frac{\frac{1}{2} (\phi')^2}{\Mp^2 H^2} < 1 \, , \label{equ:SR1}
 \eeq 
 where $\Mp =(8\pi G)^{-1/2}$ is the reduced Planck mass.}
 \end{framed}
 \noindent
 Equation (\ref{equ:SR1}) corresponds to the {\it first slow-roll condition}:
 \beq
(\phi')^2 < V\, . \label{equ:SR1X}
 \eeq
This condition alone, however, does not guarantee successful inflation.
We also need to assure that inflation does not just occur for an instant, but lasts long enough to solve the horizon problem.

\vskip 4pt
Combining the two Friedmann equations in (\ref{equ:SR1}) leads to the Klein-Gordon equation for the evolution of the scalar field
\beq
\phi'' + 3 H  \phi' = - V_{,\phi}\, , \label{equ:KG}
\eeq
where $V_{,\phi}$ denotes the derivative of the potential with respect to the field value.
To maintain the slow-roll condition (\ref{equ:SR1X}) for a sufficient period of time, we require that the acceleration of the field is small. This is quantified by the {\it second slow-roll condition}:
\beq
\phi'' < 3 H \phi'\, . \label{equ:SR2X}
\eeq
The evolution of the scalar field is then friction dominated, with the velocity of the field determined by the slope of the potential.

\vskip 4pt
The two slow-roll conditions (\ref{equ:SR1X}) and (\ref{equ:SR2X}) 
can be expressed as conditions on the shape of the inflaton potential:
\beq
\varepsilon_V \equiv \frac{\Mp^2}{2}\left(\frac{V_{,\phi}}{V}\right)^2 < 1\,,  \qquad \eta_V \equiv \Mp^2 \left|\frac{V_{,\phi\phi}}{V}\right| < 1\, .  \label{equ:PotSR}
\eeq
We will refer to the parameters in (\ref{equ:PotSR}) as the {\it (potential) slow-roll parameters}.

\begin{framed}
\noindent
{\small 
{\it Exercise.}---Applying the slow-roll conditions (\ref{equ:SR1X}) and (\ref{equ:SR2X}) to the Friedmann equation (\ref{equ:SR1}) and the Klein-Gordon equation (\ref{equ:KG}), we get
\begin{align}
3 \Mp^2 H^2 &\approx  V\, , \\
3 H  \phi' &\approx - V_{,\phi}\, . 
\end{align}
Use this to show that $\varepsilon \approx \varepsilon_V$ during slow-roll inflation.}
 \end{framed}
Inflation ends when the first slow-roll condition is violated, $\varepsilon_V(\phi_{e}) \equiv 1$.  The amount of inflation is measured in terms of `$e$-folds' of expansion, $\d N = \d \ln a$.  
The total number of $e$-folds between a point $\phi$ on the potential and the end of inflation at $\phi_e$ is
 \begin{align}
 N(\phi) \equiv \int_{a}^{a_{e}} \d \ln a =  \int_{t}^{t_e} H\, \d t &=   \int_{\phi}^{\phi_e} \frac{H}{\phi'}\, \d \phi   \nonumber \\
 &=  \int_{\phi}^{\phi_e}  \frac{1}{\sqrt{2\varepsilon}}  \frac{|\d\phi|}{\Mp} \approx   \int_{\phi}^{\phi_e}   \frac{1}{\sqrt{2\varepsilon_{V}}}
 \frac{|\d \phi|}{M_{\rm pl}} \, .
  \label{NN}
 \end{align}
 To solve the horizon problem requires at least between 40 and 60 $e$-folds (with the precise value depending on the reheating temperature).
 
 \begin{framed}
\noindent
{\small
{\it Case study: $m^2 \phi^2$ inflation.}---As an example, let us give the slow-roll analysis of arguably the simplest model of inflation:
single-field inflation driven by a mass term
\beq
V(\phi) = \frac{1}{2} m^2 \phi^2\, .
\eeq
The slow-roll parameters are
\beq
\varepsilon_{V}(\phi) = \eta_{V}(\phi) = 2 \left( \frac{M_{\rm pl}}{\phi}\right)^2\, .
\eeq
To satisfy the slow-roll conditions $\{\varepsilon_{V}, |\eta_{V}|\} < 1$, we therefore need to consider super-Planckian values for the inflaton
\beq
\phi > \sqrt{2} M_{\rm pl} \equiv \phi_{e}\, .
\eeq
The relation between the inflaton field value and the number of $e$-folds before the end of inflation is
\beq
N(\phi) = \frac{\phi^2}{4 M_{\rm pl}^2} - \frac{1}{2}\, .
\eeq
Solving the horizon problem then requires that the initial value of the field, $\phi_i$, satisfies
\beq
\phi_i > \phi_{60} \equiv 2 \sqrt{60} \, M_{\rm pl} \sim 15 \hskip 1pt M_{\rm pl}\, .
\eeq
We note that the total field excursion is super-Planckian, $\Delta \phi = \phi_i  -\phi_e \gg M_{\rm pl}$.}
\end{framed}

\subsection{Effective Field Theory}

In the absence of a complete microscopic theory of inflation, we describe inflation in the context of an effective field theory.
We are then obliged to include in the inflationary action all operators consistent with the assumed symmetries of the inflaton,
\begin{equation}
\label{equ:ODelta}
{\cal L}_{\rm eff}(\phi)\ =\ -\frac{1}{2}(\partial_\mu \phi)^2 - V(\phi) \, +\,  \sum_n c_n \hskip 1pt V(\phi) \frac{\phi^{2n}}{\Lambda^{2n}} + \sum_n d_n \hskip 1pt \frac{(\partial \phi)^{2n}}{\Lambda^{4n}} + \cdots\ .
\end{equation}
One of the remarkable features of inflation is that it is sensitive even to Planck-suppressed operators.

\paragraph{Eta problem} Successful inflation requires that the inflaton 
mass $m_\phi$ is parametrically smaller than the Hubble scale $H$:
\beq
\eta_V = \frac{m^2_\phi}{3H^2} \ll 1\, .
\eeq
It is  difficult to protect this hierarchy against high-energy corrections.
We know that some new degrees of freedom must appear at $\Lambda \lesssim M_{\rm pl}$ to give a UV-completion of gravity.
In string theory, this scale is often found to be significantly below the Planck scale, $\Lambda \lesssim M_s \ll M_{\rm pl}$.
If $\phi$ has order-one couplings to any massive fields $\psi$ (with $m_\psi \sim \Lambda$), then integrating out the fields $\psi$ yields the effective action (\ref{equ:ODelta}) for $\phi$ with order-one couplings $c_n$ and $d_n$.
The above argument makes us worry that integrating out the massive fields $\psi$ yields corrections to the potential of the form
\beq
\Delta V = c_1\, V(\phi) \frac{\phi^2}{\Lambda^2}\, ,
\eeq
where $c_1 \sim {\cal O}(1)$.
If this term arises, then the eta parameter receives the following correction
\beq
\Delta \eta_V = \frac{M_{\rm pl}^2}{V} ( \Delta V)''  \approx 2 c_1 \left( \frac{M_{\rm pl}}{\Lambda}\right)^2 \,>\, 1\, ,
\eeq
where the final inequality follows from $\Lambda \lesssim \Mp$. 
Notice that this problem is independent of the energy scale of inflation.
All inflationary models have to address the eta problem.

\paragraph{Large-field inflation} The Planck-scale sensitivity of inflation is dramatically enhanced in models with observable gravitational waves, $r \gtrsim 0.01$.
In this case, the inflaton field moves over a super-Planckian range during the last 60 $e$-folds of inflation, $\Delta \phi > M_{\rm pl}$ (see \S\ref{equ:GWs}), and an infinite number of operators contribute equally to the effective action (\ref{equ:ODelta}).
This observation makes an effective field theorist nervous and a string theorist curious~\cite{Baumann:2014nda}.

\newpage
\section{Quantum Initial Conditions}
\label{sec:Quantum}

\begin{figure}[h!]
    \centering
        \includegraphics[width=0.5\textwidth]{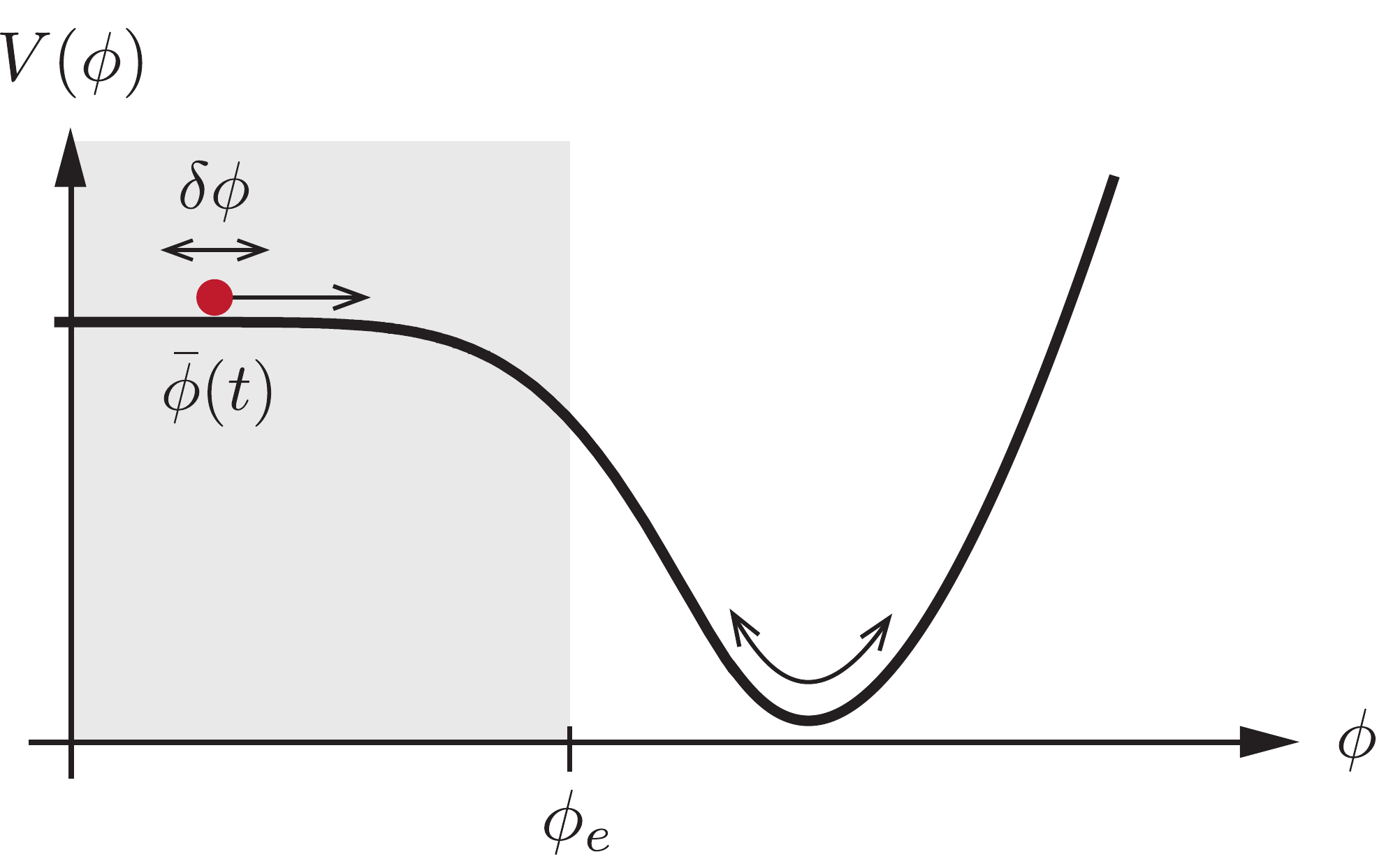}
        \vspace{-0.1cm}
    \caption{Quantum fluctuations $\delta \phi(t, {\x})$ around the classical background evolution $\bar \phi(t)$. Regions acquiring negative  fluctuations $\delta \phi$ remain potential-dominated longer than regions with positive $\delta \phi$. Different parts of the universe therefore undergo slightly different evolutions. After inflation, this induces  density fluctuations~$\delta \rho(t, {\x})$.}
    \label{fig:quantum}
\end{figure}

One of the most remarkable features of inflation is that it provides a natural mechanism for producing the initial conditions for the hot big bang.  
To see this, recall that the evolution of the inflaton field $\phi(t)$ governs the energy density of the early universe~$\rho(t)$ and, hence, controls the end of inflation  (see Fig.~\ref{fig:quantum}).
Essentially, the field $\phi$ plays the role of a ``clock" reading off the amount of inflationary expansion still to occur. By the uncertainty principle, arbitrarily precise timing is not possible in quantum mechanics.
Instead, quantum-mechanical clocks
necessarily have some variance, so the inflaton will have spatially varying fluctuations $\delta \phi(t,{\x})$.  There will therefore be local differences in the time when inflation ends, $\delta t({\x})$, so that different regions of space inflate by different amounts.  These differences in the local expansion histories lead to differences in the local densities after inflation, $\delta \rho(t, \x)$, and to curvature perturbations in comoving gauge, $\zeta(\x)$.   It is worth remarking that the theory was not engineered to produce these fluctuations, but that their origin is instead a natural consequence of treating inflation quantum mechanically.
\vspace{-0.3cm}

\subsection{Quantum Fluctuations}

\subsubsection{Free Scalar in de Sitter}

Before attacking the real problem of interest, namely the quantization of coupled inflaton-metric fluctuations during inflation, we will consider the simpler case of a free scalar field in de Sitter space.  We will assume that the scalar field carries an insignificant amount of the total energy density and, hence, doesn't backreact on the de Sitter geometry. Such a field is sometimes called a {\it spectator field}.

\vskip 4pt
The action of a massless, free scalar field in de Sitter space is
\begin{align}
S &\,=\, \frac{1}{2} \int \d^4 x \sqrt{-g}\, g^{\mu \nu} \partial_\mu \varphi \partial_\nu \varphi \nonumber \\
&\,=\, \frac{1}{2} \int \d \ct \,\d^3 x\, a^2 \Big[\dot \varphi^2 - (\partial_i \varphi)^2\Big]\, ,
\end{align}
where $a(\ct) = - 1/(H\ct)$. It is useful to define the canonically normalized field $v \equiv a \varphi$, so that
\beq
S \,=\, \frac{1}{2} \int \d \ct \,\d^3 x \left[ \dot v^2 -(\partial_i v)^2 + \frac{\ddot a}{a}v^2\right]  . \label{equ:Sv}
\eeq
This action implies the following equation of motion for the Fourier modes of the field
\beq
\boxed{ \ddot v_\k + \left( k^2 - \frac{\ddot a}{a} \right) v_\k  = 0} \ .  
 \label{equ:MS}
\eeq
Sometimes this is called the {\it Mukhanov-Sasaki (MS) equation}.

\vskip 6pt
We see that the expansion of the universe is captured by the time dependence of the effective mass of the canonically-normalized field, $m_{\rm eff}^2(\ct) \equiv \ddot a/a = 2/\ct^2$.
At early times, $- \ct \gg k^{-1}$, this mass is much smaller than the momentum $k$ of the relevant Fourier modes.  The dynamics then reduces to that of a simple harmonic oscillator in flat space,
\beq
\ddot v_\k + k^2 v_\k \approx 0\, .
\eeq
In this limit, the quantum fluctuations of the field $v$ therefore follow from the standard quantization of a simple harmonic oscillator.  

\subsubsection*{Canonical quantization}

From the action (\ref{equ:Sv}), we read off the the momentum conjugate to $v$:
\beq
\pi \equiv \frac{\partial {\cal L}}{\partial \dot v} =\dot v \, .
\eeq
We promote the fields $v(\ct,\x)$ and $\pi(\ct,\x)$ to quantum operators $\hat v(\ct,\x)$ and $\hat \pi(\ct,\x)$.   These operators satisfy the equal time commutation relation
  \beq
\label{com}
\boxed{ [\hat v(\ct, {\x}), \hat \pi(\ct, {\x^\prime})] = i \delta_D({\x} - {\x^\prime})  } \ ,  
\eeq
in units where $\hbar \equiv 1$.
The delta function is required by {\it locality}: modes at different points in space are independent and the corresponding operators therefore commute.
 In Fourier space, this becomes
 \begin{align}
 [\hat v_\k(\ct), \hat \pi_{\k'}(\ct)] &= \int \d^3 x \int \d^3 x^\prime\,  \underbrace{[\hat v(\ct, {\x}), \hat \pi(\ct, {\x^\prime})]}_{\displaystyle i  \delta_D({\x} - {\x^\prime})} \, e^{-i\k\cdot \x} e^{-i\k^\prime \cdot \x^\prime} \nonumber \\
 &= i \int \d^3 x\, e^{-i (\k + \k^\prime)\cdot \x} \nonumber \\[8pt]
 &= i\, (2\pi)^3 \delta_D(\k+\k')\, ,  \label{equ:CCRk}
 \end{align}
 where the delta function implies that modes with different wavelengths commute.
 
 \vskip 4pt 
Note that we are using the Heisenberg picture where operators vary in time while states are time independent.
The operator solution $\hat v_\k(\ct)$ is determined by two initial conditions $\hat v_\k(\ct_i)$ and $\hat \pi_\k(\ct_i) = \partial_\ct \hat v_\k(\ct_i)$. Since the evolution equation is linear, the solution is linear in these operators.
It is convenient to trade $\hat v_\k(\ct_i)$ and $\hat \pi_\k(\ct_i)$ for a single time-independent non-Hermitian operator~$\hat a_\k$, in terms of which the solution can be written as
  \beq
\boxed{ \hat v_\k(\ct) = v_k^{\vphantom{*}}(\ct) \hskip 1pt \hat a_\k^{\vphantom{\dagger}} + v^*_k(\ct) \hskip 1pt  a^\dagger_{-\k} } \ ,   \label{equ:exp2}
 \eeq
where the (complex) mode function $v_k(\ct)$ satisfies the classical equation of motion (\ref{equ:MS}).
Of course, $v_k^*(\ct)$ is the complex conjugate of $v_k(\ct)$ and $\hat a^\dagger_\k$ is the Hermitian conjugate of $\hat a_\k$.  As indicated by dropping the vector notation ${\k}$ on the subscript, the mode functions, $v_k(\ct)$ and $v_k^*(\ct)$, are the same for all Fourier modes with $k \equiv |{\k}|$.\footnote{Since the frequency $\omega_k(\ct) \equiv k^2-\ddot a/a$ in (\ref{equ:MS}) depends only on $k \equiv |{\k}|$, the evolution does not depend on direction. The constant operators $\hat a_{\k}^{\vphantom{\dagger}}$ and $\hat a^\dagger_{{\k}}$, on the other hand, define initial conditions which may depend on direction.}

\vskip 6pt
We choose the normalization of the mode functions, so that $v_k \dot v_k^* - \dot v_k v_k^* \equiv i$. 
Substituting~(\ref{equ:exp2}) into (\ref{equ:CCRk}), we then get
\beq
\label{hata}
\boxed{ [\hat a_{\k\vphantom{'}}^{\vphantom{\dagger}}, \hat a_{\k'}^\dagger] = (2\pi)^3\delta_D({\k} + {\k}') } \ , 
\eeq
which is
the standard commutation relation for the {\it raising} and {\it lowering operators} of a harmonic oscillator.
The quantum states in the Hilbert space are constructed by defining the vacuum state~$| 0 \rangle$ via
 \beq
 \label{vacuum}
 \hat a_{\k} | 0 \rangle = 0\, ,  
 \eeq
 and by producing excited states through repeated application of the creation operators $a_\k^\dagger$.

\subsubsection*{Choice of vacuum}

The most general solution of equation (\ref{equ:MS}) is
\beq
v_k(\ct) = c_1 \left(1- \frac{i}{k\ct}\right) e^{-ik\ct} + c_2 \left(1+\frac{i}{k\ct}\right)e^{ik \ct}\, ,
\eeq
where the constraint on the overall normalisation of the mode functions, $v_k \dot v_k^* - \dot v_k v_k^* \equiv i$, implies
\beq
|c_1|^2 - |c_2|^2 = \frac{1}{2k}\, . \label{equ:cons}
\eeq
At this point, we still have a one-parameter family of solutions for the mode function $v_k(\tau)$. 
A change in $v_k(\ct)$ could be accompanied by a change in $\hat a_\k$ that keeps the solution $\hat v_\k(\ct)$ unchanged.
Each such solution corresponds to a different vacuum state, cf.~eq.~(\ref{vacuum}).
However, a special choice of $v_k(\ct)$ is selected if we require the vacuum state $|0\rangle$ to be the ground state of the Hamiltonian.  

\vskip 4pt
To see this, consider the Hamiltonian operator 
\begin{align}
\hat H &=\int \d^3x \left[ \frac{1}{2}\hat \pi^2 + \frac{1}{2}(\nabla \hat v)^2 + \frac{1}{2}\frac{\ddot a}{a}\, \hat v^2 \right] .
\end{align}
Substituting the mode expansion (\ref{equ:exp2}) into this, we find
\beq
\hat H  = \frac{1}{4} \int \d^3 k \left[a_\k a_{-\k} F_k^* + a_\k^\dagger a_{-\k}^\dagger F_k + \left(2a_\k^\dagger a_\k + \delta_D(0)\right) E_k\right] ,
\eeq
where 
\begin{align}
E_k(\ct) &= |\dot v_k|^2 + \omega_k^2 |v_k|^2\, , \label{equEk0} \\
F_k(\ct) &= \dot v_k^2 + \omega_k^2 v_k^2\, .
\end{align}
The vacuum expectation value of the Hamiltonian is 
\beq
\langle 0 | \hat H | 0 \rangle = \frac{1}{4}\delta_D(0) \int \d^3 k\, E_k(\ct) \, ,
\eeq
where the divergent factor $\delta_D(0)$ is an artefact of integrating over an infinite volume. The energy density of the vacuum is
\beq
\varepsilon_k \equiv \frac{1}{4}\int \d^3 k\, E_k(\ct)\, .
\eeq
At early times, we have
\begin{align}
\lim_{\ct \to -\infty} v_k(\ct) &= c_1\, e^{-ik\ct} + c_2\, e^{ik \ct}\, , \\
\lim_{\ct \to -\infty} E_k(\ct) &= 4\, (|c_1|^2 + |c_2|^2)\,k^2\, . \label{equ:Ek}
\end{align}
Given the constraint (\ref{equ:cons}), the function in (\ref{equ:Ek}) is minimized for
\beq
|c_1| = \frac{1}{\sqrt{2k}} \, , \quad c_2=0\, . \label{equ:c11}
\eeq
Up to an irrelevant phase, this uniquely determines the {\it Bunch-Davies} mode function
\beq
\boxed{v_k(\ct) =  \frac{1}{\sqrt{2k}} \left(1- \frac{i}{k\ct}\right) e^{-ik\ct}} \ . \label{equ:BD}
\eeq
Note that (\ref{equ:c11}) implies
\begin{align}
\lim_{\ct \to -\infty} E_k(\ct) &= 2k\, , \label{equ:Ek2} \\
\lim_{\ct \to -\infty} F_k(\ct) &= 0\, ,
\end{align}
and hence
\beq
\lim_{\ct \to -\infty} \hat H = \int \d^3 k \left[a_\k^\dagger a_\k + \frac{1}{2} \delta_D(0) \right] \hbar \omega_k\, ,
\eeq
where we have reinstated Planck's constant $\hbar$.
We see that the vacuum state $|0\rangle$ is the state of minimum energy $\frac{1}{2}\hbar \omega_k$. If any function other than (\ref{equ:BD}) had been chosen as the mode function, then the state annihilated by $\hat{a}_\k$ would {\it not} be the ground state of the oscillator.

\subsubsection*{Zero-point fluctuations}

Finally, we can predict the quantum statistics of the operator
\beq
\hat v(\ct,\x) = \int \frac{\d^3 k}{(2\pi)^{3}} \left[ v_k^{\vphantom{*}}(\ct) \hskip 1pt \hat a_\k^{\vphantom{\dagger}} + v_k^*(\ct) \hskip 1pt  a^\dagger_{-\k}  \right] e^{i\k\cdot \x}\ .
\eeq
The expectation value of $\hat v$ vanishes, i.e.~$\langle \hat v \rangle \equiv \langle 0| \hat v|0\rangle = 0$.
However, the variance of inflaton fluctuations receives non-zero quantum fluctuations: 
\begin{align}
\langle |\hat v|^2 \rangle &\equiv \langle 0 | \hat v(\ct, {\bf 0})  \hat v(\ct, {\bf 0}) | 0 \rangle \nonumber \\[4pt]
&= \contraction{\int \frac{\d^3 k}{(2\pi)^{3}}  \int \frac{\d^3 k'}{(2\pi)^{3}} \ }{\langle 0 |}{ \big(  v_k^*(\eta)} {\hat a}
\contraction{\int \frac{\d^3 k}{(2\pi)^{3}}  \int \frac{\d^3 k'}{(2\pi)^{3}} \ \langle 0 | \big(  v_k^*(\ct) \hat a_\k^\dagger + v_k^{\vphantom{*}}(\ct) \hat a_\k^{\vphantom{\dagger}}\big) \big(v_{k'}^{\vphantom{*}}(\ct) }{\hat a}{_{\k'}^{\vphantom{\dagger}} + v_{k'}^*(\ct) \hat a_{\k'}^\dagger \big) |}{0} 
\int \frac{\d^3 k}{(2\pi)^{3}}  \int \frac{\d^3 k'}{(2\pi)^{3}} \ \langle 0 | \big(  v_k^*(\ct) \hat a_\k^\dagger + v_k^{\vphantom{*}}(\ct) \hat a_\k^{\vphantom{\dagger}}\big) \big(v_{k'}^{\vphantom{*}}(\ct) \hat a_{\k'}^{\vphantom{\dagger}} + v_{k'}^*(\ct) \hat a_{\k'}^\dagger \big) | 0 \rangle
 \nonumber \\[2pt]
&= \int \frac{\d^3 k}{(2\pi)^{3}}  \int \frac{\d^3 k'}{(2\pi)^{3}} \ v_{k\vphantom{'}}^{\vphantom{*}}(\ct) v_{k'}^*(\ct)\,  \langle 0 | [\hat a_{\k\vphantom{'}}^{\vphantom{\dagger}} ,\hat a_{\k'}^\dagger ] | 0 \rangle \nonumber \\[2pt]
&= \int \frac{\d^3 k}{(2\pi)^{3}}\, |v_k(\ct)|^2\nonumber \\[2pt]
&= \int \d \ln k\ \frac{k^3}{2\pi^2} |v_k(\ct)|^2\ .
\end{align}
We define the (dimensionless) {\it power spectrum} as
\beq
\boxed{ {\cal P}_v(k,\ct) \equiv \frac{k^3}{2\pi^2}|v_k(\ct)|^2}\ .   \label{equ:Pv0}
\eeq
We see that the square of the classical solution determines the variance of quantum fluctuations. 
Substituting the Bunch-Davies mode function (\ref{equ:BD}) into (\ref{equ:Pv0}), we find
\beq
{\cal P}_\varphi(k,\ct) = \frac{{\cal P}_v(k,\ct)}{a^2(\ct)} = \left(\frac{H}{2\pi}\right)^2 \Big[1+(k\ct)^2\Big] \xrightarrow{\ k \ct\to 0\ }  \left(\frac{H}{2\pi}\right)^2\, . \label{equ:Pv}
\eeq
Note that in the superhorizon limit, $k \ct \to 0$, the dimensionless power spectrum ${\cal P}_\varphi$ approaches the same constant for all momenta. This is the characteristic of a {\it scale-invariant} spectrum.

\subsubsection*{Massive fields}

The above discussion is easily generalized to massive spectator fields.
The modified equation of motion is
\beq
\ddot v_k + \left( k^2 + m^2 a^2 - \frac{\ddot a}{a}  \right) v_k  = 0\, ,
\eeq
which, in de Sitter space, becomes
\beq
\ddot v_k + \left( k^2  - \frac{\nu^2 - 1/4}{\ct^2}  \right) v_k  = 0\, , \quad {\rm where}\quad \nu^2 \equiv  \frac{9}{4} - \frac{m^2}{H^2}\, . \label{equ:EM}
\eeq
The most general solution of (\ref{equ:EM}) is
\beq
v_k(\ct) =\sqrt{-\ct}\, \left[  c_1 H_\nu^{(1)}(|k\ct|)+ c_2H_\nu^{(2)}(|k\ct|) \right]  , \label{equ:vk}
\eeq
where $H_\nu^{(1)}$ and $H_\nu^{(2)}$ are Hankel functions of the first and second kind.
Imposing the Bunch-Davies initial condition, we find
\beq
c_1 =  \frac{\sqrt{\pi}}{2} e^{i(2\nu+1)\frac{\pi}{4}}\, , \quad c_2=0\, . \label{equ:c1}
\eeq
For the moment, let us assume $m < \frac{3}{2}H$, so that $\nu$ is real.
We then find
\begin{align}
{\cal P}_\varphi(k,\ct) = \frac{{\cal P}_v(k,\ct)}{a^2(\ct)} 
&\ \xrightarrow{\ k \ct\to 0\ } \  \left(\frac{H}{2\pi}\right)^2  (k\ct)^{3/2-\nu}\, .
\end{align}
We observe that the superhorizon limit of the spectrum is scale-dependent and evolves in time.
The scale-dependence takes a power law form with the following
{\it spectral index}
\beq
n_\varphi \equiv \frac{d\ln {\cal P}_\varphi}{d \ln k}  = \frac{3}{2}-\nu  \  \xrightarrow{\ m\ll H\ } \ \frac{1}{3} \frac{m^2}{H^2}\, .
\eeq
In the limit $m\to 0$, we recover the scale-invariant spectrum of a massless field in de Sitter.

\vskip 6pt
For $m > \frac{3}{2}H$, the degree of the Hankel function becomes imaginary, $\nu \equiv i \mu$, where 
\beq
\mu \equiv \sqrt{m^2/H^2-9/4}  \ \xrightarrow{\ m\gg H\ }\ {m/H}\, .
\eeq  What used to be an irrelevant phase factor in (\ref{equ:c1}), now becomes an exponential suppression of the amplitude of the mode function
\beq
|v_k(\ct)| = \frac{\sqrt{\pi}}{2} e^{-\pi \mu /2} \sqrt{-\ct}\, |H_{i\mu}^{(1)}(|k\ct|)|\, .
\eeq
The power spectrum of very massive fields in de Sitter is therefore highly suppressed, ${\cal P}_\varphi \propto e^{-\pi m/H}$, for $m \gg \frac{3}{2}H$.

\subsubsection{Fluctuations during Inflation}
\label{ssec:fluct}

We now move to studying the fluctuations in the inflaton field during inflation. 
These fluctuations cannot be treated independently from fluctuations in the metric, since the two are coupled by the Einstein equations.  This leads to some technical complications, but conceptually the quantization of the coupled inflaton-metric fluctuations is the same as before.

\subsubsection*{Metric fluctuations}

We will treat the fluctuations of the metric in the so-called {\it ADM formalism}~\cite{ADM}.
We start by writing the perturbed line element as
\beq
\d s^2 = - N^2 \d t^2 + h_{ij} (N^i \d t+ \d x^i) (N^j \d t+ \d x^j)\, ,
\eeq
where $N \equiv N(t,\x)$ is the lapse function, $N^i\equiv N^i(t,\x)$ is the shift vector, and $h_{ij}\equiv h_{ij}(t,\x)$ is the induced metric on three-dimensional hypersurfaces of constant time $t$.
The geometry of the spatial slices is characterized by the intrinsic curvature, $R_{ij}^{(3)}$, i.e.~the Ricci tensor of the induced metric, and by the extrinsic curvature
\beq
K_{ij} \equiv \frac{1}{2N}\left(h_{ij}' -\nabla_i N_j - \nabla_j N_i\right) \, \equiv\, \frac{1}{N} E_{ij}\, .
\eeq
The four-dimensional Ricci scalar, $R$, can be written in terms of the three-dimensional Ricci scalar, $R^{(3)}$, and the extrinsic curvature tensor as
\beq
R = R^{(3)} + N^{-2} \left(E^{ij}E_{ij}-E^2\right) ,
\eeq
where indices are raised with $h^{ij}$, and $E \equiv h^{ij} E_{ij}$.

\vskip 4pt
The inflaton action can then be written as
\begin{align}
S &= \frac{1}{2} \int \d^4 x\sqrt{-g} \,\Big[R - g^{\mu \nu} \partial_\mu \phi \partial_\nu \phi - 2 V\Big]\nonumber \\
&= \frac{1}{2} \int \sqrt{h}N\left[ R^{(3)} -2 V + N^{-2}\left(E^{ij}E_{ij}-E^2\right) + N^{-2} \left(\phi'-N^i \partial_i \phi\right)^2 -  h^{ij} \partial_i \phi \partial_j \phi \right] , \label{equ:IA}
\end{align}
where for the moment we have set $M_{\rm pl} \equiv 1$.  Note that $N$ and $N_i$ do not have time derivatives acting on them and are therefore non-dynamical fields that will be fixed by constraint equations.  
Indeed, varying the action with respect to $N$ and $N^i$, we find 
\begin{align}
R^{(3)} - 2V - h^{ij}\partial_i \phi \partial_j \phi - N^2 [E_{ij}E^{ij}-E^2 -(\phi' - N^i \partial_i \phi)^2] &=0\, ,  \label{equ:CON1}\\
\nabla_i[N^{-1}(E^i_j -E\delta^i_j)] &=0\, . \label{equ:CON2}
\end{align}
Plugging the solutions for $N$ and $N^i$ back into the action leaves $\phi$ and $h_{ij}$ as the only dynamical variables.  We will perform this procedure in perturbation theory.  For the time being, we will focus on scalar perturbations.
To fix time and space reparameterizations, we will have to choose a gauge for $\phi$ and $h_{ij}$.  We will present the results for two different gauges: spatially flat gauge and comoving gauge.

\subsubsection*{Spatially flat gauge}

As the name suggests, in spatially flat gauge the induced metric is taken to be unperturbed:
\beq
h_{ij} = a^2 \delta_{ij}\, . \label{equ:SF1}
\eeq
We then consider the perturbations of the inflaton, the lapse and the shift:
\beq
\phi \equiv \bar \phi(t) + \varphi(t,\x)\,, \quad N\equiv 1 + \alpha(t,\x) \, ,\quad N_i \equiv \partial_i \beta(t,\x)\, .  \label{equ:SF2}
\eeq
Substituting (\ref{equ:SF1}) and (\ref{equ:SF2}) into the constraint equations (\ref{equ:CON1}) and (\ref{equ:CON2}), we obtain~\cite{Maldacena:2002vr}\footnote{To find the quadratic action, we only have to solve the constraints to linear order. The second order terms in $N$ and $N_i$ will multiply the zeroth order constraints which vanish when the background equations of motion are imposed.}
\beq
\alpha = \frac{\bar \phi^{\hskip 1pt \prime}}{2H} \, \varphi\, ,  \quad \partial^2\beta = \frac{(\bar\phi')^2}{2H^2} \frac{d}{dt}\left(-\frac{H}{\bar \phi'} \,\varphi\right) .
\eeq
Plugging this solution into the action (\ref{equ:IA}), expanding to second order and performing a few
 integrations by parts, we find
\beq
S_2 = \frac{1}{2}\int \d t\,\d^3 x \ a^3 \left[(\varphi')^2 - \frac{1}{a^2} (\partial\varphi)^2 -\left[V'' -2(3\varepsilon-\varepsilon^2 + \varepsilon \eta) H^2\right] \varphi^2\right] .
\eeq
Switching to conformal time, the equation of motion for the Fourier components of the canonically normalized field $v\equiv a \varphi$ is
\beq
\ddot v_k + \left(k^2 + a^2 \left[V'' - 6\varepsilon H^2\right] - \frac{\ddot a}{a}\right) v_k = 0 \, , \label{equ:747}
\eeq
where we have dropped terms in the effective mass that are higher order in slow-roll parameters. 
Note that the term $a^2 \left[V'' - 6\varepsilon H^2\right] \sim {\cal O}(\varepsilon, \eta) (aH)^2$ is always smaller than $\ddot a/a = 2(aH)^2$. However, it is the only source of time evolution on superhorizon scales and even a small evolution can accumulate over time.
In order not to have to follow the evolving field on superhorizon scales, it is useful to evaluate the solution at horizon crossing, $k=aH$, and then map it to a field that is known not to evolve outside the horizon. This constant mode is the {\it comoving curvature perturbation}, which in spatially flat gauge is defined as
\beq
\zeta = - \frac{H}{\bar \phi^{\hskip 1pt \prime}} \, \varphi\, .
\eeq
The power spectrum of $\zeta$ is
\beq
{\cal P}_\zeta(k) = \left( \frac{H}{\bar \phi^{\hskip 1pt \prime}}\right)^2 {\cal P}_\varphi(k, \ct) \bigg|_{k=aH}\, .
\eeq
Dropping the slow-roll suppressed terms in (\ref{equ:747}), the equation of motion reduces to that of a massless field in de Sitter space. The power spectrum ${\cal P}_\varphi(k, \eta)$ is therefore given by (\ref{equ:Pv}) and we get
\beq
\boxed{{\cal P}_\zeta(k) = \left( \frac{H}{\bar \phi^{\hskip 1pt \prime}}\right)^2 \left(\frac{H}{2\pi}\right)^2 \bigg|_{k=aH}}\ . \label{equ:Pzeta1}
\eeq
\vskip 4pt
\noindent
Another advantage of using $\zeta$ rather than $\varphi$ to characterize the initial conditions is that $\zeta$ remains well-defined after inflation.

\subsubsection*{Comoving gauge}

In comoving gauge the inflaton field is taken to be unperturbed:
\beq
\phi = \bar \phi(t)\, . \label{equ:CG1}
\eeq
All perturbations are then carried by the metric:
\beq
h_{ij} = a^2 (1+2\zeta(t,\x))\delta_{ij}\, , \quad N\equiv 1 + \alpha(t,\x) \, ,\quad N_i \equiv \partial_i \beta(t,\x)\, , \label{equ:CG2}
\eeq
where $\zeta$ is the comoving curvature perturbation. 
Substituting this ansatz into the constraint equations (\ref{equ:CON1}) and (\ref{equ:CON2}), we obtain~\cite{Maldacena:2002vr}
\beq
\alpha = \frac{\zeta'}{H}\, , \quad \partial^2 \beta = - \frac{\partial^2\zeta}{H}+ a^2  \frac{(\bar 
\phi')^2}{2H^2} \zeta'  \, .  \label{equ:alpha2}
\eeq
Plugging this solution into the action (\ref{equ:IA}), expanding to second order, performing integrations by parts and using the background equations of motion, we get a remarkably simple result
\beq
S_2 = \int \d t \, \d^3 x\, a^3 \varepsilon \left((\zeta')^2 - \frac{1}{a^2} (\partial\zeta)^2\right) .
\eeq
Note that there is no mass term, so $\zeta$ is conserved outside the horizon.
Consider the equation of motion
\beq
\zeta_k'' + (3+\eta)H \zeta_k' + \frac{k^2}{a^2} \zeta_k = 0\, .
\eeq
On super-Hubble scales, $k \ll aH$, this becomes $\zeta_k'' + (3+\eta)H \zeta_k' \approx 0$, which clearly has a constant mode as a solution.

\vskip 4pt
The equation of motion of the canonically normalized field, $v \equiv a \sqrt{2\varepsilon}\, \zeta \equiv z\, \zeta$, is
\beq
\ddot v_k + \left(k^2  - \frac{\ddot z}{z}\right) v_k = 0 \, , \label{equ:vEoM}
\eeq
where the effective mass can be written as
\beq
\frac{\ddot z}{z} = \frac{\nu^2-1/4}{\ct^2} \, ,  \quad {\rm with} \quad \nu \approx \frac{3}{2} +  \varepsilon+\frac{\eta}{2}\, .
\eeq
Recall that we have seen eq.~(\ref{equ:vEoM}) before when we considered a massive scalar field in de Sitter space, cf.~eq.~(\ref{equ:EM}). 
The Bunch-Davies mode function therefore is
\beq
|v_k(\ct)| = \frac{\sqrt{\pi}}{2}  \sqrt{-\ct}\, |H_{\nu}^{(1)}(-k\ct)|\ \xrightarrow{\ k\ct \to 0\ } \ \frac{2^\nu \Gamma(\nu)}{2\sqrt{\pi}} \,\frac{1}{\sqrt{k}}\, (-k\ct)^{-\nu+1/2}  \, , \label{equ:vk}
\eeq
and the superhorizon limit of the power spectrum of $\zeta$ is
\begin{align}
{\cal P}_\zeta(k) \equiv \lim_{k\ct\to 0} \frac{k^3}{2\pi^2}|\zeta_k(\ct)|^2  &= \frac{1}{z^2(\ct)} \lim_{k\ct\to 0}  \frac{k^3}{2\pi^2} |v_k(\ct)|^2 \nonumber \\[4pt]
&=  \frac{1}{2 a^2 \varepsilon} \, \frac{k^2}{4\pi^2} \, (-k\ct)^{-2\nu+1}  \,=\, \frac{1}{16\pi^2}\frac{H^2(\ct)}{\varepsilon(\ct)}  \,(-k\ct)^{-2\nu+3}\, . \label{equ:spec0}
\end{align}
Note that the time dependence of $H(\ct)$ and $\varepsilon(\ct)$ precisely cancels the time dependence of the final factor in (\ref{equ:spec0}), so that the power spectrum is time independent.
Let $k_*$ be a reference scale that exits the horizon at time $\ct_*=-1/k_*$. Equation (\ref{equ:spec0}) can then be written as 
\beq
\boxed{{\cal P}_\zeta(k) =  \frac{1}{8\pi^2} \frac{1}{\varepsilon_*}  \frac{H_*^2}{M_{\rm pl}^2}  \,(k/k_*)^{-2\varepsilon_*-\eta_*}}\ , \label{equ:Pzeta2}
\eeq
where $H_*\equiv H(\ct_*)$ and $\varepsilon_* \equiv \varepsilon(\ct_*)$.

\subsection{Curvature Perturbations}

For ease of reference, we briefly summarize the results of the previous section.
The power spectrum of the curvature perturbation $\zeta$, cf.~eqs.~(\ref{equ:Pzeta1}) and (\ref{equ:Pzeta2}), takes a power law form
\beq
{\cal P}_\zeta(k)  = A_{\rm s} \left(\frac{k}{k_*}\right)^{n_{\rm s}-1}\, ,
\eeq
where the amplitude and the spectral index are
\begin{align}
A_{\rm s} &\equiv \frac{1}{8\pi^2} \frac{1}{\varepsilon_*}\frac{H^2_*}{M_{\rm pl}^2}\, , \label{equ:As}\\
n_{\rm s} &\equiv 1  - 2\varepsilon_* - \eta_*\, . \label{equ:ns}
\end{align}
The observational constraint on the scalar spectral index is $n_{\rm s} =0.9603\pm0.0073$. The observed percent-level deviation from the scale-invariant value, $n_{\rm s}=1$, are the first direct measurement of time dependence in the inflationary dynamics.

\begin{framed}
{\small \noindent {\it Exercise.}---Show that for slow-roll inflation, the results~(\ref{equ:As}) and (\ref{equ:ns}) can be written as
\begin{align}
A_{\rm s} &= \frac{1}{24\pi^2} \frac{1}{\varepsilon_V} \frac{V}{M_{\rm pl}^4}\, , \\
n_{\rm s}  &=  1 - 6 \varepsilon_V + 2 \eta_V\, ,
\end{align}
where $\varepsilon_V$ and $\eta_V$ are the potential slow-roll parameters defined in
\eqref{equ:PotSR}.
This expresses the amplitude of curvature perturbations and the spectral index in terms of the shape of the inflaton potential.}
\end{framed}

\subsection{Gravitational Waves}
\label{equ:GWs}

\begin{figure}[h!]
\begin{center}
\includegraphics[scale=0.75]{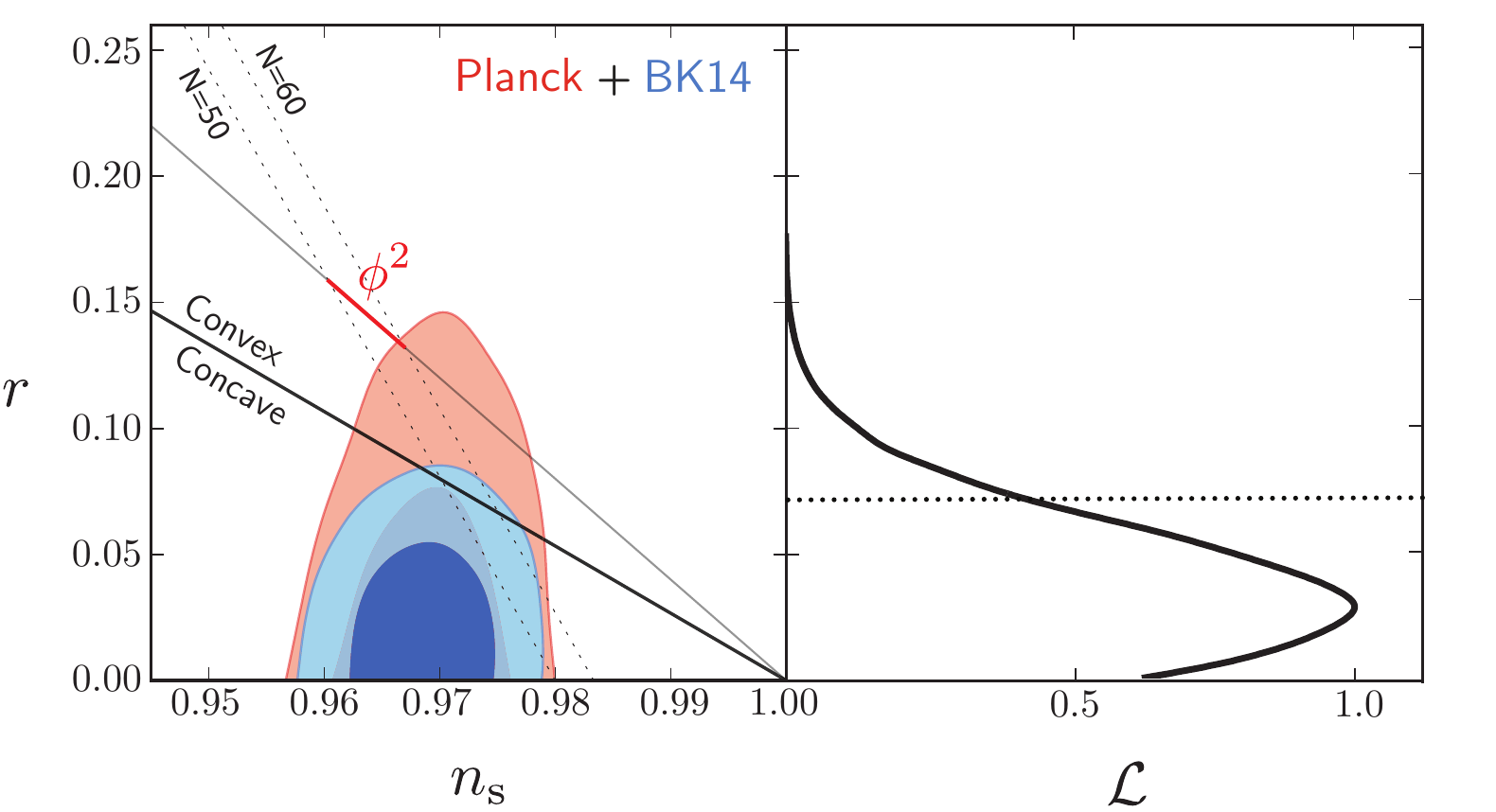}
\caption{Current constraints on $n_{\rm s}$ and $r$ from CMB measurements of Planck and BICEP~\cite{Ade:2015tva}.}
\label{fig:nsr}
\end{center}
\end{figure}

 Arguably the cleanest prediction of inflation is a  spectrum of primordial gravitational waves. These are tensor perturbations to the spatial metric,
  \beq
  \d s^2 = a^2(\ct) \left[ -\d \ct^2 +(\delta_{ij} + 2\gamma_{ij}) \d x^i \d x^j \right] , \label{equ:tensors}
  \eeq
  where $\gamma_{ij}$ is transverse and traceless.  Substituting (\ref{equ:tensors}) into the Einstein-Hilbert action and expanding to second order gives
  \beq
 S = \frac{M_{\rm pl}^2}{2}\int \d^4 x \sqrt{-g} \, R \qquad \Rightarrow \qquad S_2 = \frac{M_{\rm pl}^2}{8} \int \d \ct \,\d^3 x\, a^2 \left[ \dot \gamma_{ij}^2 - (\partial\gamma_{ij})^2 \right]  . \label{equ:h2}
  \eeq
\noindent
It is convenient to use rotational symmetry to align the $z$-axis of the coordinate system with the momentum of the mode, i.e.~$\k \equiv (0,0,k)$, and write
  \beq
 \frac{M_{\rm pl}}{2} \, a \gamma_{ij} \equiv \frac{1}{\sqrt{2}}\left(\begin{array}{ccc} v_+ & v_\times & 0 \\ v_\times & - v_+ & 0 \\ 0 & 0 & 0 \end{array} \right) .
  \eeq
The action (\ref{equ:h2}) then becomes
  \beq
  S_2 =  \frac{1}{2} \sum_{\lambda = +,\times} \int \d \ct  \,\d^3 x   \left[ \dot v_{\lambda}^2 - (\partial v_{\lambda})^2  + \frac{\ddot a}{a} v_{\lambda}^2\right] ,
  \eeq
which is just two copies of the action of a massless scalar field, one for each polarization mode of the gravitational wave, $v_{+,\times} $.
The equation of motion for each polarization is
\beq
\ddot v_{k} + \left(k^2 -\frac{\ddot a}{a}\right) v_k =0\, ,
\eeq
where the effective mass can be written as
\beq
\frac{\ddot a}{a} = \frac{\nu^2-1/4}{\ct^2} \, ,  \quad {\rm with} \quad \nu \approx \frac{3}{2} +  \varepsilon\, .
\eeq
The Bunch-Davies mode function is then given by (\ref{equ:vk}).
The superhorizon limit of the power spectrum of the tensor flucutations then is
\begin{align}
{\cal P}_\gamma(k) = 2\times \lim_{k\ct\to 0} \frac{k^3}{2\pi^2} |\gamma_k(\ct)|^2 &=   2 \times \left(\frac{2}{a M_{\rm pl}}\right)^2 \lim_{k\ct\to 0} \frac{k^3}{2\pi^2} |v_k(\ct)|^2  \nonumber \\[4pt]
&=   \frac{2}{\pi^2}\frac{H^2(\ct)}{M_{\rm pl}^2}  \,(-k\ct)^{-2\nu+3}\, , \label{equ:spec}
\end{align}
where the factor of 2 accounts for the sum over the two polarization modes.
Introducing the reference scale $k_*$, this can be written as

\beq
\fbox{$\displaystyle {\cal P}_\gamma(k) = \frac{2}{\pi^2} \frac{H^2_*}{M_{\rm pl}^2} (k/k_*)^{-2\varepsilon_*} $} \ .
\eeq
This result is arguably the most robust and model-independent prediction of inflation. 
 We see that the form of the tensor power spectrum is again a power law, 
 \beq
{\cal P}_\zeta(k)  = A_{\rm t} \left(\frac{k}{k_*}\right)^{n_{\rm t}}\, ,
\eeq
where the amplitude and the spectral index are
\begin{align}
A_{\rm t} &\equiv \frac{2}{\pi^2} \frac{H^2_*}{M_{\rm pl}^2}\, , \label{equ:At}\\
n_{\rm t} &\equiv  - 2\varepsilon_* \, . \label{equ:nt}
\end{align}
 Notice that the tensor amplitude is a direct measure of the expansion rate $H$ during inflation.  This is in contrast to the scalar amplitude which depends on both $H$ and $\varepsilon$.
 The tensor tilt is a direct measure of $\varepsilon$, whereas the scalar tilt depends both on $\varepsilon$ and $\eta$.
Observationally, a small value for $n_{\rm t}$ is hard to distinguish from zero.
The tensor amplitude is often normalized with respect to the measured scalar amplitude, $A_{\rm s}=(2.196\pm0.060)\times 10^{-9}$ (at $k_*=0.05$ Mpc$^{-1}$): 
\beq
\boxed{r \equiv \frac{A_{\rm t}}{A_{\rm s}} = 16\hskip 1pt\varepsilon_*}\ , \label{equ:r}
\eeq
where $r$ is the {\it tensor-to-scalar ratio}.
Inflationary models make predictions for $(n_{\rm s},r)$.  The latest observational constraints on these parameters are shown in Fig.~\ref{fig:nsr}.
  
 \begin{framed}
\noindent
{\small
{\it Case study: $m^2 \phi^2$ inflation.}---In Section~\ref{sec:Inflation}, we showed that the slow-roll parameters of $m^2 \phi^2$ inflation are 
\beq
\varepsilon_V(\phi) = \eta_V(\phi) = 2 \left( \frac{M_{\rm pl}}{\phi}\right)^2\, ,
\eeq
and the number of $e$-folds before the end of inflation is
\beq
N(\phi) = \frac{\phi^2}{4 M_{\rm pl}^2} - \frac{1}{2} \approx \frac{\phi^2}{4 M_{\rm pl}^2}\, .
\eeq
At the time when the CMB fluctuations crossed the horizon at $\phi = \phi_*$, we  have
\beq
\varepsilon_{{V}, *} = \eta_{{V}, *} \approx \frac{1}{2N_*} \, .
\eeq
The spectral tilt and the  tensor-to-scalar ratio therefore are
\begin{align}
n_{\rm s} &\equiv 1 - 6\hskip 1pt \varepsilon_{{V},*} + 2\hskip 1pt \eta_{{V},*} = 1 - \frac{2}{N_*} \approx 0.97\, ,  \\
r &\equiv 16\hskip 1pt \varepsilon_{{V},*} = \frac{8}{N_*} \approx 0.13\, , 
\end{align}
where the final equalities are for $N_* \approx 60$. Comparison with Fig.~\ref{fig:nsr} shows that these predictions are in conflict with the latest CMB  data.}
\end{framed}

 \begin{framed}
\noindent
{\small
{\it The Lyth bound.}---During $m^2 \phi^2$ inflation, the inflaton field value varies by $\Delta \phi \approx 15\,M_{\rm pl}$ from the time the mode observed in the CMB exited the horizon until the end of inflation. This super-Planckian field variation is, in fact, a general feature of all inflationary models with observable gravitational waves.

\vskip 4pt
\noindent
Combining (\ref{equ:SR1}) and (\ref{equ:r}), the tensor-to-scalar ratio can be related  to the evolution of the inflaton:
\beq
r = \frac{8}{M_{\rm pl}^2} \left(\frac{d \phi}{d N}\right)^2\, ,
\eeq
where $\d N = H \d t$.
The total field excursion between the time when CMB fluctuations exited the horizon at $N_*$ and the end of inflation at $N_e$ can therefore be written as the  integral
\beq
\frac{\Delta \phi}{M_{\rm pl}} = \int_{N_e}^{N_*} \d N\, \sqrt{\frac{r}{8}} \, .
\eeq
During slow-roll evolution, $r(N)$ doesn't evolve much and one may obtain the following approximate relation, called the {\it Lyth bound}\hskip 1pt:
\beq
\frac{\Delta \phi}{M_{\rm pl}} = {\cal O}(1) \times \left( \frac{r}{0.01}\right)^{1/2} \, ,
\eeq
where $r \equiv r(N_*)$ is the tensor-to-scalar ratio on CMB scales.
Large values of the tensor-to-scalar ratio, $r > 0.01$, therefore correlate with super-Planckian field excursions, $\Delta \phi > \Mp$. }
\end{framed}



A major goal of current efforts in observational cosmology is to detect the tensor component of the primordial fluctuations. 
Its amplitude depends on the energy scale of inflation and is therefore not predicted (i.e.~it varies between models). 
While this makes the search for primordial tensor modes difficult, it is also what makes it exciting. Detecting tensors would reveal the energy scale at which inflation occurred, providing an important clue about the physics driving the inflationary expansion.

Most searches for tensors focus on the imprint that tensor modes leave in the polarization of the CMB.  Polarization is generated through the scattering of the anisotropic radiation field off the free electrons just before decoupling.  The presence of a gravitational wave background creates an anisotropic stretching of the spacetime which induces a special type of polarization pattern, the so-called B-mode pattern (a pattern whose ``curl" doesn't vanish). 
Such a pattern cannot be created by scalar (density) fluctuations and is therefore a unique signature of primordial tensors (gravitational waves).
A large number of ground-based, balloon and satellite experiments are currently searching for the B-mode signal predicted by inflation.   A B-mode detection would be a milestone towards a complete understanding of the origin of all structure in the universe.

\newpage
\section{Primordial Interactions}
\label{sec:NG}

In the previous section, we have computed the two-point function of primordial curvature perturbations, or its Fourier equivalent, the power spectrum.
If the initial conditions are drawn from a Gaussian distribution function, then the power spectrum contains all of the information about the primordial perturbations. In general, however, higher-order correlations can encode a significant amount of new information.  In particular, these correlations are sensitive to nonlinear interactions while the power spectrum only probes the free theory. 

\vskip 4pt
In this section, I will introduce the basic formalism for computing the non-Gaussianity produced by inflation. I will apply this to single-field slow-roll inflation and show that gravitational nonlinearities produce a robust (but small) non-Gaussian signature.
In Section~\ref{sec:HeavyRelics}, I will show that extra massive fields can get excited during inflation and that their decays lead to distinctive signatures in cosmological correlation functions. 

\vskip 4pt
For further reading I highly recommend the classic papers by Maldacena~\cite{Maldacena:2002vr} and Weinberg~\cite{Weinberg:2005vy}, as well as the wonderfully clear reviews by Chen~\cite{Chen:2010xka}, Wang~\cite{Wang:2013eqj}, Lim~\cite{Lim} and Komatsu~\cite{Komatsu:2010hc}.

\subsection{Non-Gaussianity}

The leading non-Gaussian signature is the three-point correlation function,
or its Fourier equivalent, the bispectrum
\beq
B_\zeta({\k}_1, {\k}_2, {\k}_3) \equiv \langle \zeta_{\k_1} \zeta_{{\k}_2} \zeta_{{\k}_3} \rangle \, .
\eeq
For perturbations around an FRW background, the momentum dependence of the bispectrum simplifies considerably:
Because of homogeneity, or {\it translation invariance}, the bispectrum is proportional to a delta function of the sum of the momenta, $B_\zeta({\bf k}_1, {\bf k}_2, {\bf k}_3) \propto \delta_D({\bf k}_1 + {\bf k}_2 + {\bf k}_3)$, i.e.~the sum of the momentum three-vectors must form a closed triangle.
Because of
isotropy, or {\it rotational invariance}, the bispectrum only depends on the magnitudes of the momentum vectors, but not on their orientations,  
\beq
B_\zeta({\bf k}_1, {\bf k}_2, {\bf k}_3) = (2\pi)^3 \delta_D({\bf k}_1 + {\bf k}_2 + {\bf k}_3) \,B_\zeta(k_1,k_2,k_3)\, . \label{equ:Bzeta}
\eeq
It is convenient to define the dimensionless bispectrum as
\beq
{\cal B}_\zeta(k_1,k_2,k_3) \equiv  \frac{(k_1k_2k_3)^2}{(2\pi^2)^2} B_\zeta(k_1,k_2,k_3)\, .
\eeq
The {\it amplitude} of the non-Gaussianity is then defined as the size of the bispectrum
in the equilateral momentum configuration:
\beq
f_{\rm NL}(k) \equiv \frac{5}{18} \frac{{\cal B}_\zeta(k,k,k)}{{\cal P}_\zeta^2(k)}\, ,
\eeq
where we have indicated that $f_{\rm NL}$ can in general depend on the overall momentum.
On the other hand, if the fluctuations are {\it scale-invariant}, then $f_{\rm NL}$ is a constant and we can write
\beq
{\cal B}_\zeta(k_1,k_2,k_3) \equiv  \frac{18}{5} f_{\rm NL} \times {\cal S}(x_2,x_3)\times {\cal P}_\zeta^2\, ,
\eeq
where $x_2 \equiv k_2/k_1$ and $x_3\equiv k_3/k_1$.  The shape function ${\cal S}(x_2,x_3)$ is normalized so that ${\cal S}(1,1) \equiv 1$.

\begin{figure}[h!]
    \centering
        \includegraphics[width=.72\textwidth]{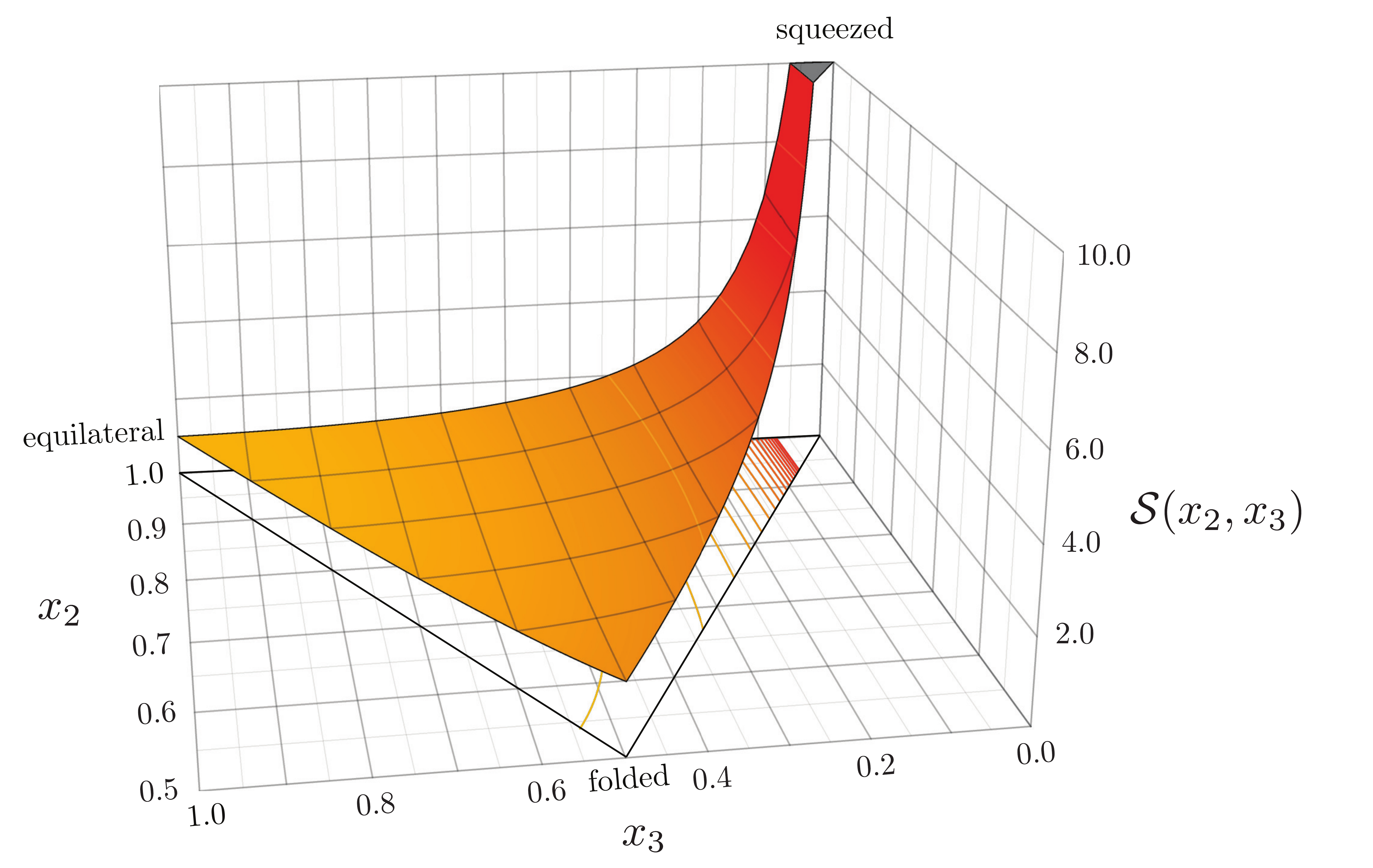}
    \caption{Bispectrum of local non-Gaussianity. The signal is peaked for squeezed
triangles.}
    \label{fig:local}
\end{figure}

\begin{figure}[h!]
    \centering
        \includegraphics[width=.72\textwidth]{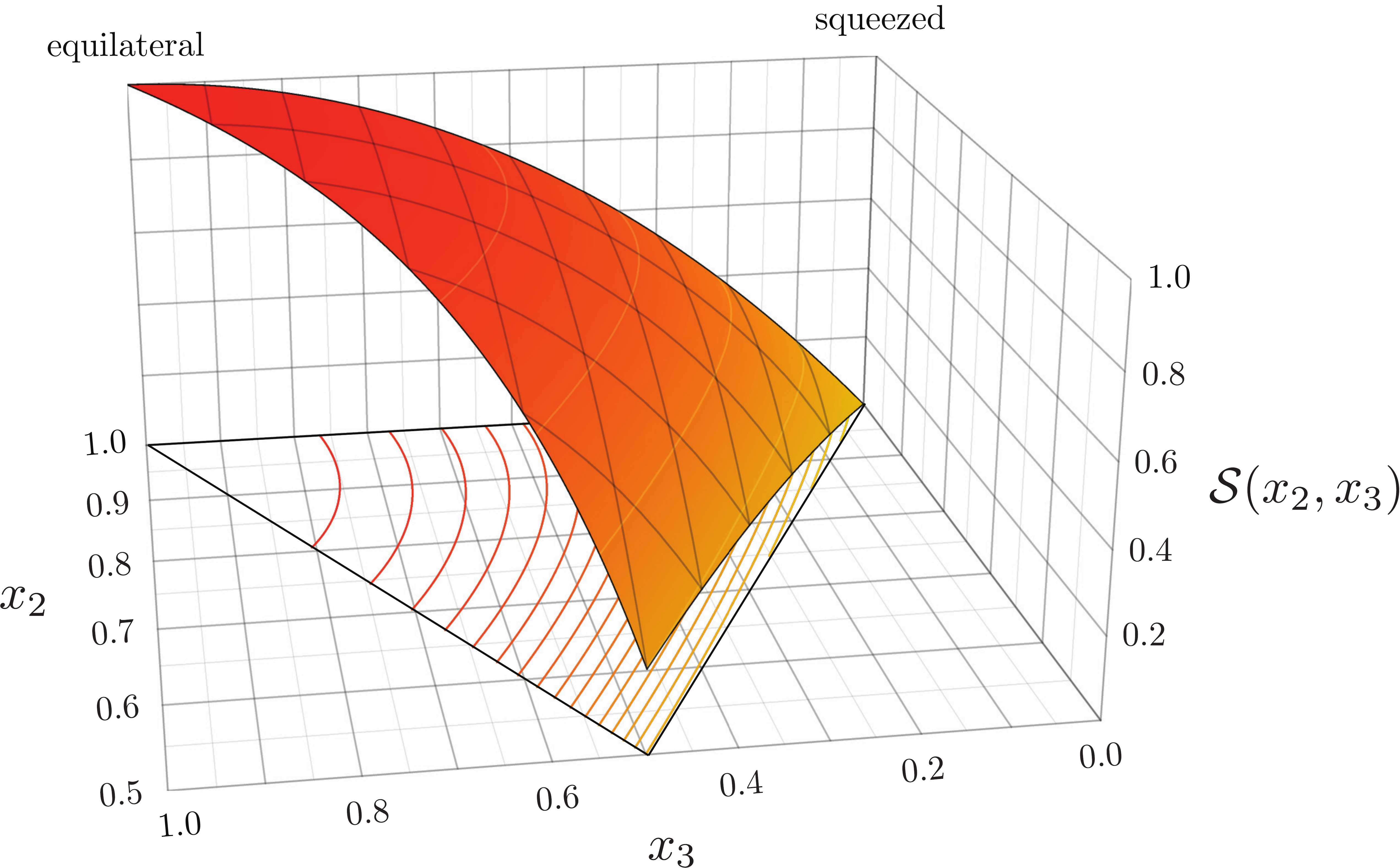}
    \caption{Bispectrum for the interaction $\dot \zeta^3$. The signal is peaked for
equilateral triangles.}
    \label{fig:equilateral}
\end{figure}

\vskip 4pt
As we will discuss in detail below, the {\it shape} of the non-Gaussianity contains a lot of information about the microphysics of inflation.  A divergence of the signal for squeezed
triangles is a signature for extra degrees of freedom during inflation (see Fig.~\ref{fig:local} and Sec.~\ref{sec:HeavyRelics}), while a peak in the signal for equilateral triangles arises from higher-derivative inflaton self-interactions (see Fig.~\ref{fig:equilateral}).

\subsection{In-In Formalism}

The problem of computing correlation functions in cosmology differs in important ways from the corresponding analysis of quantum field theory applied to particle physics.

\vskip 4pt
In particle physics, the main observable is the $S$-matrix, i.e.~the transition probability for a state\ $|{in}\rangle$\ in the far past to become some state\ $|
{out} \rangle $\ in the far future,
\begin{equation} \langle {out} | \,S | {in} \rangle = \langle {out}(+{\infty}) | {in}(- {\infty}) \rangle\ .\label{apC:eqnC.60}
\end{equation}
The scattering particles are taken to be non-interacting at very early and very late times, when they are far from the interaction region, and the asymptotic states can be taken to be vacuum states of the free Hamiltonian~$H_0$.

  \begin{figure}[h!]
    \centering
        \includegraphics[width=.4\textwidth]{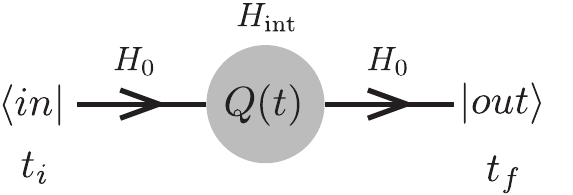}
    \caption{Particle physicists compute in-out transition amplitudes.}
    \label{fig:inout}
\end{figure}

\vskip 4pt
In cosmology, on the other hand, the task is to determine  the
expectation values of products of operators \textit{at a fixed time}.
Boundary conditions are
only imposed at very early times when their wavelengths are much
smaller than the horizon and the interaction picture
fields should have the same form as in Minkowski space. As we have
seen in the previous section, this leads to the definition of the
Bunch--Davies vacuum. In this
section, we will describe the $in$-$in$ formalism\footnote{This is
also referred to as the Schwinger--Keldysh formalism
\cite{Schwinger:1960qe}.\index{Schwinger--Keldysh formalism} The use
of the $in$-$in$ formalism in cosmology was pioneered
in~\cite{Calzetta:1986ey,Jordan:1986ug,Maldacena:2002vr,Weinberg:2005vy}
(see also \cite{Seery:2007we, Adshead:2009cb}) and is reviewed in
\cite{Chen:2010xka, Wang:2013eqj}.} to compute cosmological
correlation functions as expectation values in two\ $|{in} \rangle$\
states.

  \begin{figure}[h!]
    \centering
        \includegraphics[width=.4\textwidth]{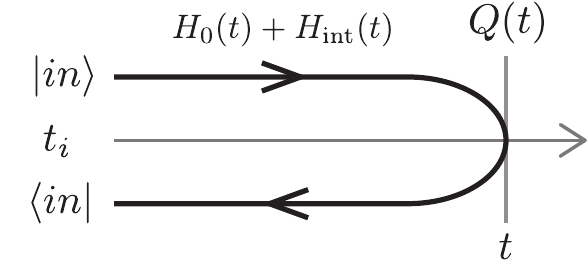}
    \caption{Cosmologists compute in-in expectation values.}
    \label{fig:inin}
\end{figure}

\vskip 4pt
Our goal is to compute $n$-point functions of the curvature perturbation $\zeta$ or the gravitational wave polarization modes $\gamma^\times$ and $\gamma^+$.
We will collectively denote these fluctuations by the field $\psi =\{ \zeta, \gamma^\times, \gamma^+\}$ and consider expectation values of operators such as $\hat Q= \hat \psi_{{\k}_1} \hat \psi_{{\k}_2} \cdots \hat \psi_{{\k}_n}$,
\beq
\label{equ:QQ}
\langle \hat Q(\ct) \rangle = \langle in | \, \hat Q(\ct)\,  | in \rangle\, ,
\eeq
where $|in\rangle$
is the vacuum of the interacting theory at some moment ${\ct}_i$ in
the far past, and $\ct > \ct_i$ is some later time, such as
horizon crossing or the end of inflation.  To compute the matrix element in (\ref{equ:QQ}) we evolve $Q(\ct)$ back to $\ct_i$ using the perturbed Hamiltonian $\delta H$. 
Computing this time evolution is complicated by the interactions inside of $\delta H =  H_0 + H_{\rm int}$, which lead to nonlinear equations of motion.
We therefore introduce the {\it interaction picture} in which the leading time dependence of the fields is determined by the quadratic Hamiltonian $H_0$ (or linear equations of motion). Corrections arising from the interactions are then treated as a power series in $H_{\rm int}$.

\vskip 4pt
This leads to the following important result:
\beq
\label{equ:InIn}
\fbox{$\displaystyle \langle \hat Q(\ct) \rangle \ =\ \langle 0 |\ \bar T e^{ i \int_{-\infty(1-i\epsilon)}^\ct  \hat H_{\rm int}^I(\ct') \,\d \ct' }\, \hat Q^I(\ct) \  T e^{ -i \int_{-\infty(1+i\epsilon)}^\ct \hat H_{\rm int}^I(\ct'') \,\d \ct'' }  \, | 0 \rangle $} \ ,
\eeq
where $T$ ($\bar T$) is the (anti-)time-ordering symbol.
Note that both $\hat Q^I$ and $\hat H_{\rm int}^I$ are evaluated using interaction picture operators. 
The standard $i \epsilon$ prescription has been used to effectively turn off the interaction in the far past and project the interacting $|in \rangle$ state onto the free vacuum $|0\rangle$.
By expanding the exponentials, we can compute the correlation function perturbatively in $\hat H_{\rm int}$. For example, at leading order, we find
\beq
\fbox{$\displaystyle \langle \hat Q(\ct) \rangle \ =\ -i \int_{-\infty}^\ct \d \ct' \ \langle 0 |\, [\hat Q^I(\ct) ,  \hat H_{\rm int}^I(\ct') ]\, | 0 \rangle $} \ . \label{equ:InInMaster}
\eeq
We can use Feynman diagrams to organize the power series, drawing interaction vertices for every power of $\hat H_{\rm int}$.

\vskip 4pt
In the following insert, I will derive the $in$-$in$ master formula~(\ref{equ:InIn}).
Readers who are more interested in applications of the result, may skip this part.

\begin{framed}
\noindent
{\small {\it Derivation.}---The time evolution of the operators in the Heisenberg picture is determined by 
\begin{equation}
\frac{d\hat \psi}{d{\ct}} = i[\hat H, \hat \psi] \, , \quad  \frac{d\hat p_\psi}{d{\ct}} = i [\hat H, \hat p_\psi]\, , \label{equ:H1}
\end{equation}
where $\hat H =  \hat H_0 + \hat H_{\rm int}$ is the perturbed Hamiltonian. This time evolution is complicated by the interactions inside $\hat H$, which lead to nonlinear equations of motion.
We therefore introduce the \textit{interaction picture} in which the leading time dependence of the fields is determined by the quadratic Hamiltonian $\hat H_0$ (or, equivalently, by the linear equations of motion):
\begin{equation}
\frac{d\hat \psi^I}{d\ct} = i[\hat H_0, \hat \psi^I] \ , \quad  \frac{d\hat p_\psi^I}{d\ct} = i [\hat H_0, \hat p_\psi^I]\, . \label{equ:H2}
\end{equation}
The solution to these equations can be written as
\begin{equation}
\hat \psi^I_{\k}({\ct}) = \psi^I_k({\ct})\, \hat a_{\k} + h.c.\, , \label{equ:PiI}
\end{equation}
where $\psi^I_k({\ct})$ is the solution to the free-field equation
and the operators $\hat a_{\k}$ define the free-field vacuum~$|0\rangle$. 
Corrections to the evolution of the operators can then be treated perturbatively in $\hat H_{\rm int}$.
Relatively straightforward algebraic manipulations of (\ref{equ:H1}) and (\ref{equ:H2}) allow us to express an operator in the Heisenberg picture in terms of operators in the interaction picture~\cite{Weinberg:2005vy},
\begin{equation}
\hat Q({\ct}) = \hat F^{-1}({\ct},{\ct}_i)\, \hat Q^I({\ct})\,  \hat F({\ct},{\ct}_i)\, ,
\label{apC:eqnC.66}
\end{equation}
where
\begin{equation}
\hat F({\ct},{\ct}_i) \equiv T e^{ -i \int_{{\ct}_i}^{\ct} \hat H_{\rm int}^I({\ct}'') \,\d {\ct}'' }\, .
\label{apC:eqnC.67}
\end{equation}
We can think of $\hat F({\ct},{\ct}_i)$ as an operator evolving quantum states in the interaction picture,
\begin{equation}
|\Omega({\ct}) \rangle = \hat F({\ct},{\ct}_i) | \Omega({\ct}_i) \rangle\, ,
\label{apC:eqnC.68}
\end{equation}
where $| \Omega({\ct}_i) \rangle \equiv | \Omega \rangle$.  We would like to relate the vacuum of the interacting theory, $| \Omega \rangle$, to the vacuum of the free theory, $|0\rangle$.
Inserting a complete set of energy eigenstates $\{|\Omega \rangle, |n\rangle\}$ of the full theory, where $|n\rangle$ are the excited states, we have
\begin{equation}
| 0\rangle = |\Omega \rangle \langle \Omega|0\rangle + \sum_{n}  |n \rangle \langle n|0\rangle  \, ,
\label{apC:eqnC.69}
\end{equation} and correspondingly
\begin{align}
e^{-i \hat H({\ct} -{\ct}_i)}| 0\rangle  &= e^{-i\hat H({\ct} - {\ct}_i)} |\Omega \rangle \langle \Omega|0\rangle + \sum_{n} e^{-iE_n({\ct} - {\ct}_i)} |n \rangle \langle n|0\rangle \, .
\label{apC:eqnC.70}
\end{align}
Adding a small imaginary part to the initial time,
${\ct}_i \to -{\infty} (1-i\epsilon) \equiv - {\infty}^-$, will project out the excited states,
$e^{-iE_n({\ct}-{\ct}_i)} \to e^{- {\infty} \times \epsilon E_n}(\cdots)  \to 0$.
We are then left with
\begin{equation} \label{equ:fhatomega}
\hat F({\ct}, -{\infty}^-) |\Omega\rangle = \frac{\hat F({\ct}, -{\infty}^-) |0\rangle}{\langle \Omega | 0 \rangle}\, .
\end{equation}
The $i \epsilon$ prescription has effectively turned off the interactions in the far past and projected the interacting vacuum $|\Omega \rangle$ onto the free vacuum $|0\rangle$.
Setting \hbox{$|\langle \Omega | 0 \rangle|\rightarrow 1$,} we arrive at the
$in$-$in$ master formula
\begin{equation}
\label{equ:InInMaster}
\langle \hat Q({\ct}) \rangle \ =\ \langle 0 |\ \bar T e^{ i \int_{-{\infty}^+}^{\ct} \hat H_{\rm int}^I({\ct}') \,\d {\ct}' }\, \hat Q^I({\ct}) \  T e^{ -i \int_{-{\infty}^-}^{\ct} \hat H_{\rm int}^I({\ct}'') \,\d {\ct}'' }  \, | 0 \rangle \, ,
\end{equation}
where ${\infty}^{\pm} \equiv {\infty}(1{\pm} i\epsilon)$. The integration contour goes from $-{\infty}(1-i\epsilon)$ to ${\tau}$ (where the correlation function is evaluated) and back to $-{\infty}(1+i\epsilon)$.}
\end{framed}

\subsection{Gravitational Floor}
\label{ssec:floor}

Let us now apply this formalism to the calculation of the bispectrum of curvature perturbations in single-field slow-roll inflation~\cite{Maldacena:2002vr}.
The nonlinearities of the gravitational evolution produce a minimal amount of non-Gaussianity which we will call the ``gravitational floor".  Unfortunately, the amplitude of the signal is too small to be detectable in the foreseeable future.  

\vskip 4pt
To compute the bispectrum we need to expand the inflationary action (\ref{equ:IA}) to third order in perturbations.
It is convenient to perform the computation in comoving gauge, cf.~eqs.~(\ref{equ:CG1}) and (\ref{equ:CG2}).  In \S\ref{ssec:fluct}, we solved the lapse  $N$ and the shift $N_i$ to first order in $\zeta$. This is sufficient also for the cubic action since second-order and third-order perturbations in $N$ and $N_i$  will multiply the first-order and zeroth-order constraint equation, respectively. 
Substituting (\ref{equ:alpha2})  into (\ref{equ:IA}) and expanding to third order, we find~\cite{Maldacena:2002vr}
\begin{align}
S_3 &= \frac{1}{2}\int \d t \,\d^3x\, a^3 \Big(2\varepsilon^2 \,\zeta(\zeta')^2 - 4 a^{-2} \varepsilon^2 \,\zeta' (\partial\zeta) (\partial \zeta) + 2 a^{-2} \varepsilon^2\, \zeta(\partial \zeta)^2 \nonumber \\
 &\hspace{3cm} + a^{-2} \varepsilon \eta'\, \zeta^2 \zeta' + a^{-2}\varepsilon\, \partial \zeta \partial \beta \partial^2 \beta + \frac{1}{2} a^{-2} \varepsilon\, \partial^2 \zeta (\partial \beta)^2 + 2 f(\zeta) \frac{\delta L_2}{\delta \zeta} \Big)\, , \label{equ:S3}
\end{align}
where we have defined
\begin{align}
\frac{\delta L_2}{\delta \zeta} &\equiv (a^3 \varepsilon\, \zeta')' - a \varepsilon \partial^2 \zeta\, ,\\
f(\zeta) &\equiv \frac{\eta}{4}\zeta^2 + \frac{\zeta \zeta'}{H}  + \frac{-(\partial \zeta)^2 + \partial^{-2}(\partial_i \partial_j(\partial_i \zeta \partial_j \zeta))}{4a^2 H^2}  + \frac{\partial\zeta \partial \beta - \partial^{-2} (\partial_i \partial_j(\partial_i\zeta\partial_j \beta))}{2a^2H^2} \, .
\end{align}
Maldacena showed that the term proportional to $f(\zeta)$ in (\ref{equ:S3}) can be removed by a field redefinition,
\beq
\zeta \to \tilde \zeta + f(\tilde \zeta)\, .
\eeq
This field redefinition has the following effect on the correlation function:
\beq
\label{equ:red}
\langle \zeta({\bf x}_1) \zeta({\bf x}_2) \zeta({\bf x}_3) \rangle = \langle \tilde \zeta({\bf x}_1) \tilde \zeta ({\bf x}_2) \tilde \zeta ({\bf x}_3) \rangle  + \frac{\eta}{2} \left( \langle \tilde \zeta({\bf x}_1) \tilde \zeta ({\bf x}_2) \rangle \langle \tilde \zeta({\bf x}_1) \tilde \zeta ({\bf x}_3) \rangle  + {\rm cyclic} \right) + \cdots\, . \eeq
The term proportional to $f(\zeta)$ in~(\ref{equ:S3}) therefore leads to a contribution to $f_{\rm NL}$ of order $\eta \ll 1$.
The effect of the remaining interactions is computed by
expanding the $in$-$in$ master formula (\ref{equ:InIn}) to first order in $H_{\rm int} = - L_3 + {\cal O}(\zeta^4)$,
\beq
\label{equ:inte}
\langle \zeta_{{\bf k}_1} \zeta_{{\bf k}_2} \zeta_{{\bf k}_3} \rangle = - i \int_{-\infty}^0 \d \ct \ \langle 0 | \big[\hat \zeta_{{\bf k}_1} \hat \zeta_{{\bf k}_2} \hat \zeta_{{\bf k}_3}(0), \hat H_{\rm int}(\ct)\big] | 0 \rangle\, ,
\eeq
where we have switched to conformal time and taken the superhorizon limit $k\ct \to 0$.

\vskip 4pt
\begin{framed}
{\small
\noindent
{\it Back-of-the-envelope estimate.}---Before embarking on a lengthy calculation of the bispectrum, it is often advisable to perform an order-of-magnitude estimate of the expected size of the signal, i.e.~to estimate (\ref{equ:inte}) without explicitly performing the integral.
For example, the first term in (\ref{equ:S3}), can be written as 
\beq
\int \d \tau \, H_{\rm int}(\tau) \subset
 - \int  \d\tau \d^3 x \, a^2 \varepsilon^2 \zeta \dot \zeta^2 \ .
\eeq
 We only need to keep track of factors of $H$ and $\varepsilon$. Any time- and momentum-dependence will work itself out and only contributes to the shape function. Using $a \propto H^{-1}$ and $\zeta \propto \dot \zeta \propto {\cal P}_\zeta^{1/2} \sim H/\sqrt{\varepsilon}$, we estimate that the contribution from the three-point vertex is $\sim H\sqrt{\varepsilon}$. 
Combining this with estimates for the size of the three external legs, $\zeta^3 \sim H^3 \varepsilon^{-3/2}$, we find
\beq
\langle \zeta^3 \rangle = - i \int \d \tau\, \langle[\hat \zeta^3, \hat H_{\rm int}(\tau)] \rangle \propto \frac{H^4}{\varepsilon} \propto {\cal O}(\varepsilon) \,{\cal P}_\zeta^2 \sim f_{\rm NL}\,{\cal P}_\zeta^2\ .
\eeq
Similar results are obtained for the other interactions in (\ref{equ:S3}), $f_{\rm NL} \sim {\cal O}(\varepsilon)$. 
We also include the contribution from the field redefinition in (\ref{equ:red}), $f_{\rm NL} \sim {\cal O}(\eta)$. We conclude that the non-Gaussianity in slow-roll inflation is slow-roll suppressed,
\beq
f_{\rm NL} \sim  {\cal O}(\varepsilon, \eta) \ll 1\, .
\eeq
This small amount of non-Gaussianity will be unobservable in the CMB.}
\end{framed}

To get the full momentum-dependence of the bispectrum, we actually have to do some real work and compute the integral in (\ref{equ:inte}) using the free-field mode functions for $\zeta$.
The final result for the dimensionless bispectrum is~\cite{Maldacena:2002vr}
\begin{align}
\frac{{\cal B}_\zeta(k_1,k_2,k_3)}{{\cal P}_\zeta^2} &\,=\,  \frac{\varepsilon}{2} \left[-\left(\frac{k_1^2}{k_2k_3} + 2\,{\rm perms.}\right) + \left(\frac{k_1}{k_2} + 5\,{\rm perms.}\right) + \frac{8}{k_t} \left(\frac{k_1k_2}{k_3} + 2\,{\rm perms.} \right)\right] \nonumber \\
 &\hspace{0.6cm}+\frac{\eta}{2}\left(\frac{k_1^2}{k_2k_3} + 2\, {\rm perms.}\right) \, ,
\end{align}
where $k_t\equiv k_1+k_2+k_3$. It is interesting to take the {\it squeezed limit}, $k_3 \ll k_1\approx k_2$, of the result:
\beq
\lim_{k_3 \to 0} \frac{{\cal B}_\zeta(k_1,k_2,k_3)}{{\cal P}_\zeta^2}  = (2\varepsilon+\eta)\, \frac{k_1}{k_3}\, .
\eeq
Note that the coefficient in the squeezed limit equals the deviation from scale invariance of the power spectrum, $1-n_{\rm s} = 2\varepsilon+\eta$.  In fact, this is a general result that applies to all models of single-field inflation, not necessarily just to slow-roll models.
The {\it single-field consistency relation} states that~\cite{Maldacena:2002vr, Creminelli:2004yq}
\beq
\boxed{\lim_{k_3\to 0}  B_\zeta(k_1,k_2,k_3) = (1-n_{\rm s})\, P_\zeta(k_1) P_\zeta(k_3)}\ ,
\eeq
i.e.~the squeezed limit of the three-point function is suppressed by
$(1-n_{\rm s})$ and vanishes for perfectly scale-invariant perturbations.

\begin{framed}
{\small
\noindent
{\it Proof.}---The squeezed triangle correlates one long-wavelength mode, $k_{\rm L} = k_3$, to two short-wavelength modes, $k_{\rm S}=k_1 \approx k_2$,
\beq
\label{equ:sq}
\langle \zeta_{{\bf k}_1} \zeta_{{\bf k}_2} \zeta_{{\bf k}_3} \rangle  \ \to \ \langle (\zeta_{{\bf k}_{\rm S} })^2 \zeta_{{\bf k}_{\rm L}}\rangle\, .
\eeq
Modes with longer wavelengths freeze earlier. The long mode $\zeta_{{\bf k}_{\rm L}}$ will therefore already be frozen and act as a classical background wave when the two short modes $\zeta_{{\bf k}_{\rm S} }$ exit the horizon.  This provides an intuitive way to study the correlations between long and short modes.

Why should $(\zeta_{{\bf k}_{\rm S} })^2$ be correlated with  $\zeta_{{\bf k}_{\rm L}}$?
The theorem says that it isn't correlated if $\zeta_{\bf k}$ is precisely scale-invariant, but that the short scale power does get modified by the long-wavelength mode if $n_{\rm s} \ne 1$. Let's see why.
We decompose the evaluation of (\ref{equ:sq}) into two steps: 
\begin{enumerate}
\item[i)] we calculate the power spectrum of short fluctuations $\langle  \zeta_{{\rm S}}^2\rangle_{\zeta_{{\rm L}}}$ in the presence of a long mode $\zeta_{{\rm L}}$;
\item[ii)] we then calculate the correlation $\langle (\zeta_{{\rm S} })^2 \zeta_{{\rm L}}\rangle$, i.e.~average the short-scale power spectrum over realizations of the long modes.
\end{enumerate}
The calculation of $\langle  \zeta_{{\rm S}}^2\rangle_{\zeta_{{\rm L}}}$ is simplest in real space:
When the background mode is homogeneous, $\zeta_{\rm L}(x) \equiv \bar \zeta_{\rm L}$, it can be reabsorbed simply by a rescaling of the spatial coordinates, $\tilde x^i = e^{\bar \zeta_{\rm L}} x^i$ (recall that $\d s^2 =  - \d t^2 + a^2(t) e^{2\zeta({\bf x},t)} \d {\bf x}^2$).
 After this rescaling, $\bar \zeta_{\rm L}$ no longer appears in the action, so that the two-point function in the new coordinates is the same as in the absence of $\bar \zeta_{\rm L}$. In other words, in the limit of constant $\bar \zeta_{\rm L}$, we have
\beq
\langle \zeta_{\rm S}({\bf x}_1) \zeta_{\rm S}({\bf x}_2) \rangle_{\bar \zeta_{\rm L}} = \langle \zeta_{\rm S}(\tilde{\bf x}_1) \zeta_{\rm S}(\tilde{\bf x}_2) \rangle\, .
\eeq
When $\zeta_{\rm L}$ is slowly varying, we can evaluate it at the middle point ${\bf x}_+ \equiv \frac{1}{2}({\bf x}_1 + {\bf x}_2)$ to get
\beq
\tilde{\bf x}_- \simeq {\bf x}_- + \zeta_{\rm L}({\bf x}_+) \cdot {\bf x}_-  + \cdots\, ,
\eeq
where we defined ${\bf x}_- \equiv {\bf x}_1 -{\bf x}_2$.
The two-point function at linear order in $\zeta_{\rm L}$ therefore is
\beq
\langle \zeta_{\rm S}({\bf x}_2) \zeta_{\rm S}({\bf x}_3) \rangle_{\zeta_{\rm L}(x)} \simeq \xi_{\rm S}(|{\bf x}_-|) + \zeta_{\rm L}({\bf x}_+) [{\bf x}_- \cdot \nabla \xi_{\rm S}(|{\bf x}_-|)]\ ,
\eeq
where 
\beq
\xi_{\rm S}(|{\bf x}_-|) \equiv \int \frac{\d^3 k_{\rm S}}{(2\pi)^3}\, P_\zeta(k_{\rm S}) \,e^{i\k_{\rm S} \cdot {\bf x}_-} \, .
\eeq
The three-point function then is
\begin{align}
\langle \zeta_{\rm S}({\bf x}_1)\zeta_{\rm S}({\bf x}_2) \zeta_{\rm L}({\bf x}_3) \rangle &\ \simeq\ \langle \zeta_{\rm L}({\bf x}_3) \zeta_{\rm L}({\bf x}_+)\rangle [{\bf x}_- \cdot \nabla \xi_{\rm S}(|{\bf x}_-|)]\, , \nonumber \\[4pt]
&\ = \ \int \frac{\d^3 k_{\rm L}}{(2\pi)^3} \int \frac{\d^3 k_{\rm S}}{(2\pi)^3} \, e^{i \k_{\rm L} \cdot (\x_3-\x_+)} P_\zeta(k_{\rm L}) P_\zeta(k_{\rm S}) \left[ {\bf k}_{\rm S} \cdot \frac{\partial}{\partial {\bf k}_{\rm S}}\right] e^{i {\bf k}_{\rm S} \cdot {\bf x}_-}\, .
\end{align}
Integrating by parts, inserting $1 =\int \d^3 k_3\, \delta_D({\bf k}_3 + {\bf k}_{\rm L})$ and using
\beq
\frac{\partial}{\partial {\bf k}_{\rm S}}\cdot [{\bf k}_{\rm S} P_\zeta(k_{\rm S})] = P_\zeta( k_{\rm S}) \, \frac{d \ln(k_{\rm S}^3 P_\zeta(k_{\rm S}))}{d \ln k_{\rm S}}\, ,
\eeq
we get
\begin{align}
\langle \zeta_{\rm S}({\bf x}_1)\zeta_{\rm S}({\bf x}_2) \zeta_{\rm L}({\bf x}_3) \rangle & =  - \int \frac{\d^3 k_3}{(2\pi)^3}
\int \frac{\d^3 k_{\rm L}}{(2\pi)^3} \int \frac{\d^3 k_{\rm S}}{(2\pi)^3}\ e^{- i {\bf k}_1 \cdot {\bf x}_1 - i {\bf k}_{\rm L} \cdot {\bf x}_+ + i {\bf k}_{\rm S} \cdot {\bf x}_-}  \nonumber \\
&\hspace{1cm} \times \left[ (2\pi)^3 \delta_D({\bf k}_1 + {\bf k}_{\rm L}) P_\zeta(k_1) P_\zeta(k_{\rm S}) \, \frac{d \ln(k_{\rm S}^3 P_\zeta(k_{\rm S}))}{d \ln k_{\rm S}}\right] .
\end{align}
Letting ${\bf k}_{\rm L} = {\bf k}_2 + {\bf k}_3$ and ${\bf k}_{\rm S} = \frac{1}{2}({\bf k}_2 -{\bf k}_3)$, we get $ - i {\bf k}_{\rm L} \cdot {\bf x}_+ + i {\bf k}_{\rm S} \cdot {\bf x}_- =  - i {\bf k}_2 \cdot {\bf x}_2 - i {\bf k}_3 \cdot {\bf x}_3$.
Changing variables in the integration and Fourier transforming, we get 
\begin{align}
\lim_{k_3 \to 0} \langle \zeta_{{\bf k}_1} \zeta_{{\bf k}_2} \zeta_{{\bf k}_3} \rangle &\ =\ - (2\pi)^3 \delta_D({\bf k}_1 + {\bf k}_2 + {\bf k}_3)\, P_\zeta(k_1) P_\zeta(k_3)\, \frac{d \ln k_3^3 P_\zeta(k_3)}{d \ln k_3}\, ,  \nonumber \\
&\ = \ (2\pi)^3 \delta_D({\bf k}_1 + {\bf k}_2 + {\bf k}_3) \, (1-n_{\rm s}) \, P_\zeta(k_1) P_\zeta(k_3) \, .
\end{align}
This completes the proof.}
\end{framed}

\noindent
The fact that the squeezed limit of the bispectrum for single-field inflation vanished implies that it can be used as a clean diagnostic for extra fields during inflation. In the next section, we will expand on this view of the squeezed limit as a particle detector.

\newpage
\section{Heavy Relics}
\label{sec:HeavyRelics}

As we have seen in the previous sections, computing the quantum correlations in single-field slow-roll inflation is a well-defined problem leading to clean predictions for the cosmological correlation functions.  In this final section, we will look for deviations from those predictions associated with the existence of extra particles during inflation. 

\vskip 4pt
We will begin, in \S\ref{ssec:massive}, with a review of the allowed spectrum of particles in de Sitter space, highlighting the qualitative differences to the corresponding results in flat space.  In \S\ref{ssec:VR} and \S\ref{ssec:collider}, we derive the basic observational imprints that these massive fields create when they are excited during inflation. 
For further details we refer the reader to the vast literature on the subject, e.g.~\cite{Chen:2009zp, Baumann:2011nk, Chen:2012ge, Assassi:2012zq, Noumi:2012vr, Baumann:2014nda, Arkani-Hamed:2015bza, Lee:2016vti, Bordin:2016ruc, Flauger:2016idt, Gong:2013sma, Baumann:2017jvh}.

\subsection{Massive Fields in de Sitter}
\label{ssec:massive}


Particles in Minkowski space are classified as unitary irreducible representations of the Poincar\'e group~\cite{Wigner:1939cj, Bargmann:1948ck}. 
The eigenvalues of the Casimir operators of the Poincar\'e group are related as follows to the mass $m$ and the spin $s$ of the particles:
\begin{align}
{\cal C}_1 &\equiv P_\mu P^\mu = m^2\, , \\
 {\cal C}_2 &\equiv  W_\mu W^\mu = -s(s+1)\hskip 1pt m^2\, , 
\end{align}
where $P_\mu$ is the four-momentum and $W_\mu$ is the Pauli-Lubanski pseudovector. We distinguish between massive and massless particles. Massive particles carry $2s+1$ degrees of freedom (transverse and longitudinal polarizations), while massless particles have only two transverse polarizations.

\begin{figure}[t!]
    \centering
      \includegraphics[width=.9\textwidth]{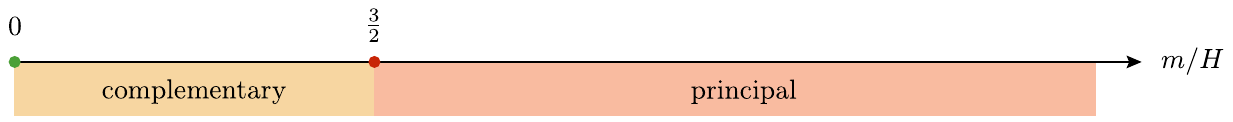}\\[10pt]
         \includegraphics[width=.9\textwidth]{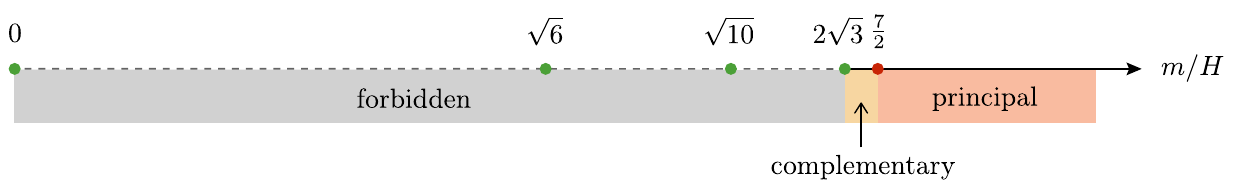}
         \\
    \caption{Spectrum of spin-0 (top) and spin-4 fields (bottom) in de Sitter space. The green points correspond to masses in the discrete series.}
    \label{fig:spectrum}
\end{figure}

\vskip 4pt
Similarly, particles in de Sitter space are classified as unitary irreducible representations of the de Sitter group $SO(1,4)$. 
The Casimir operators of the de Sitter group have the following eigenvalues~\cite{Deser:2003gw, deWit:2002vz}:
\begin{align}
	{\cal C}_{1} &\equiv \frac{1}{2}M_{AB}M^{AB} = m^2-2(s-1)(s+1)H^2 \, , \label{equ:C1} \\
	{\cal C}_{2} &\equiv W_A W^A = -s(s+1)\left[m^2-\left(s^2+s-\frac{1}{2}\right)H^2\right]  ,
\end{align}
where $M_{AB}$ are the generators of $SO(1,4)$, with $A,B = \{0,\ldots, 4\}$. Notice that this time the Casimir (\ref{equ:C1}) is not just proportional to the mass alone, but also has a term that depends on spin. For $s\ge 1$, the representations of de Sitter space fall into three distinct categories~\cite{Thomas, Newton, Garidi:2003ys}: 
\begin{center}
\begin{tabular}{ccc}
{\color{NewRed}principal series} & {\color{NewOrange}complementary series} & {\color{NewGreen}discrete series} \\[4pt]
\ \ $\displaystyle \frac{m^2}{H^2} \ge \left(s-\frac{1}{2}\right)^2$ \quad & \quad $\displaystyle s(s-1) < \frac{m^2}{H^2} < \left(s-\frac{1}{2}\right)^2$ \quad & \quad $\displaystyle \frac{m^2}{H^2} = s(s-1)-t(t+1)$\, , 
\end{tabular}
\end{center}
\vskip 4pt
for $t=0,1,2,...,s\!-\!1$, which is called the depth of the field. Masses that are not associated with one of the above categories correspond to non-unitary representations and are therefore not allowed in the spectrum (see Fig.~\ref{fig:spectrum}). At the specific mass values corresponding to the discrete series, the system gains an additional gauge invariance and some of the lowest helicity modes become pure gauge modes; this phenomenon is called partial masslessness~\cite{Deser:2001pe}. We see that unitarity demands the existence of a lower bound
\begin{align}
	m^2\ge s(s-1)H^2\, ,
\end{align}
on the masses of fields except those that belong to the discrete series. For $s=2$, this is known as the Higuchi bound \cite{Higuchi:1986py}.

\vskip 4pt
The Lagrangian for massive fields with arbitrary spin in flat space was constructed by Singh and Hagen in~\cite{Singh:1974qz, Singh:1974rc}, and generalized to (A)dS spaces in~\cite{Zinoviev:2001dt}. 
For massive fields with spin greater than~2, the action is rather complex and requires introducing auxiliary fields of lower spins. An alternative, which we will follow, is to use a group theoretical approach to find the equations of motion directly~\cite{Deser:2003gw}. In four spacetime dimensions, a massive bosonic spin-$s$ field is described by a totally symmetric rank-$s$ tensor, $\sigma_{\mu_1\cdots\mu_s}$, subject to the constraints
\begin{align}
\nabla^{\mu_1}\sigma_{\mu_1\cdots\mu_s}=0 \, , \quad {\sigma^{\mu}}_{\mu\mu_3\cdots\mu_s}=0\, .\label{spinscond}
\end{align}
The conditions in (\ref{spinscond}) project out the components of the tensor which transform as fields with lower spins. The Casimir eigenvalue equation of the de Sitter group then gives the on-shell wave equation satisfied by these fields:
\begin{align}
\big(\Box - m_s^2\big)\sigma_{\mu_1\cdots\mu_s} = 0\, ,\label{eom}
\end{align}
where $\Box \equiv \nabla_\mu \nabla^\mu$ is the Laplace-Beltrami operator on ${\rm dS}_4$ and $m_s^2 \equiv m^2-(s^2-2s-2)H^2$. The shift in the mass arises from the mismatch between the Casimir and Laplace-Beltrami operators in de Sitter space and is necessary to describe the correct representations for massless fields.
Explicit solutions to the equation of motion (\ref{eom}) can be found in Appendix~A of~\cite{Lee:2016vti}. The spatial components contain all the physical degrees of freedom, while the other components are constrained variables or pure gauge modes.\footnote{This can be seen by solving the equation of motion \eqref{eom} explicitly, where the normalizations of non-physical components become singular~\cite{Lee:2016vti}.} 
The late-time behaviour of the solution is  
\beq
\lim_{\ct \to 0}\sigma_{i_1 \cdots i_s}(\ct,\x) = \sum_\pm \ct^{\Delta_\pm-s} {\sigma}^{\pm}_{i_1 \cdots i_s}(\x)  \, ,
\eeq
where $\Delta_\pm$ is the conformal dimension of the field
\beq
\Delta_\pm = \frac{3}{2} \pm i\mu \, , \quad \mu \equiv \sqrt{ \frac{m^2}{H^2} -\left(s-\frac{1}{2}\right)^2}\, .
\eeq
Note that, for $s=0$, the value $m=0$ corresponds to a conformally-coupled scalar field.
In the limit $m\gg sH$, the parameter $\mu$ becomes the mass of the particle in Hubble units, $\mu \to m/H$.
Particles belonging to the principal series correspond to $\mu\ge 0$, which covers the largest mass range. For real $\mu$, the asymptotic scaling is given by a complex-conjugate pair, resulting in a wavefunction that oscillates logarithmically in conformal time. The complementary series has imaginary $\mu$ and corresponds to the interval $-i\mu = (0,1/2)$. In that case, only the growing mode survives in the late-time limit.

  \begin{figure}[h!]
    \centering
        \hspace{2cm}\includegraphics[width=.67\textwidth]{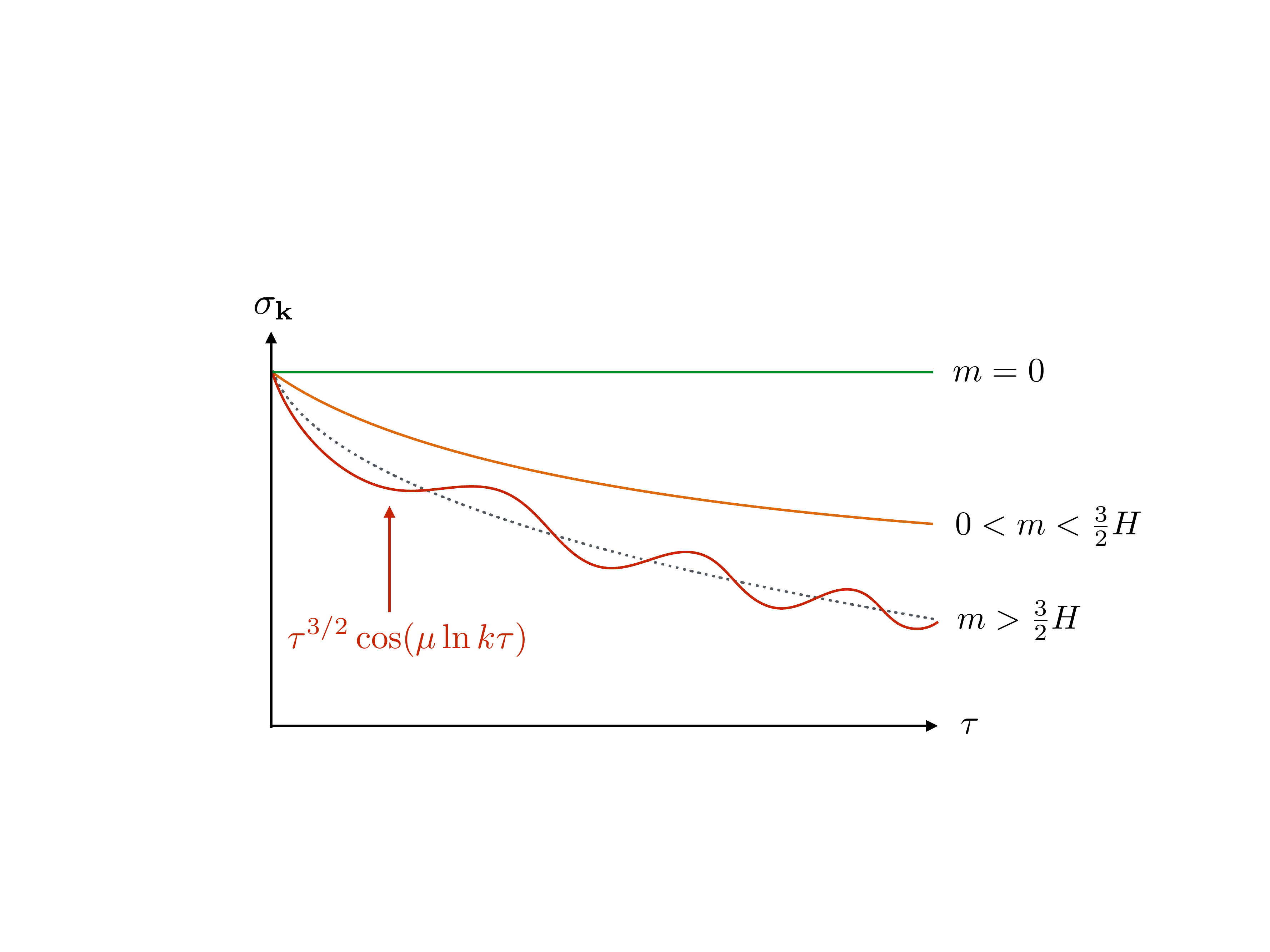}
    \caption{Superhorizon evolution of a massive scalar field. The decay of the field amplitude is determined by the mass of the field.}
    \label{fig:Spin0Evolution}
\end{figure}

\subsection{Local and Non-Local}
\label{ssec:VR}

In the remainder of this section, we will study the imprints of massive particles on inflationary correlations functions.
Massive particles during inflation can have two types of effects: 
\begin{itemize}
\item Particles with masses $M \gg H$ can be integrated out, leading to an EFT of inflaton interactions:
  \begin{figure}[h!]
    \centering
     \includegraphics[width=.7\textwidth]{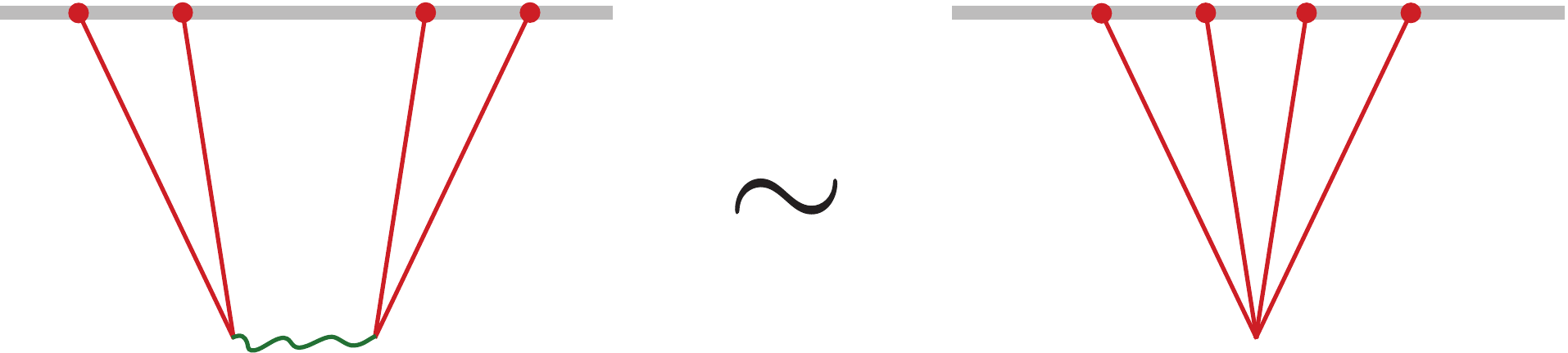}
\end{figure}

\vspace{-0.65cm}
\noindent
These {\it local} interactions create characteristic imprints
in the non-Gaussianity of the inflaton fluctuations. In particular, the soft limits of the resulting correlation functions must satisfy the single-field consistency relation discussed in the previous section.

\item Particles with masses $M \lesssim {\rm few} \times H$ cannot be integrated out completely, but are produced nonperturbatively by the expanding spacetime. When these particles decay, they produce characteristic {\it non-local} signatures in cosmological correlators. 
The single-field consistency relation will be violated.

\end{itemize}

\subsubsection*{Local effects}

To illustrate the local effects arising from massive particles, we consider the following two-field example: 
\beq
{\cal L} \,=\, - \frac{1}{2}(\partial_\mu \phi)^2 - V(\phi) -  \frac{1}{2}(\partial_\mu \sigma)^2 - \frac{1}{2}M^2 \sigma^2 + \frac{\sigma (\partial_\mu \phi)^2}{\tilde \Lambda}\, , \label{equ:LL}
\eeq
where $M \gg H$.
Integrating out the massive field $\sigma$ gives
\begin{align}
{\cal L}_{\rm eff} &\,=\, - \frac{1}{2} (\partial_\mu \phi)^2  - V(\phi) + \frac{1}{8 \tilde \Lambda^2} (\partial_\mu \phi)^2 \, \frac{1}{\Box+M^2}\,(\partial_\mu \phi)^2 + \cdots \nonumber \\
&\,=\, - \frac{1}{2} (\partial_\mu \phi)^2  - V(\phi) +  \frac{1}{8\tilde\Lambda^2 M^2} \left( (\partial_\mu \phi)^4 +  (\partial_\mu \phi)^2\, \frac{\Box}{M^2}\, (\partial_\mu \phi)^2 + \cdots \right) , 
\end{align}
where the second line is written as an expansion in $(H/M)^2$, with $H$ being the Hubble scale during inflation. Truncating the expansion of EFT operators at the lowest order, we obtain the Lagrangian~\cite{Creminelli:2003iq}
\beq
{\cal L}_{\rm eff} = {\cal L}_0(\phi) + \frac{(\partial_\mu \phi)^4}{8 \Lambda^4}\, , \label{equ:Creminelli}
\eeq
where ${\cal L}_0$ is the canonical slow-roll Lagrangian and $\Lambda^2 \equiv \tilde \Lambda M$.  Following~\cite{Creminelli:2003iq}, we will compute the bispectrum induced by the interaction in (\ref{equ:Creminelli}).

\vskip 4pt 
This time it is useful to work in spatially flat gauge.
At leading order in the slow-roll parameters, the action for the inflaton perturbation $\varphi \equiv \phi - \bar \phi(t)$ can then be calculated from (\ref{equ:Creminelli}) without including the contributions from $\sqrt{-g}R$.  The intuitive reason for this simplification is that in the limit of a flat potential the perturbation $\varphi$ does not induce a perturbation in the spacetime curvature.
Evaluating one of the legs of the interaction $(\partial_\mu \phi)^4$ on the background solution, 
we get
\beq
\includegraphicsbox[scale=.5]{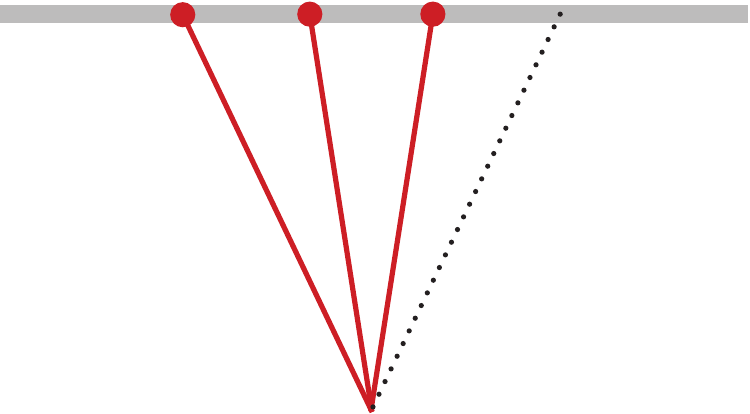} \qquad \Leftrightarrow \qquad {\cal L}_{{\rm eff}, 3} = - \frac{ \phi^{\hskip 1pt \prime}}{4\Lambda^4} \, \varphi' (\partial_\mu \varphi)^2\, .
\eeq

\noindent
Using this interaction, we can compute the bispectrum $\langle \varphi \varphi \varphi \rangle$. We will relate the result to $\langle \zeta \zeta \zeta \rangle$ at horizon crossing to avoid the error that would be induced by the superhorizon evolution of~$\varphi$.
To perform this matching, we, in principle, need to know the relation between $\varphi$ and $\zeta$ up to second order.
However, the quadratic term is slow-roll suppressed, so it suffices to use the linear relationship
\beq
\zeta = - \frac{H}{ \phi^{\hskip 1pt \prime}}\, \varphi\, .
\eeq
Feeding ${\cal H}_{\rm int} = - {\cal L}_{{\rm eff},3}$ into the $in$-$in$ master formula (\ref{equ:InInMaster}), we find
\begin{align}
\frac{{\cal B}_\zeta(k_1,k_2,k_3)}{{\cal P}_\zeta^2} 
&\,=\, - \frac{4i}{k_1k_2k_3} \, \frac{(\phi^\prime)^2}{\Lambda^4}  \int_{-\infty}^0 \d \ct\left(-k_1^2k_2^2k_3^3 \, \ct^2 - (\k_1\cdot \k_2) k_3^2 (1-ik_1\ct)(1-ik_2\ct) \right) e^{ik_t\tau}  \nonumber \\[4pt]
&\qquad + {\rm perms.} + c.c.\, , \\[-1cm]
&\hspace{5.4cm} \uparrow \hspace{3cm} \uparrow \nonumber \\
&\hspace{5.2cm} {(\varphi')^3} \hspace{2cm} (\varphi')(\partial_i \varphi)^2 \nonumber
\end{align}
where $k_t \equiv k_1+k_2+k_3$.
Evaluating the integral, we get~\cite{Creminelli:2003iq}
\beq
\frac{{\cal B}_\zeta(k_1,k_2,k_3)}{{\cal P}_\zeta^2}  = \frac{8}{k_1k_2k_3} \frac{(\phi^\prime)^2}{\Lambda^4} \, \frac{1}{k_t^2} \Bigg(\sum_i k_i^5 + \sum_{i\ne j} (2k_i^4 k_j - 3 k_i^3 k_j^2) + \sum_{i\ne j\ne l} (k_i^3 k_j k_l - 4 k_i^2 k_j^2 k_l)\Bigg)\, . 
\eeq

\vspace{0.25cm}
\noindent
The following features of this result are worth noting:
\begin{itemize}
\item The signal peaks in the equilateral configuration (cf.~Fig.~\ref{fig:equilateral}).
\item 
The squeezed limit, $\lim_{k_{\rm L} \to 0} \langle \zeta_{\k_{\rm S}}  \zeta_{\k_{\rm S}}  \zeta_{\k_{\rm L}}\rangle$, is an analytic function of $k_{\rm L}/k_{\rm S}$.
This is a consequence of the interaction being local.
\item The amplitude is bounded,   
\beq
f_{\rm NL} = \frac{35}{108} \frac{(\phi^\prime)^2}{\Lambda^4} < 1\, ,
\eeq
where the final inequality follows from $\phi^\prime < \Lambda^2$, as required by consistency of the truncation used in (\ref{equ:Creminelli}).  
\end{itemize}

\subsubsection*{Nonlocal effects}

For $M \lesssim {\rm few} \times H$, there are effects of massive particles that cannot be captured by local inflaton interactions, i.e.~the field cannot be integrated out completely. 
The non-locality of the effective inflaton interactions will be reflected in a characteristic non-analyticity in its correlation functions.

\vskip 4pt
As a concrete example, we consider a massive scalar field~$\sigma$.
Its two-point function in de Sitter space is 
\begin{align}
\langle\sigma_\k(\ct)\sigma_{\k'}(\ct')\rangle' = \frac{\pi}{4}H^2(\ct\ct')^{3/2}e^{-\pi\mu}H_{i\mu}(-k\ct)H^*_{i\mu}(-k\ct')\, ,\label{scalar2pt0}
\end{align}
where $H_{i\mu}\equiv H_{i\mu}^{(1)}$ is the Hankel function of the first kind and $\mu\equiv \sqrt{m^2/H^2-9/4}$. 
For now, let us focus on particles belonging to the principal series, so that $\mu$ is real.  
The local part of the two-point function has support only at coincident points in position space, while the non-local part describes correlations over long distances. 
In Fourier space, the local and non-local parts of the two-point function are analytic and non-analytic in the momentum~$k$, respectively. In the late-time limit, we can split (\ref{scalar2pt0}) into its local and non-local parts
\begin{align}
\lim_{\ct, \ct'\to 0}\langle\sigma_\k(\eta)\sigma_{\k'}(\ct')\rangle'_{\rm local}&= \frac{H^2(\ct\ct')^{3/2}}{4\pi}\Gamma(-i\mu)\Gamma(i\mu)\left[e^{\pi\mu}\Big(\frac{\ct}{\ct'}\Big)^{i\mu}+e^{-\pi\mu}\Big(\frac{\ct}{\ct'}\Big)^{-i\mu}\right] ,\label{scalar2ptlocal}\\
\lim_{\ct, \ct'\to 0}\langle\sigma_\k(\ct)\sigma_{\k'}(\ct')\rangle'_{\text{non-local}}&= \frac{H^2(\ct\ct')^{3/2}}{4\pi}\left[\Gamma(-i\mu)^2\Big(\frac{k^2\ct\ct'}{4}\Big)^{i\mu}+\Gamma(i\mu)^2\Big(\frac{k^2\ct\ct'}{4}\Big)^{-i\mu}\right] .\label{scalar2ptnonlocal}
\end{align}
Note that $\Gamma(\pm i\mu) \to e^{-\pi\mu/2}$ for large $\mu$, resulting in an overall suppression of $e^{-\pi\mu}$ of the non-local contribution (\ref{scalar2ptnonlocal}).
In the next section, we will see how the non-analyticity of the non-local part of the two-point function of $\sigma$ is reflected in the inflationary correlators.

\subsection{Cosmological Collider Physics}
\label{ssec:collider}

The correlations in the massive field $\sigma$ become observable when they get converted to inflaton fluctuations through interactions such as the mixing term in (\ref{equ:LL}):
\beq
{\cal L} \,\supset\, \frac{\sigma(\partial_\mu \phi)^2}{\Lambda}\, . \label{equ:Lmix}
\eeq
Evaluating one of the legs of the inflaton field on the background solution, $\bar \phi$, leads to a bispectrum for the inflaton fluctuations $\varphi$ (and hence of the curvature perturbations $\zeta$); cf.~Fig.~\ref{equ:collider}. The derivation of the bispectrum of curvature perturbations can be found in~\cite{Arkani-Hamed:2015bza, Lee:2016vti}. Here, we summarize the main features of the result:

\begin{figure}[t!]
    \centering
     \includegraphics[width=.55\textwidth]{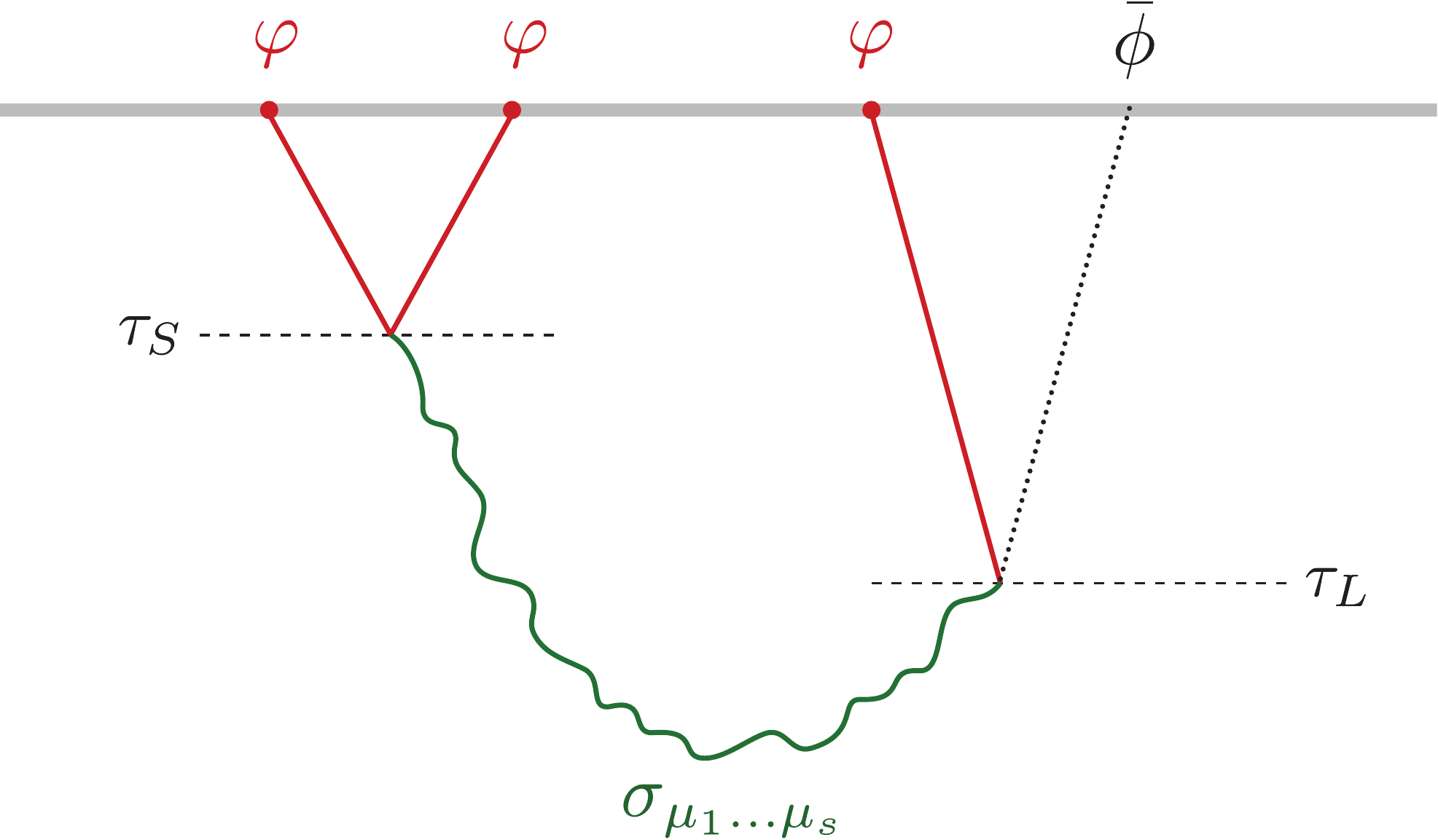}
   \caption{The spontaneous creation and decay of a massive particle $\sigma$ produces a four-point function for the inflaton field $\phi$. Evaluating one of the legs on the background solution, $\bar \phi$, leads to a three-point function for the inflaton fluctuations $\varphi$.}
   \label{equ:collider}
\end{figure}

\begin{figure}[t!]
\centering
 \includegraphics[scale=0.85]{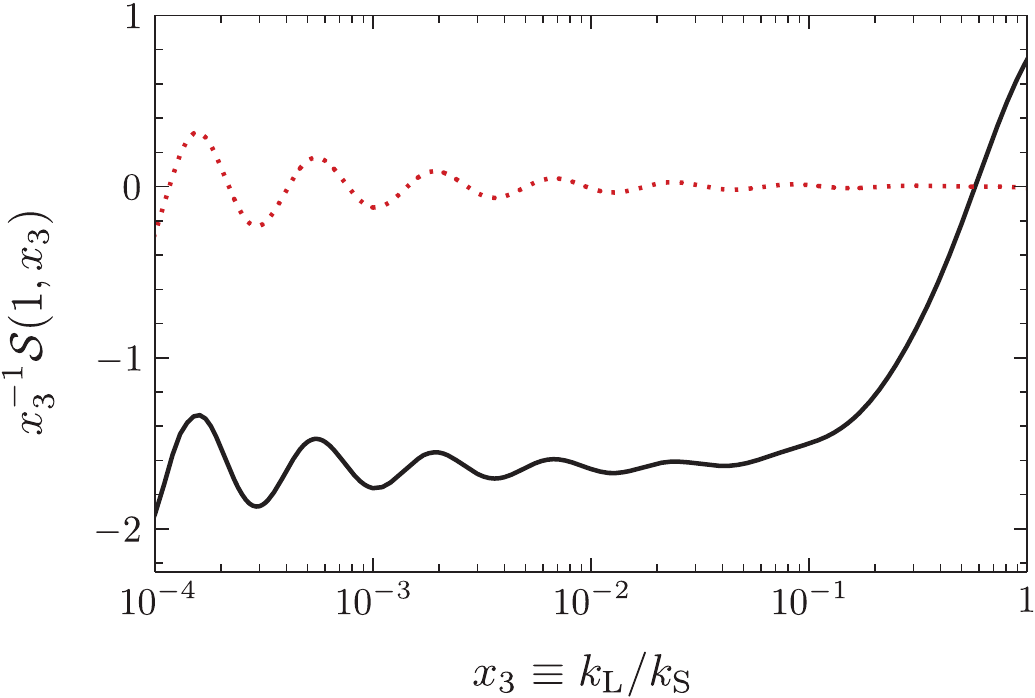}
\caption{Shape function for the bispectrum arising from the exchange of a spin-2 particle with $\mu=5$, evaluated in the isosceles-triangle configuration, $x_2 \equiv k_2/k_1=1$ (figure adapted from~\cite{Lee:2016vti}). The dotted line is the non-local part of the signal. }
\label{fig:spectroscopy}
\end{figure}
\begin{figure}[t!]
\centering
\hspace{1.6cm} \includegraphics[scale=0.87]{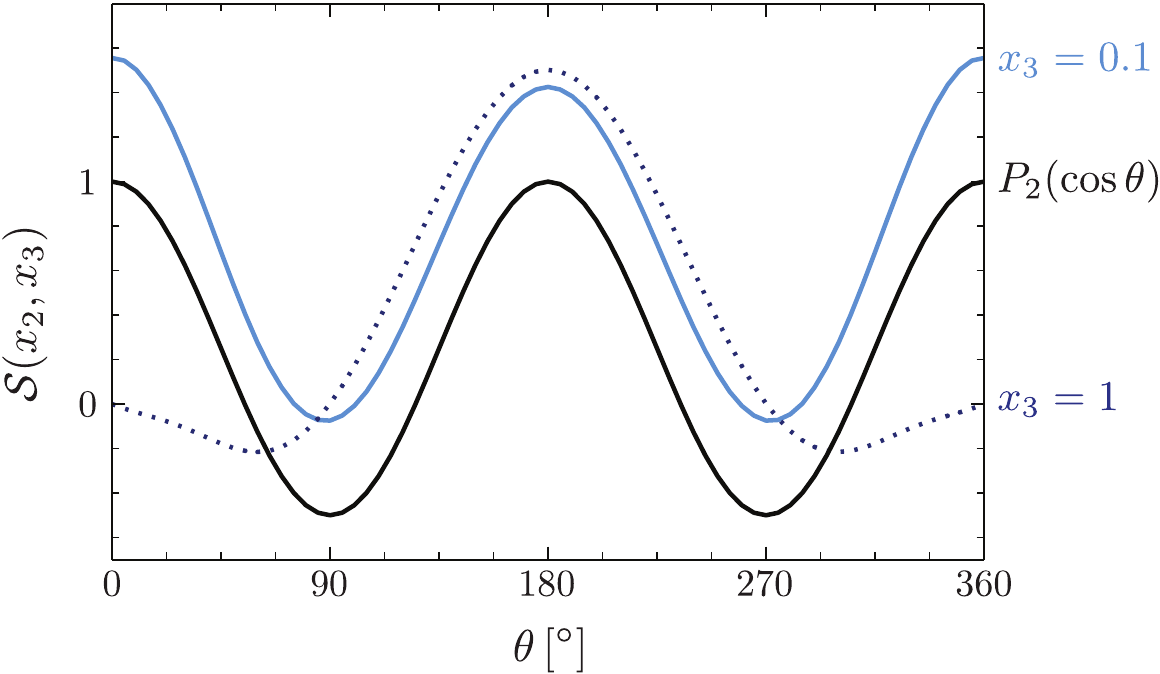}
\caption{Shape function for the bispectrum arising from the exchange of a spin-2 particle with $\mu=5$, as a function of the base angle $\theta=\cos^{-1}(\hat\k_1\cdot\hat\k_3)$ for fixed ratios of $x_3 \equiv k_3/k_1$ (figure adapted from~\cite{Lee:2016vti}). }
\label{fig:spin}
\end{figure}

\begin{itemize}
\item The non-local effect associated with the massive particle exchange shows up as a non-analyticity in the squeezed limit, $\lim_{k_{\rm L} \to 0} \langle \zeta_{\k_{\rm S}}\zeta_{-\k_{\rm S}} \zeta_{\k_{\rm L}} \rangle$.  Between the horizon crossing times of the long mode $\zeta_{\rm L}$ and the short modes $\zeta_{\rm S}$ the amplitude of the massive field oscillates with a frequency set by the mass of the field.   This leads to distinct oscillations in the bispectrum of curvature perturbations:
\beq
\lim_{k_{\rm L} \to 0} \langle \zeta_{\k_{\rm S}}\zeta_{-\k_{\rm S}} \zeta_{\k_{\rm L}} \rangle \,\propto\, \left(\frac{k_{\rm L}}{k_{\rm S}}\right)^{3/2}\cos\left[\frac{M}{H} \ln\left(\frac{k_{\rm L}}{k_{\rm S}}\right) + \delta \right] ,
\eeq
where $\delta$ is a computable phase~\cite{Arkani-Hamed:2015bza}. We see that the mass $M$ of the field is encoded in the frequency of the oscillations (see Fig.~\ref{fig:spectroscopy}).
\item Particles with spin $s$ lead to a unique angular dependence (see Fig.~\ref{fig:spin}):
\beq
\lim_{k_{\rm L} \to 0} \langle \zeta_{\k_{\rm S}}\zeta_{-\k_{\rm S}} \zeta_{\k_{\rm L}} \rangle \,\propto\,  P_s(\cos\theta)\, ,
\eeq
where $\cos \theta \equiv \k_{\rm S} \cdot \k_{\rm L}$ and $P_s$ is a Legendre polynomial of degree $s$. 
\end{itemize}

\noindent
There are several effects that determine the amplitude of the signal:
\begin{itemize}
\item If we make the conservative assumption that the interaction in (\ref{equ:Lmix}) is only of gravitational strength~\cite{Arkani-Hamed:2015bza}, then the scale determining the mixing between $\sigma$ and $\phi$ is~$\Lambda = M_{\rm pl}$. In that case, the amplitude of the bispectrum is at most of order the gravitational floor, $f_{\rm NL} \lesssim {\cal O}(\varepsilon)$.   However, in principle, the mixing can be much larger. In particular, the mixing interaction in (\ref{equ:Lmix}) remains perturbative as long as $\Lambda \ge (\phi')^{1/2}$, allowing for non-Gaussianity of order $f_{\rm NL} \lesssim {\cal O}(1)$~\cite{Chen:2009zp, Baumann:2011nk, Assassi:2013gxa, Lee:2016vti}.  Finally, in the EFT for the inflationary fluctuations~\cite{Cheung:2007st} the cutoff $\Lambda$ can even be below $(\phi')^{1/2}$ and the non-Gaussianity can be much larger, $f_{\rm NL} \lesssim {\cal O}({\cal P}_\zeta^{-1/2})$.

\item In addition, the spontaneous production of massive particles in the principal series is exponentially suppressed. This Boltzmann suppression is inherited by the amplitude of the bispectrum, $f_{\rm NL} \propto e^{-\pi M/H}$.
The creation of particles in the complementary series, on the other hand, does not receive the same suppression and the signal can be large~\cite{Chen:2009zp, Baumann:2011nk}. Instead of oscillations, the squeezed limit in that case has a monotonic scaling $(k_{\rm L}/k_{\rm S})^\Delta$, where the scaling dimension $\Delta $ carries the information about the mass of the particle.

\item For particles in the principal series, the squeezed limit is suppressed by a factor of $(k_{\rm L}/k_{\rm S})^{3/2}$. This is to be compared to the squeezed limit for equilateral non-Gaussianity which scales as $(k_{\rm L}/k_{\rm S})^{2}$. The signal from particles in the complementary series can be larger in the squeezed limit, $(k_{\rm L}/k_{\rm S})^\Delta$, where $0<\Delta < 3/2$. Finally, partially massless particles do not decay on superhorizon scales and may therefore induce signals that are not suppressed in the squeezed limit~\cite{Baumann:2017jvh}.
 \end{itemize}
 
\begin{figure}[h!]
\centering
\quad \includegraphics[scale=0.9]{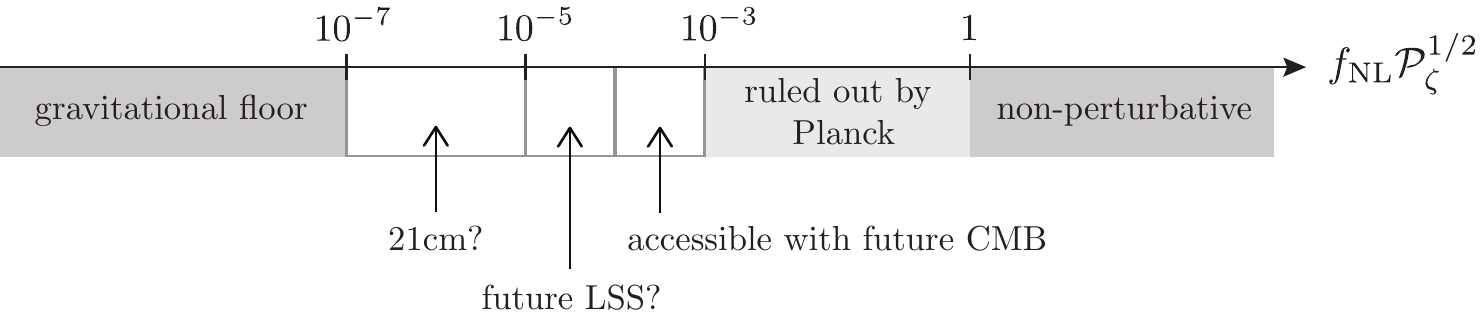}
\caption{\label{fig:fNLconstraints} Schematic illustration of current and future constraints on (scale-invariant) primordial non-Gaussianity. The ``gravitational floor" denotes the minimal level of non-Gaussianity created by purely gravitational interactions during inflation~\cite{Maldacena:2002vr} (see \S\ref{ssec:floor}). }
\end{figure}

 \noindent
Figure~\ref{fig:fNLconstraints} is a schematic illustration of current and future constraints on (scale-invariant) primordial non-Gaussianities. We see that the perturbatively interesting regime spans about seven orders of magnitude in $f_{\rm NL}$. Of this regime, three orders of magnitude have been ruled out by current CMB observations, leaving a window of opportunity of about four orders of magnitude to be explored. Accessing these low levels of non-Gaussianity will be challenging.  Even optimistic projections for future CMB observations won't reduce the constraints by more than an order of magnitude. Digging deeper will require new cosmological probes, such as observations of the large-scale structure of the universe~\cite{Alvarez:2014vva} and the tomography of the 21cm transition of neutral hydrogen gas~\cite{Loeb:2003ya}. 

\newpage
\addcontentsline{toc}{section}{References}
\bibliographystyle{utphys}
\bibliography{TASI-Refs} 

\end{document}